\documentclass[a4paper,11pt]{article}
\pdfoutput=1
\usepackage{jheppub}

\usepackage{graphicx}
\usepackage[figuresright]{rotating}
\usepackage{bm,amsmath,amssymb}
\usepackage[mathscr]{eucal}
\usepackage{multirow}
\usepackage{xcolor}
\usepackage{mathrsfs}
\usepackage{mathtools}
\usepackage{slashed}
\usepackage{graphics}
\usepackage{graphicx}
\usepackage{subfigure}
\usepackage{dsfont}
\usepackage{longtable}
\usepackage{bbm} 
\usepackage{float}
\usepackage{tcolorbox}
\usepackage{lipsum}
\usepackage{cancel}
\usepackage{enumerate}
\usepackage{makecell}

\definecolor{lcolor}{rgb}{0.,0.0,0.}
\definecolor{citcolor}{rgb}{0,0.,0.5}

\newcommand{\Hcal}{\mathcal{H}}

\newcommand{\Ccal}{\mathcal{C}}
\newcommand{\Pcal}{\mathcal{P}}
\newcommand{\Rcal}{\mathcal{R}}

\newcommand{\Jcal}{\mathcal{J}}
\newcommand{\vect}[1]{\boldsymbol{#1}_{\perp}}
\newcommand{\kt}{\vect{k}}
\newcommand{\kgt}{\boldsymbol{k_{g\perp}}}

\newcommand{\Pt}{\vect{P}}

\newcommand{\qt}{\vect{q}}
\newcommand{\lt}{\vect{l}}

\newcommand{\at}{\vect{a}}
\newcommand{\bt}{\vect{b}}
\newcommand{\Kt}{\vect{K}}

\newcommand{\ktone}{\boldsymbol{k_{1\perp}}}
\newcommand{\kttwo}{\boldsymbol{k_{2\perp}}}

\newcommand{\xt}{\vect{x}}
\newcommand{\yt}{\vect{y}}
\newcommand{\zt}{\vect{z}}
\newcommand{\ut}{\vect{u}}

\newcommand{\Xt}{\vect{X}}

\newcommand{\rt}{\vect{r}}

\newcommand{\rxyt}{\boldsymbol{r}_{xy}}

\newcommand{\rzxt}{\boldsymbol{r}_{zx}}
\newcommand{\rzyt}{\boldsymbol{r}_{zy}}

\newcommand{\rzzpt}{\boldsymbol{r}_{zz'}}

\newcommand{\rzytp}{\boldsymbol{r}_{z'y'}}

\newcommand{\rxytp}{\boldsymbol{r}_{x'y'}}
\newcommand{\rxxtp}{\boldsymbol{r}_{xx'}}
\newcommand{\ryytp}{\boldsymbol{r}_{yy'}}
\newcommand{\rxypt}{\boldsymbol{r}_{xy'}}
\newcommand{\ryxpt}{\boldsymbol{r}_{yx'}}
\newcommand{\rzpxt}{\boldsymbol{r}_{z'x}}
\newcommand{\rzpyt}{\boldsymbol{r}_{z'y}}

\newcommand{\QV}{\bar{Q}_{\rm V3}}

\newcommand{\rzxpt}{\boldsymbol{r}_{zx'}}
\newcommand{\rzxtp}{\boldsymbol{r}_{z'x'}}
\newcommand{\rxpyt}{\boldsymbol{r}_{x'y}}
\newcommand{\rwytp}{\boldsymbol{r}_{w'y'}}

\newcommand{\rwbxtp}{\boldsymbol{r}_{\bar{w}'x'}}
\newcommand{\rxwbtp}{\boldsymbol{r}_{x'\bar{w}'}}

\newcommand{\RtS}{\boldsymbol{R}_{\rm SE1}}
\newcommand{\RtV}{\boldsymbol{R}_{\rm V}}

\newcommand{\RtR}{\boldsymbol{R}_{\rm R}}
\newcommand{\RtRb}{\overline{\boldsymbol{R}}_{\rm R}}

\newcommand{\der}{\mathrm{d}}

\newcommand{\Tr}{\mathrm{Tr}}

\title{Semi-inclusive single-jet production in DIS at next-to-leading order in the Color Glass Condensate}
\author[a]{Paul Caucal,}
\emailAdd{caucal@subatech.in2p3.fr}
\author[a,b]{Elouan Ferrand,}
\emailAdd{elouan.ferrand@universite-paris-saclay.fr}
\author[c,d,e,f,g]{Farid Salazar }
\emailAdd{faridsal@uw.edu}

 \affiliation[a]{SUBATECH UMR 6457 (IMT Atlantique, Université de Nantes, IN2P3/CNRS), 4 rue Alfred Kastler, 44307 Nantes, France}
\affiliation[b]{Université Paris-Saclay, Faculté des sciences d'Orsay, Magistère de Physique Fondamentale d'Orsay, Orsay, 91405, France}
\affiliation[c]{Institute for Nuclear Theory, University of Washington, Seattle WA 98195-1550, USA}
\affiliation[d]{Nuclear Science Division, Lawrence Berkeley National Laboratory, Berkeley, California 94720, USA}
\affiliation[e]{Physics Department, University of California, Berkeley, California 94720, USA}
\affiliation[f]{Department of Physics and Astronomy, University of California, Los Angeles, California 90095, USA}
\affiliation[g]{Mani L. Bhaumik Institute for Theoretical Physics, University of California, Los Angeles, California 90095, USA}

\abstract{Within the Color Glass Condensate (CGC) effective field theory, we derive the next-to-leading order (NLO) cross-section for the single-jet semi-inclusive cross-section in deep inelastic scattering (DIS) at small $x$, for both longitudinally and transversely polarized virtual photons. We provide analytic expressions, valid at finite $N_c$ and suitable for numerical evaluation, for both the cross-section differential in rapidity and transverse momentum and the cross-section differential in rapidity only. Our NLO formulae demonstrate that the very forward rapidity regime is plagued by large double logarithmic corrections coming from phase space constraints on soft gluons close to the kinematic threshold for jet production. A joint resummation of small-$x$ and threshold logarithms at single logarithmic accuracy is proposed to remedy the instability of the cross-section in this regime. By integrating over the single-jet phase space, we recover known results for the NLO DIS structure functions at small $x$, previously obtained using the optical theorem.}

\begin{document}
\maketitle
\newpage 

\section{Introduction}

The dipole picture~\cite{Kopeliovich:1981pz, Bertsch:1981py, Mueller:1989st,Nikolaev:1990ja} of Deep Inelastic Scattering (DIS) of electron off protons or large nuclei has emerged as a natural framework to study the high energy or Regge limit \cite{Regge:1959mz} of this process where the Bjorken variable $x_{\rm Bj}\sim Q^2 /s$  goes to 0, for fixed $Q^2$ and large squared center-of-mass energies $s$. In this regime, the fast increase in the number of gluons populating the proton or nucleus incoming wave-function is tamed by non-linear recombination effects, a phenomenon known as gluon saturation \cite{Gribov:1984tu,Mueller:1985wy}. This phenomenon is characterized by an emergent semi-hard transverse momentum scale, the saturation scale $Q_s$, which grows with energy \cite{Mueller:2002zm,Iancu:2002tr,Munier:2003vc} and the nucleus mass number~\cite{McLerran:1993ka,McLerran:1993ni,McLerran:1998nk}. Unraveling the impact of gluon saturation in high energy DIS is one of the primary objectives of the future Electron-Ion Collider \cite{Accardi:2012qut,AbdulKhalek:2021gbh}. In the color dipole picture at leading order in perturbative QCD (pQCD), the ``incoming" virtual photon splits into a quark-antiquark pair which subsequently probes the dense gluon system of the nucleus, which is effectively treated as a highly occupied classical gluon field generated by the large $x$ valence partons. 

Advancing the understanding of the dipole picture of DIS in the high energy limit has been the subject of intense investigations over the past several years. These developments have followed multiple directions, with a special emphasis on the calculation of higher order quantum corrections for several DIS processes such as the total cross-section (structure functions \cite{Balitsky:2007feb,Beuf:2011xd,Beuf:2016wdz,Beuf:2017bpd,Hanninen:2017ddy,Beuf:2021qqa,Beuf:2021srj,Beuf:2022ndu}, including the diffractive ones \cite{Beuf:2022kyp}), exclusive vector meson production \cite{Boussarie:2016bkq,Mantysaari:2021ryb,Mantysaari:2022bsp}, exclusive dijet/dihadron production \cite{Boussarie:2016ogo,Fucilla:2022wcg}, inclusive dijet plus photon production \cite{Roy:2019cux,Roy:2019hwr}, inclusive quarkonium plus gluon production \cite{Kang:2023doo}, and the inclusive dijet/dihadron cross-section \cite{Caucal:2021ent,Bergabo:2022tcu,Bergabo:2023wed,Taels:2022tza,Iancu:2022gpw}. The seminal work in \cite{Dominguez:2011wm} established the connection between the dipole picture and transverse momentum-dependent factorization framework, which has resulted in several ongoing efforts to elucidate further this correspondence~\cite{Mueller:2012uf,Mueller:2013wwa,Balitsky:2015qba,Xiao:2017yya,Altinoluk:2019fui,Altinoluk:2019wyu,Altinoluk:2020qet,Boussarie:2020vzf,Iancu:2021rup,Iancu:2022lcw,Caucal:2022ulg,Caucal:2023nci,Caucal:2023fsf,Hatta:2022lzj,Altinoluk:2023qfr,Mukherjee:2023snp}. 

In this paper, we compute the next-to-leading order (NLO) corrections to the semi-inclusive single-jet production cross-section (SIDIS) in the high energy limit of DIS\footnote{Although the single-jet cross-section computed in this paper is inclusive over the final state particles,  occasionally this process is referred to in the literature as ``semi-inclusive". We shall use inclusive and semi-inclusive interchangeably.}. We perform the calculation in the color dipole picture and within the Color Glass Condensate effective field theory (CGC EFT)~\cite{McLerran:2001sr,Iancu:2002xk,Gelis:2010nm,Gelis:2012ri,Morreale:2021pnn}. Our starting point for the calculation of the NLO corrections to semi-inclusive single-jet production in DIS is the analytic expressions for inclusive dijet production in DIS obtained in \cite{Caucal:2021ent}. At NLO, the quark-antiquark dipole is accompanied by either a real or virtual gluon. To obtain the SIDIS cross-section, one must integrate over the unmeasured partons in the final state. After integration, the resulting expression for a given NLO diagram may display new ultraviolet divergences in coordinate space. One of the main objectives of this work is to show that these divergences cancel out amongst each other and to provide analytic expressions for the NLO impact factor. We have computed the cross-section differential in jet transverse momentum $\kt$ and jet rapidity $\eta$, and for the cross-section differential in $\eta$ only. The latter also involves the appearance of new UV divergences that emerge after integration over $\kt$, which must be treated with care to demonstrate the finiteness of the cross-section.

Our work has been motivated on several fronts. The first one is related to the connection with TMD factorization. The SIDIS process computed at leading order in the dipole picture is known to admit a factorization in terms of the quark TMD distribution at small $x$ \cite{McLerran:1998nk,Venugopalan:1999wu,Mueller:1999wm} in the regime $Q^2\gg \kt^2$. Remarkably, this factorization holds both in the dilute regime where $\kt^2\gg Q_s^2$ and in the non-linear regime with $\kt^2\sim Q_s^2$ \cite{Marquet:2009ca,Zhou:2013gsa}. This factorization framework has been recently employed in phenomenological studies of lepton-jet correlations \cite{Tong:2022zwp,Tong:2023bus} and transverse energy-energy correlators \cite{Kang:2023oqj}. To elucidate these studies one must determine if TMD factorization persists at NLO. The first step to address this question is to perform the full NLO computation in the CGC EFT, and then examine the TMD limit $\kt^2, Q_s^2  \ll Q^2$. 

Another motivation is to assess the sensitivity of this process to gluon saturation effects. Recently, it has been shown in \cite{Iancu:2020jch} that the very forward jet regime has great potential to reveal saturation effects at large $Q^2$ where non-perturbative contamination is expected to be suppressed. Let us briefly review the physical explanation behind this enhanced sensitivity in the forward regime. The transverse size of the quark-antiquark pair is determined by the scale $1/\bar Q$, with $\bar Q^2=z(1-z)Q^2$ and $z=k^-/q^-$ the relative light-cone longitudinal momentum of the jet with respect to that of the virtual photon. Thus, even at large $Q^2$, the size of the dipole can be comparable to $1/Q_s$ if $z$ is sufficiently close to 1. For $1-z\lesssim Q_s^2/Q^2\ll 1$, the dipole undergoes strong scattering and therefore probes the gluon-saturated wave-function of the nucleus. In terms of concrete observables, the onset of saturation and high energy evolution is signaled by the disappearance of the Cronin-like peak~\cite{Antreasyan:1978cw,Jalilian-Marian:2003rmm} and the suppression in the nuclear modification factor for the SIDIS cross-section differential in $\kt$ for fixed $\bar Q^2\ll Q^2$ as $x_{\rm Bj}$ gets smaller (or as the energy of the collision increases) \cite{Iancu:2020jch}. The nuclear modification factor for the $\kt$-integrated SIDIS cross-section differential in $z$ also undergoes a strong suppression for $z$ close to 1 \cite{Iancu:2020jch}. 

Based on these promising phenomenological results at leading order in the dipole picture, we would like to address the impact of QCD radiative corrections. Owing to our analytic expressions for the NLO corrections to SIDIS, we find large double and single logarithms of $1-z$ in the forward rapidity regime, even in the case of a jet measurement, contrary to what is claimed in~\cite{Iancu:2020jch}. The origin of these large logarithms is clear, as also stated in \cite{Iancu:2020jch}: they arise from the mismatch between real and virtual soft gluon corrections when reaching the edge of the real emission phase space. We shall call these logarithms ``threshold logarithms" as they come from soft gluons which are strongly suppressed near the kinematic threshold $z=1$. This is in direct analogy with the large logarithms which arise in very forward hadron production in $pA$ collisions \cite{Xiao:2018zxf,Liu:2020mpy,Shi:2021hwx, Wang:2022zdu,Liu:2022ijp} when the longitudinal momentum fraction of the produced parton with respect to the incoming one approaches unity. 

The coefficients of the double and single logarithms are computed explicitly. In particular, the dominant double logarithmic correction $-\alpha_sC_F/(2\pi)\ln^2(1-z)$ is negative, which results in the suppression of the cross-section in the forward rapidity regime, similar to the suppression induced by gluon saturation. This is reminiscent of the competing effect between saturation and Sudakov suppression for inclusive back-to-back dijet production in DIS \cite{Mueller:2012uf,Mueller:2013wwa,Caucal:2022ulg,Taels:2022tza}. Yet, soft gluon radiation for SIDIS at very forward rapidities should be less unfavorable than for inclusive back-to-back dijet in regards to distinguishing it from saturation effects. Indeed, by exponentiating the threshold double logarithms, one realizes that the Sudakov suppression for $z$ close to one completely factorizes and is independent of the nuclear target (contrary to Sudakov suppression for back-to-back dijet which is a convolution in coordinate space). Consequently, we expect the exact cancellation (at least to double logarithmic accuracy) of the Sudakov suppression in a nuclear modification factor, leaving only the imprint of saturation in such a ratio.

The NLO corrections to semi-inclusive single hadron production cross-section at small $x$ have been computed in \cite{Bergabo:2022zhe}, while the NLO impact factor for \textit{diffractive} single hadron production is computed in \cite{Fucilla:2023mkl}. Our present work differs from these studies in several aspects. First, we consider single \textit{jet} production instead of single hadron production. While jet measurements will be challenging at the EIC~\cite{Dumitru:2014vka,Dumitru:2018kuw,Zhao:2021kae,Mantysaari:2019hkq,Boussarie:2021ybe,vanHameren:2021sqc}, jet cross-sections are theoretically more robust as they do not suffer from the uncertainties introduced by fragmentation functions. Unlike \cite{Bergabo:2022zhe}, we explicitly regulate and isolate divergences in the transverse space using dimensional regularization. This method also allows us to calculate explicitly the NLO impact for both photon polarizations (longitudinal and transverse). Furthermore, we work at finite $N_c$ and include the contributions where the tagged jet is sourced by a gluon. Hence, most of the results presented in this paper are new. One of the surprises lies in the $1/N_c^2$ suppressed corrections to the NLO impact factor: we find that they depend on the quadrupole CGC correlator (CGC average of a trace of four Wilson lines). This dependence is caused by the color exchange in real and virtual final state gluon emissions between the quark and the antiquark and is not present in the diffractive case \cite{Fucilla:2023mkl}. It would have not been possible to anticipate the appearance of the quadrupole in the NLO impact factor based on the high-energy evolution of the LO cross-section alone, as the latter only involves dipole correlators. 

By integrating the real and virtual cross-section for inclusive dijet production over the full phase space (3-body phase space in the case of the real cross-section), we recover the NLO corrections to the total DIS cross-section at small $x$ previously obtained in \cite{Beuf:2017bpd,Hanninen:2017ddy,Balitsky:2007feb} using the optical theorem. 
We observe subtle cancellations among diagrams in covariant perturbation theory that are necessary to obtain the expressions in \cite{Hanninen:2017ddy}. These cancellations rely on properly identifying the physical impact parameter vector of the virtual photon in the dipole frame. These results are important consistency checks of our formulae in this paper and the results in \cite{Caucal:2021ent} and of the validity of the dipole picture of DIS 
within the CGC EFT at NLO. 

Our paper is divided as follows. In section~\ref{sec:LO-review}, we define the semi-inclusive single-jet cross-section and review the calculation of this process in the dipole picture at leading order, starting from the inclusive dijet case in the CGC EFT. Section \ref{sec:NLO-calc} details the calculation of the NLO diagrams contributing to the SIDIS cross-section. We further show that the cross-section is finite and that its high energy evolution is given by the BK-JIMWLK equation. Our final results for the NLO impact factor are presented in subsection \ref{sub:final-dblediff}. We demonstrate the emergence of large threshold logs in the very forward jet limit in section \ref{sec:discuss}. We present the results for rapidity-only differential jet cross-section in section~\ref{sec:eta-diff}. Finally, we recover the fully inclusive DIS cross-section in section~\ref{sec:total-xs}. 

The paper is supplemented by four appendices. Appendix~\ref{app:transverse} gathers the analytic formula for the SIDIS cross-section for transversely polarized virtual photons. In appendix~\ref{app:single-log}, we provide explicit expressions for the single threshold logarithms. Appendix~\ref{app:useful-id} includes useful mathematical identities in distribution theory. Finally,  appendix~\ref{app:final-state-cancellation} presents a short proof of the cancellation of the final state gluon emissions in fully inclusive DIS.

\section{Review of the SIDIS cross-section at leading order}
\label{sec:LO-review}

This first section is a brief derivation of the semi-inclusive single-jet cross-section at leading order in the dipole picture, starting from the LO expression for inclusive dijet production in the CGC. We fix our notations and properly define the semi-inclusive single-jet cross-section.

\subsection{Dijet cross-section at leading order}

The calculation of the inclusive \textit{dijet} cross-section in DIS at leading order in pQCD is now a textbook exercise \cite{Gelis:2002nn,Dominguez:2011wm,Mantysaari:2019hkq}. For a detailed derivation within the CGC effective field theory and in standard covariant perturbation theory, we refer the reader to \cite{Caucal:2021ent}. In the dipole frame, the virtual photon with polarization $\lambda$ has a four-momentum $q^\mu=(-Q^2/(2q^-),q^-,\boldsymbol{0}_\perp)$ in light cone coordinates, with $Q^2=-q^2$ the spacelike virtuality of the photon. The nucleus or proton target has instead only a large $P^+$ component, such that its four-momentum reads $P^\mu=(P^+,0,\boldsymbol{0}_\perp)$, where we neglected the nucleus mass. We denote $\ktone$ and $\kttwo$ as the transverse momenta of the quark and antiquark jets $j_1$ and $j_2$, and $\eta_1$, $\eta_2$ are their rapidities, which are related to the longitudinal momentum fraction $z_i= k_i^-/q^-$ with respect to the virtual photon as
\begin{equation}
    \eta_i=\ln\left(\sqrt{2}z_iq^-/k_{i\perp}\right)\,,\quad i=1,2 \,.
\end{equation}
With these notations, the hadronic component of the inclusive dijet cross-section at leading order in $\alpha_s$ reads
\begin{align}
    \left.\frac{\der \sigma^{\gamma_{\lambda}^{\star}+A\to j_1j_2+X}}{ \der^2 \ktone \der^2 \kttwo \der \eta_1 \der \eta_{2}}\right|_{\rm LO}  \!\!\!\!\!\!\! &=  \frac{\alpha_{\mathrm{em}} e_f^2 N_c}{(2\pi)^6}\delta(1-z_1-z_2)\nonumber\\
    &\times\int \der^2\xt\der^2\yt\der^2\xt'\der^2\yt' \ e^{-i\ktone\cdot\rxxtp}e^{-i\kttwo\cdot\ryytp}\Rcal_{\mathrm{LO}}^{\lambda}(\rxyt,\rxytp)\nonumber\\
    &\times\left\langle1-D_{xy}-D_{y'x'}+Q_{xy;y'x'}\right\rangle \label{eq:dijet-LO-cross-section} \,.
\end{align}
In this equation, $\alpha_{\rm em}$ refers to the fine structure constant, $e_f^2$ is the sum of the squares of the fractional electric charge of the light quarks, and the perturbative factor $\Rcal$, associated with the decay of the virtual photon into the quark-antiquark pair, is defined to be 
\begin{align}
    \Rcal_{\mathrm{LO}}^{\mathrm{L}}(\rxyt,\rxytp) &=  8 z_1^3 z_2^3  Q^2 K_0\left(\sqrt{z_1z_2}Q r_{xy}\right) K_0\left(\sqrt{z_1z_2}Q r_{x'y'}\right) \,, \label{eq:dijet-NLO-LLO} \\
    \Rcal_{\mathrm{LO}}^{\mathrm{T}}(\rxyt,\rxytp) &=  2 z_1^2z_2^2 \left[z_1^2 +z_2^2 \right]  \frac{\rxyt \cdot \rxytp}{r_{xy} r_{x'y'}}  Q^2K_1\left(\sqrt{z_1z_2}Q r_{xy}\right) K_1\left(\sqrt{z_1z_2}Qr_{x'y'}\right)\label{eq:dijet-NLO-TLO} \,.
\end{align}
Throughout this paper, we shall use the following notation for the difference of any transverse vectors $\at$ and $\bt$:
\begin{equation}
    \boldsymbol{r}_{ab}\equiv\at -\bt\,.
\end{equation}
The CGC color correlator in the last line of Eq.\,\eqref{eq:dijet-LO-cross-section} encodes the interaction of the color dipole with the CGC ``shockwave", where the average $\langle ...\rangle$ over stochastic large $x$ color sources $\rho$ with weight functional $\mathcal{W}_{Y}[\rho]$ is taken at some rapidity scale $Y$ which is arbitrary at this perturbative order. It depends both on the dipole and on the quadrupole, defined as the trace of the product of two and four Wilson lines:
\begin{align}
    D_{xy}&\equiv\frac{1}{N_c}\textrm{Tr}\left(V(\xt)V^\dagger(\yt)\right) \,,\\
    Q_{xyy'x'}&\equiv \frac{1}{N_c}\textrm{Tr}\left(V(\xt)V^\dagger(\yt)V(\yt')V^\dagger(\xt')\right) \,.
\end{align}

\subsection{Semi-inclusive single-jet cross-section at LO}

We define the semi-inclusive single-jet cross-section in terms of the inclusive cross-section for producing $N$ jets 
\begin{equation}
    \der \sigma^{\gamma_{\lambda}^{\star}+A\to j_1j_2,...,j_N+X} \,,
\end{equation}
as
\begin{align}
     \frac{\der \sigma^{\gamma_{\lambda}^{\star}+A\to j+X}}{\der^2 \kt\der\eta}=\sum_{N=2}^{\infty}&\int\der^2\ktone...\der^2\boldsymbol{k}_{N\perp}\int\der\eta_1...\der \eta_N \  \der \sigma^{\gamma_{\lambda}^{\star}+A\to j_1,...,j_N+X}\nonumber\\
     &\times F_N\left[\kt,\eta; \ktone,...,\boldsymbol{k}_{N\perp},\eta_1,...,\eta_N\right] \,, 
\end{align}
where $F_N$ is a measurement function whose standard definition reads
\begin{align}
    F_N\left[\kt,\eta;\ktone,...,\boldsymbol{k}_{\perp,N},\eta_1,...,\eta_N\right]=\sum_{i=1}^{N}\delta^{(2)}(\kt-\boldsymbol{k}_{i\perp})\delta(\eta-\eta_i) \,.
\end{align}
With this definition, each jet in the final state is accounted for in the semi-inclusive single-jet cross-section. Note that this definition violates unitarity, namely the integral over $\kt$ and $\eta$ of the single-jet semi-inclusive cross-section does not yield the total inclusive cross-section since a given event may be counted multiple times. For alternative choices of measurement functions $F_N$ which satisfy unitarity, see \cite{Cacciari:2019qjx}.

At leading order, only the $N=2$ term contributes to the semi-inclusive single-jet cross-section. Throughout the paper, we use $k_1^\mu$ and $k_2^\mu$ for the 4-vectors of the quark and antiquark jet respectively, and we reserve the notation $k^\mu$ for the 4-vector of the tagged jet (which can either be the quark, antiquark or gluon jet at NLO). Integrating over the ``antiquark-jet" $j_2$ four-momentum yields a Dirac delta which fixes $\yt=\yt'$, so that the resulting $\yt'$ integral is straightforward:
\begin{align}
    \left.\frac{\der \sigma^{\gamma_{\lambda}^{\star}+A\to q+X}}{ \der^2 \ktone  \der \eta_1 }\right|_{\rm LO} &=\frac{\alpha_{\mathrm{em}} e_f^2 N_c}{(2\pi)^4}\int\der^2\xt\der^2\xt'\der^2\yt e^{-i\ktone\cdot\rxxtp}\Hcal_{\mathrm{LO}}^{\lambda}(z_1,Q^2,\rxyt,\rxpyt)\nonumber\\
    &\times\Xi_{\rm LO}(\xt,\yt,\xt')\label{eq:quarkjet-LO-cross-section} \,.
\end{align}
The SIDIS hard factors in coordinate space are given by
\begin{align}
    \Hcal_{\mathrm{LO}}^{\mathrm{L}}(z,Q^2,\rxyt,\rxpyt) &=  8 z^3(1-z)^2 Q^2 K_0(\bar{Q} r_{xy}) K_0(\bar{Q} r_{x'y}) \,, \label{eq:SIDIS-NLO-LLO} \\
    \Hcal_{\mathrm{LO}}^{\mathrm{T}}(z,Q^2,\rxyt,\rxpyt) &=  2 z\left[z^2 +(1-z)^2 \right]  \frac{\rxyt \cdot \rxpyt}{r_{xy} r_{x'y}}  \bar{Q}^2K_1(\bar{Q} r_{xy}) K_1(\bar{Q}r_{x'y})\label{eq:SIDIS-NLO-TLO} \,,
\end{align}
for a longitudinally and transversely polarized photon respectively. We define $\bar Q^2=z(1-z)Q^2$.
The color structure is much simpler, as only dipole operators are involved when $\yt=\yt'$,
\begin{align}
    \Xi_{\mathrm{LO}}(\xt,\yt,\xt') & =\left \langle D_{xx'} - D_{xy} -  D_{yx'} + 1 \right \rangle \,.
\end{align}
To get the contribution where the measured jet is $j_2$, one notices that it is sufficient to make the following change
\begin{equation}
    \Xi_{\rm LO}(\xt,\yt,\xt')\to\Xi^*_{\mathrm{LO}}(\xt,\yt,\xt') \,,
\end{equation}
in the previous expression.
This property also holds for the NLO corrections: the contribution where the anti-quark jet is measured can be obtained from the contribution where the quark jet is measured by taking the complex conjugate of the color structure in the formula. As expected, the SIDIS jet cross-section is not sensitive to the imaginary part of the dipole scattering amplitude (usually named $C$-odd component or "odderon" \cite{Lukaszuk:1973nt,Joynson:1975az,Hatta:2005as}).
In the end, the semi-inclusive single-jet cross-section at leading order is given by
\begin{align}
    \left.\frac{\der \sigma^{\gamma_{\lambda}^{\star}+A\to j+X}}{ \der^2 \kt  \der \eta }\right|_{\rm LO} &=\frac{\alpha_{\mathrm{em}} e_f^2 N_c}{(2\pi)^4}\int\der^2\xt\der^2\xt'\der^2\yt e^{-i\kt\cdot\rxxtp}\Hcal_{\mathrm{LO}}^{\lambda}(z,Q^2,\rxyt,\rxpyt)\nonumber\\
    &\times2\mathfrak{Re}\left \langle D_{xx'} - D_{xy} -  D_{yx'} + 1 \right \rangle \label{eq:SIDIS-LO-cross-section} \,.
\end{align}
Since we will also consider the transverse momentum integrated cross-section, we quote here the leading order result for the $\eta$-differential cross-section:
\begin{align}
    \left.\frac{\der \sigma^{\gamma_{\lambda}^{\star}+A\to j+X}}{  \der \eta }\right|_{\rm LO} &=\frac{\alpha_{\mathrm{em}} e_f^2 N_c}{(2\pi)^2}\int\der^2\xt\der^2\yt \Hcal_{\mathrm{LO}}^{\lambda}(z,Q^2,\rxyt,\rxyt)\times4\mathfrak{Re}\left\langle 1-D_{xy}\right\rangle \,,
\end{align}
which is a particularly simple result. If one further integrates over $\eta$, one gets twice the fully inclusive DIS cross-section at small $x$ at leading order. As previously mentioned, this is because each dijet event is counted twice due to our particular definition of the inclusive single-jet cross-section.

\section{Calculation of the double differential SIDIS cross-section at NLO}
\label{sec:NLO-calc}
\begin{figure}
    \centering
    \includegraphics[width=\textwidth]{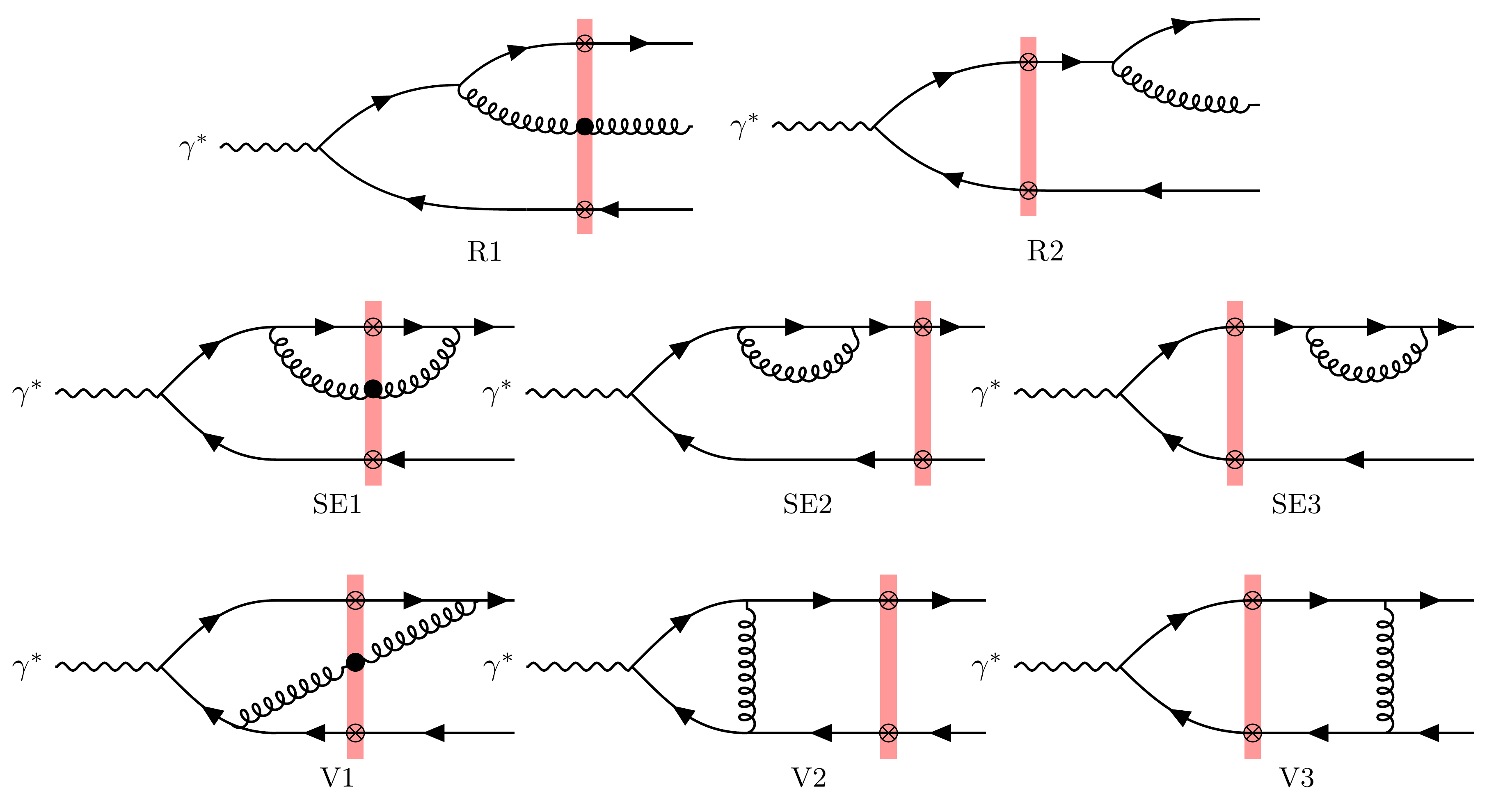}
    \caption{Feynman diagrams contributing to the NLO cross-section. The quark-antiquark exchanged diagrams, named with an additional line over the diagram's label, are not shown. Also not shown is the diagram where the virtual photon creates a $q\bar q$ loop annihilating into a gluon, which finally generates a $q\bar q$ pair. This diagram vanishes when summing over light quark charges \cite{Taels:2022tza}.}
    \label{fig:feynman-diagrams}
\end{figure}

Our starting point is the inclusive dijet cross-section computed at NLO in \cite{Caucal:2021ent}. The Feynman diagrams that contribute to the cross-section are displayed in Fig.\,\ref{fig:feynman-diagrams}.
In this figure, the diagrams in which the quark and antiquark are interchanged are not shown. Throughout this paper, they will be noted with an additional line over the label of the graph, e.g. $\rm V1\to \overline{V1}$.
We shall also adopt the following notation. The contribution from two given Feynman graphs $\rm G1$ and $\rm G2$ to the SIDIS cross-section is noted $\der\sigma_{\rm G1\times G2^*}$, where the graph $\rm G1$ is in the amplitude and $\rm G2$ is in the complex conjugate amplitude. For instance, the vertex correction with gluon crossing the shock-wave $\rm V1$ (see Fig.\,\ref{fig:feynman-diagrams}) gives two terms to the SIDIS cross-section, $\der\sigma_{\rm V1\times LO^*} $ and $\der\sigma_{\rm LO\times V1^*}$.

We will detail the interesting features of the calculation for the simpler case of a longitudinally polarized virtual photon but when possible we will show results that are applicable for both polarizations. The results for the SIDIS cross-section for transversely polarized virtual photons are given in the appendix~\ref{app:transverse}. For the reader who is not interested in the derivation of the cross-section but only in the final result, we refer to section \ref{sub:final-dblediff} where only the expressions for the NLO SIDIS cross-section are provided in their final form.

From the expressions obtained in \cite{Caucal:2021ent} for the inclusive dijet cross-section, it is relatively easy to get the semi-inclusive single-jet cross-section. The main difficulty comes from the emergence of new divergences when one of the final state parton is integrated out. In the virtual cross-section, there are only two jets in the final state. Consequently, we will simply integrate over the anti-quark jet phase space. As discussed in the previous section, the contribution where the antiquark jet is tagged can be obtained by taking the complex conjugate of the color structure in the final result, leaving invariant the rest of the expression. For the real cross-section, we will have to consider two cases: either the fermionic jet or the gluonic jet is tagged. This will be done in subsection \ref{sub:real-graphs}.

Like in the calculation of inclusive dijet production in the CGC \cite{Caucal:2021ent}, we employ dimensional regularization in the transverse plane to isolate the UV and collinear divergences of a given diagram. To regulate the light cone divergences as the longitudinal momentum fraction of the gluon $z_g=k_g^-/q^-$ goes to zero, we use a cut-off regulator $\Lambda^- \equiv z_0 q^-$.

\subsection{Cancellations between virtual and real diagrams}
\label{sub:cancellation}

Before computing explicitly these phase-space integrals, we point out, following \cite{Bergabo:2022zhe}, some nice cancellations between virtual and real diagrams that share the same topology at the cross-section level. In particular, we note that 
\begin{align}
    \left.\frac{\der \sigma^{\gamma_{\lambda}^{\star}+A\to q+X}}{ \der^2 \ktone  \der \eta_1 }\right|_{\rm \overline{SE1}\times LO^*}+\left.\frac{\der \sigma^{\gamma_{\lambda}^{\star}+A\to q+X}}{ \der^2 \ktone  \der \eta_1 }\right|_{\rm \overline{R1}\times \overline{R2}^*}&=0 \,, \\
     \left.\frac{\der \sigma^{\gamma_{\lambda}^{\star}+A\to q+X}}{ \der^2 \ktone  \der \eta_1 }\right|_{\rm \overline{V1}\times LO^*}+\left.\frac{\der \sigma^{\gamma_{\lambda}^{\star}+A\to q+X}}{ \der^2 \ktone  \der \eta_1 }\right|_{\rm R1\times \overline{R2}^*}&=0 \,,
\end{align}
and likewise for their complex conjugate. We have explicitly checked that these identities are true for both photon polarizations $\lambda=\rm L,T$.
Although the real terms $\rm \overline{R1}\times \overline{R2}^*$ and $\rm R1\times \overline{R2}^*$ do not need to be calculated when the fermionic jet is tagged, they will contribute when the gluon jet is measured in the final state.

In addition to the cancellations above, our particular choice of the dimensional regularization scheme for the transverse UV and IR divergences implies that
\begin{align}
    \left.\frac{\der \sigma^{\gamma_{\lambda}^{\star}+A\to q+X}}{ \der^2 \ktone  \der \eta_1 }\right|_{\rm SE3\times LO^*}&=0 \,, \\
    \left.\frac{\der \sigma^{\gamma_{\lambda}^{\star}+A\to q+X}}{ \der^2 \ktone  \der \eta_1 }\right|_{\rm \overline{SE3}\times LO^*}&=0 \,, \\
    \left.\frac{\der \sigma^{\gamma_{\lambda}^{\star}+A\to q+X}}{ \der^2 \ktone  \der \eta_1 }\right|_{\rm \overline{R2}\times\overline{R2}^*}&=0 \,, \label{eq:R2'xR2'}
\end{align}
in contrast to the results in \cite{Bergabo:2022zhe} obtained without using dimensional regularization. 

\subsection{Virtual diagrams}

In this subsection, we discuss the virtual diagrams that contribute to the NLO SIDIS cross-section. In the end, thanks to the results of the previous subsection, we only have to compute the contributions from diagrams $\rm SE1$, $\rm SE2$, $\rm \overline{SE2}$, $\rm V1$, $\rm V2$ and $\rm V3$.

\subsubsection{Vertex and self-energy corrections with gluon crossing the shock-wave}

Diagrams $\rm V1\times LO^*$ and $\rm SE1\times LO^*$ have been computed in 
\cite{Caucal:2021ent} (see also Eqs.\,(3.17) in \cite{Caucal:2022ulg}). In particular, the contribution from $\rm SE1\times LO^*$ has a UV divergence in coordinate space which is regularized by adding a suitable counter-term. The UV counter-term is then computed in dimensional regularization and added to the other UV divergent contributions such as $\rm V2\times LO^*$, $\rm SE2\times LO^*$ and $\rm\overline{SE2}\times LO^*$. Those diagrams, including the UV counter-term of $\rm SE1\times LO^*$, will be discussed in the next paragraph. Integrating over the anti-quark phase space the inclusive dijet cross-section coming from $\rm V1\times LO^*$ yields
\begin{align}
    &\left.\frac{\der \sigma^{\gamma_{\rm L}^{\star}+A\to q+X}}{  \der^2 \ktone \der \eta_1}\right|_{\rm V1\times LO^*} =-\frac{\alpha_{\rm em}e_f^2N_c}{(2\pi)^4} \int\der^2\xt \der^2\xt' \der^2\yt \ e^{-i\ktone \cdot \rxxtp} 8z_1^3 (1-z_1)^2Q^2 K_0(\bar{Q}r_{x'y})\nonumber\\
     & \times\frac{\alpha_s}{\pi}\int_{z_0}^{z_1} \frac{\der z_g}{z_g}\left(1-\frac{z_g}{z_1} \right) \left(1+\frac{z_g}{1-z_1} \right) \left(1-\frac{z_g}{2z_1} - \frac{z_g}{2(1-z_1+z_g)} \right) \nonumber\\
     & \times\int\frac{\der^2\zt}{\pi}  \frac{\rzxt\cdot \rzyt}{\rzxt^2 \rzyt^2}  e^{-i\frac{z_g}{z_1}\ktone \cdot \rzxt} K_0( QX_{\rm V})  \Xi_{\rm NLO,1}(\xt,\yt,\zt,\xt') \,,
     \label{eq:V1LO}
\end{align}
with
\begin{align}
     X_{\rm V}^2 &= (1-z_1)(z_1-z_g)\rxyt^2 + z_g(z_1-z_g)\rzxt^2 +(1-z_1)z_g\rzyt^2\,,\\
    \Xi_{\rm NLO,1} &= \frac{N_c}{2}\left\langle 1-D_{yx'}-D_{xz}D_{zy}+D_{zx'}D_{xz} \right\rangle- \frac{1}{2N_c}\left\langle 1-D_{xy}-D_{yx'}+D_{xx'} \right\rangle \,.
\end{align}
The color structure $\Xi_{\rm NLO1}$ only involves dipole correlators. 
Similarly, the UV regular part of $\rm SE1\times LO^*$ is given by
\begin{align}
    &\left.\frac{\der \sigma^{\gamma_{\rm L}^{\star}+A\to q+X}}{  \der^2 \ktone \der \eta_1}\right|_{\rm SE1\times LO^*,reg} =\frac{\alpha_{\rm em}e_f^2N_c}{(2\pi)^4} \int\der^2\xt \der^2\xt' \der^2\yt \ e^{-i\ktone \cdot \rxxtp}
    8 z_1^3 (1-z_1)^2Q^2 K_0(\bar{Q}r_{x'y})\nonumber\\
     & \times\frac{\alpha_s}{\pi}\int_{z_0}^{z_1} \frac{\der z_g}{z_g} \left(1-\frac{z_g}{z_1}+\frac{z_g^2}{2z_1^2} \right) \int\frac{\der^2\zt}{\pi} \frac{1}{\rzxt^2}\left[e^{-i\frac{z_g}{z_1}\ktone \cdot \rzxt}  K_0(QX_V)  \Xi_{\rm NLO,1}(\xt,\yt,\zt,\xt')\right.\nonumber\\
     &-\left.e^{-\frac{\rzxt^2}{e^{\gamma_E}\rxyt^2}}K_0(\bar Qr_{xy})C_F\Xi_{\rm LO}(\xt,\yt,\xt')\right] \,,
     \label{eq:SE1LOreg}
\end{align}
where the last line in the above expression corresponds to the UV counter-term such that the $\zt$ integral is well-defined as $\rzxt=\zt-\xt\to0$. Indeed, in this limit, the expression in the square bracket vanishes like $\mathcal{O}(r_{zx})$.

\subsubsection{The pole term}
\label{subsub:pole-term}

We define the pole term of the virtual cross-section as the sum of diagrams $\rm V2\times LO^*$, $\rm SE2\times LO^*$, $\rm\overline{SE2}\times LO^*$ and the UV singular part of $\rm SE1\times LO^*$:
\begin{align}
    \left.\frac{\der \sigma^{\gamma_{\rm L}^{\star}+A\to q+X}}{  \der^2 \ktone \der \eta_1}\right|_{\rm UV}&\equiv \left.\frac{\der \sigma^{\gamma_{\rm L}^{\star}+A\to q+X}}{  \der^2 \ktone \der \eta_1}\right|_{\rm V2\times LO^*}+\left.\frac{\der \sigma^{\gamma_{\rm L}^{\star}+A\to q+X}}{  \der^2 \ktone \der \eta_1}\right|_{\rm SE2\times LO^*}\nonumber\\
    &+\left.\frac{\der \sigma^{\gamma_{\rm L}^{\star}+A\to q+X}}{  \der^2 \ktone \der \eta_1}\right|_{\rm \overline{SE2}\times LO^*}+\left.\frac{\der \sigma^{\gamma_{\rm L}^{\star}+A\to q+X}}{ \der^2 \ktone \der \eta_1}\right|_{\rm SE1\times LO^*,UV} \,. \label{eq:UV-term}
\end{align}
Using Eqs.\,(5.37),\,(5.40) and (5.78) from \cite{Caucal:2021ent} and integrating over the anti-quark phase space, we get
\begin{align}
    &\left.\frac{\der \sigma^{\gamma_{\rm L}^{\star}+A\to q+X}}{ \der^2 \ktone  \der \eta_1 }\right|_{\rm UV}=\frac{\alpha_{\rm em}e_f^2N_c}{(2\pi)^4}\int\der^2\xt\der^2\xt'\der^2\yt e^{-i\ktone\cdot\rxxtp}\Hcal_{\mathrm{LO}}^{\lambda=\rm L}(\rxyt,\rxpyt)\Xi_{\rm LO}(\xt,\yt,\xt')\nonumber\\
    &\times\frac{\alpha_sC_F}{2\pi}\left\{\left(\ln\left(\frac{z_1}{z_0}\right)-\ln\left(\frac{z_2}{z_0}\right)\right)\left(\frac{2}{\varepsilon}+\ln(e^{\gamma_E}\pi\mu^2\rxyt^2)\right)+\frac{1}{2}\ln^2\left(\frac{z_2}{z_1}\right)-\frac{\pi^2}{6}+\frac{5}{2}\right\}.\,
    \label{eq:dijet-NLO-finite-SE1-V1-SE2uv}
\end{align}
In dimensional regularization, finite terms may arise from the product between the UV pole in $1/\varepsilon$ and $\mathcal{O}(\varepsilon)$ terms from the Dirac traces appearing in the numerator of the Feynman amplitude. These finite pieces have been explicitly computed in \cite{Caucal:2021ent}. For the specific combination given by Eq.\,\eqref{eq:UV-term}, these finite terms cancel among each other.

\subsubsection{Explicit calculation of the final state vertex correction}

The final state vertex correction corresponding to diagram $\rm V3$ needs more work. This is because in \cite{Caucal:2021ent}, we were not able to find a closed analytic expression for this diagram. Fortunately, the integration over the anti-quark phase space allows many simplifications. 

The formula obtained in \cite{Caucal:2021ent} for the contribution of $\rm V3\times LO^*$ to the inclusive dijet cross-section is
\begin{align}
    &\left.\frac{\der \sigma^{\gamma^{\star}_{\rm L}+A \rightarrow q\bar q  + X}}{\der^2\ktone \der\eta_1 \der^2\kttwo \der \eta_2}\right|_{\rm V3\times LO^*} =\frac{ \alpha_{em}e^{2}_{f}N_c}{(2\pi)^6}\delta(1-z_1-z_2) \int\der^2\xt\der^2\xt'\der^2\yt\der^2\yt'8z_{1}^{3}z_{2}^{3}Q^2\nonumber\\
    &\times e^{-i\ktone\cdot\rxxtp-i\kttwo\cdot\ryytp}K_0(\bar{Q}r_{x'y'}) \frac{\alpha_s}{\pi} \int_{z_0}^{z_1} \frac{\der z_g}{z_g} K_0(\bar{Q}_{\rm V3}r_{xy})  \left\{ \left(1-\frac{z_g}{z_1}\right)^2 \left(1+\frac{z_g}{z_2}\right) (1+z_g) \right.\nonumber \\ 
    &\times e^{i(\Pt+z_g\qt)\cdot \rxyt}K_0(-i \Delta_{\rm V3}r_{xy}) - \left(1-\frac{z_g}{z_1}\right) \left(1+\frac{z_g}{z_2}\right)\left(1-\frac{z_g}{2z_1} + \frac{z_g}{2z_2} - \frac{z_g^2}{2z_1z_2} \right) \nonumber \\ 
    &\left.  \times e^{i\frac{z_g}{z_1}\ktone\cdot \rxyt}\Jcal_{\odot}\left(\rxyt,\left(1-\frac{z_g}{z_1}\right)\Pt,\Delta_{\rm V3}\right) \right\}  \Xi_{\rm NLO,3}(\xt,\yt;\xt',\yt') +(1\leftrightarrow 2) \,, \label{eq:V3-dijet}
\end{align} 
where the $(1\leftrightarrow 2)$ notation corresponds to quark-antiquark interchange. The kinematic variables in this formula are defined by $\Pt=z_2\ktone-z_1\kttwo$, $\qt=\ktone+\kttwo$, $\bar Q_{\rm V3}^2=(z_1-z_g)(z_2+z_g)Q^2$ and $\Delta_{\rm V3}^2=\left(1-\frac{z_g}{z_1} \right)\left(1+\frac{z_g}{z_2} \right)\Pt^2$. The CGC color correlator of this diagram is
\begin{align}
    \Xi_{\rm NLO,3}&=\frac{N_c}{2}\left\langle 1-D_{xy}-D_{y'x'}+D_{xy}D_{y'x'}\right\rangle\nonumber\\
    &-\frac{1}{2N_c}\left\langle1-D_{xy}-D_{y'x'}+Q_{xy;y'x'}\right\rangle \,,
\end{align}
and depends on the quadrupole in the $1/N_c^2$ suppressed term. The function $\Jcal_{\odot}$ defined by
\begin{align}
    \Jcal_{\odot}(\rt,\Kt,\Delta)\equiv\int\frac{\der^2\lt}{(2\pi)}\frac{2\lt\cdot\Kt e^{i\lt\cdot\rt}}{\lt^2\left[(\lt-\Kt)^2-\Delta^2-i\epsilon\right]} \,, 
\end{align}
is not known analytically. However, another expression for this function in terms of a single scalar integral over a Feynman parameter $u$ was found in \cite{Caucal:2021ent}. This expression turns out to be key to simplifying Eq.\,\eqref{eq:V3-dijet} once one integrates over the antiquark phase space. As shown in appendix F.1 of \cite{Caucal:2021ent}, we have indeed
\begin{align}
    \Jcal_{\odot}(\rt,\Kt,\Delta)=& \ i\rt\cdot\Kt\int_0^1\der u \ e^{iu\Kt \cdot \rt}K_0\left(-i \delta_{\mathrm{V}3} r_\perp\right)\nonumber\\
    &+\Kt^2\int_0^1\der u \ e^{iu\Kt \cdot \rt}\frac{u r_\perp K_1\left( -i \delta_{\mathrm{V}3} r_\perp\right)}{(-i\delta_{\mathrm{V}3})}\,, \label{eq:dijet-NLO-app-Jdotfinal2}
\end{align}
with $\delta_{\rm V3}= \sqrt{|u(1-u)\Kt^2-u\Delta^2|}$.
Using this representation of the $\Jcal_{\odot}$ function and after integration over the anti-quark phase space, one can perform the integration over the remaining Feynman parameter $u$ analytically. The final result for the diagram $\rm V3\times LO^*$ reads
\begin{align}
    &\left.\frac{\der \sigma^{\gamma_{\rm L}^{\star}+A\to q+X}}{ \der^2 \ktone  \der \eta_1 }\right|_{\rm V3\times LO^*} =\frac{\alpha_{\mathrm{em}} e_f^2 N_c}{(2\pi)^4}\int\der^2\xt\der^2\xt'\der^2\yt\der^2\yt' e^{-i\ktone\cdot\left(\rxxtp+\frac{z_2}{z_1}\ryytp\right)}\nonumber\\
    &\times8z_1^3z_2^2Q^2K_0(\bar Qr_{x'y'})\frac{\alpha_s}{2\pi^2}\int_{z_0}^{1}\frac{\der z_g}{z_g}\left\{K_0(\bar Q_{\rm V3}r_{xy})\frac{\left(1-\frac{z_g}{z_1}\right)^2\left(1+\frac{z_g}{z_2}\right)e^{i\frac{z_g}{z_1}\ktone\cdot\rxyt}\Theta(z_1-z_g)}{\ryytp^2\left[\ryytp^2+(z_1-z_g)\left(2\ryytp\cdot\rxyt-\frac{z_g}{z_2}\rxyt^2\right)\right]}\right.\nonumber\\
    &\left[(1+z_g)\ryytp^2+2z_1\left(1-\frac{z_g}{2z_1}+\frac{z_g}{2z_2}-\frac{z_g^2}{2z_1z_2}\right)\rxyt\cdot\ryytp\right]\nonumber\\
    &+K_0(\bar Q_{\rm R2}r_{xy})\frac{\left(1-\frac{z_g}{z_2}\right)^2\left(1+\frac{z_g}{z_1}\right)e^{-i\frac{z_g}{z_1}\ktone\cdot\rxyt}\Theta(1-z_1-z_g)}{\left(\ryytp+\frac{z_g}{z_2}\rxyt\right)^2\left[\ryytp^2+(z_1+z_g)\left(2\ryytp\cdot\rxyt+\frac{z_g}{z_2}\rxyt^2\right)\right]}\nonumber\\
    &\left.\left[(1+z_g)\left(\ryytp+\frac{z_g}{z_2}\rxyt\right)^2+2z_1\left(1-\frac{z_g}{2z_2}+\frac{z_g}{2z_1}-\frac{z_g^2}{2z_1z_2}\right)\left(\rxyt\cdot\ryytp+\frac{z_g}{z_2}\rxyt^2\right)\right]\right\}\nonumber\\
    &\times \left[\frac{N_c}{2}\left\langle 1-D_{xy}-D_{y'x'} +D_{xy}D_{y'x'} \right\rangle- \frac{1}{2N_c}\left\langle 1-D_{xy}-D_{y'x'}+Q_{xyy'x'} \right\rangle\right]\,, \label{eq:V3xLO-final}
\end{align}
with $z_2=1-z_1$ and $\bar Q_{\rm R2}^2=(1-z_1-z_g)(z_1+z_g)Q^2$. Although this expression has still four integrals over 2-dimensional transverse coordinates and one integral over the longitudinal momentum fraction $z_g$ of the gluon, it is a much simpler formula than Eq.\,\eqref{eq:V3-dijet} because the integrant has an explicit analytic expression. 
This concludes our calculation of the virtual corrections to the semi-inclusive single-jet cross-section in DIS in the saturation formalism.

\subsection{Real diagrams}
\label{sub:real-graphs}

For the real diagrams, the implementation of the jet definition is more complicated than for the virtual diagrams, as two partons can be clustered together to form a single jet. Also, in the case of a three-jet event, each jet can be tagged, in particular the one created by the additional gluon. We will first consider the case where the measured jet is fermionic, including the case where the gluon and the fermion are clustered together. In the last paragraph of this subsection, we will compute the semi-inclusive single-gluon production cross-section, excluding the cases where the gluon lies in the same jet as the quark or the antiquark. The use of the small-$R$ or narrow jet approximation simplifies the treatment of overlapping contributions: if the diagram does not have a collinear singularity, the phase space where the gluon and the fermion or the two fermions are clustered together in the same jet is systematically power of $R^2$ suppressed \cite{boussarie:tel-01468540}. For these diagrams, one can simply integrate over the two untagged partons to get the single-jet semi-inclusive cross-section.

Jets are defined from the partonic final state using a procedure common to all clustering algorithms within the generalized $k_t$ family \cite{Salam:2010nqg,Cacciari:2011ma}, including the famous anti-$k_t$ \cite{Cacciari:2008gp} or Cambridge-Aachen \cite{Wobisch:1998wt} algorithms. The condition for any pair of partons labeled $i$ and $k$ to be clustered together within the same jet with transverse momentum $\kt$ and longitudinal momentum fraction $z$ reads, up to power of $R^4$ corrections,
\begin{align}
    \Ccal_{ik}^2\le R^2\kt^2\frac{z_i^2z_k^2}{z^4}\,,\quad \Ccal_{ik}=\frac{z_i}{z}\boldsymbol{k}_{\perp,k}-\frac{z_k}{z}\boldsymbol{k}_{\perp,i}\,,\label{eq:collinearity-def}
\end{align}
with $\Ccal_{ik}$ the collinearity vector between the two partons (which goes to $0$ as the two partons become collinear). If they are clustered together, the four momentum of the jet is simply the sum of the four momenta of the two partons $i$ and $k$. In particular $\kt =\boldsymbol{k}_{\perp,i}+\boldsymbol{k}_{\perp,k} $ and $z=z_i+z_k$.

We shall divide the discussion of the real cross-section depending on whether the measured jet is fermionic (by that, we mean that it contains at least a quark or an anti-quark) or gluonic.

\subsubsection{Fermion-jet contributions with collinear divergence}

We start with the diagram $\rm R2\times \rm R2^*$ which is the only diagram with a collinear divergence when the gluon and the quark become collinear to each other. By quark-antiquark symmetry, the diagram $\rm\overline{R2}\times\overline{R2}^*$ is also collinearly divergent when the gluon and the anti-quark are collinear. However, if the quark jet is measured, then this contribution to the real NLO single-jet semi-inclusive cross-section is identically zero in dimension regularization according to Eq.\,\eqref{eq:R2'xR2'}. If the antiquark jet is measured, one can simply relate via complex conjugation of the CGC color correlator the contribution of diagram $\rm\overline{R2}\times\overline{R2}^*$ with that of diagram $\rm R2\times \rm R2^*$ when the quark jet is measured.

The implementation of the jet definition gives two terms from $\rm R2\times \rm R2^*$. The first one, labeled ``in", corresponds to the case where the gluon and the quark are clustered within the same jet. For jets defined with generalized $k_t$ algorithms in the small $R$ approximation, one imposes the condition
\begin{align}
    \Ccal_{qg}^2\le R^2\kt^2\frac{z_1^2z_g^2}{z^4}\,,
\end{align}
on the phase-space integration, with $\kt=\ktone+\kgt$ and $z=z_1+z_g$.
Then the quark-jet is measured and the anti-quark jet is integrated over. This contribution is computed in dimensional regularization to regulate the collinear singularity. We have
\begin{align}
    &\left.\frac{\der \sigma^{\gamma_{\lambda}^{\star}+A\to q +X}}{ \der^2 \ktone  \der \eta_1 }\right|_{\rm R2\times R2^*,in}=\left.\frac{\der \sigma^{\gamma_{\lambda}^{\star}+A\to q +X}}{ \der^2 \ktone  \der \eta_1 }\right|_{\rm LO}\times \frac{\alpha_sC_F}{\pi}\left\{\left(\frac{3}{4}-\ln\left(\frac{z_1}{z_0}\right)\right)\frac{2}{\varepsilon}+\ln^2(z_1)\right.\nonumber\\
    &\left.-\ln^2(z_0)+\left(\ln\left(\frac{z_1}{z_0}\right)-\frac{3}{4}\right)\ln\left(\frac{R^2\ktone^2}{\tilde{\mu}^2z_1^2}\right)-\frac{\pi^2}{2}+\frac{3}{2}\left(1-\ln\left(2z_1\right)\right)+3+\frac{1}{4}+\mathcal{O}(R^2)\right\}\,,
\label{eq:dijet-NLO-real-collinardiv}
\end{align}
with the $\overline{\rm MS}$ $\tilde\mu^2=4\pi e^{-\gamma_E}\mu^2$.
In this expression, the $+1/4$ finite term comes from the product between the $1/\varepsilon$ collinear pole and the $\mathcal{O}(\varepsilon)$ term in the $\varepsilon$-expansion of the Dirac trace in $4-\varepsilon$ dimension of diagram $\rm R2 \times \rm R2^*$.

The second term comes from three-jet configurations, with the quark, the gluon, and the anti-quark forming three well-separated jets. The gluon and anti-quark jet are then integrated over. The case where the gluon is measured and the other two jets are integrated will be considered in section \ref{eq:gluon-jets}. Further, it is convenient to decompose this contribution into two terms, labeled ``soft" and ``reg", by isolating the soft divergent contribution as $z_g\to 0$.  We have then
\begin{align}
    &\left.\frac{\der \sigma^{\gamma_{\lambda}^{\star}+A\to q+X}}{ \der^2 \ktone  \der \eta_1 }\right|_{\rm R2\times R2^*,soft}=\frac{\alpha_{\rm em}e_f^2N_c}{(2\pi)^4}\int\der^2\xt\der^2\xt'\der^2\yt e^{-i\ktone\cdot\rxxtp}\Hcal_{\mathrm{LO}}^{\lambda}(z_1,Q^2,\rxyt,\rxpyt)\nonumber\\
    \times &\Xi_{\rm LO}(\xt,\yt,\xt')\times \frac{\alpha_sC_F}{\pi}\left\{\ln^2\left(\frac{z_1}{z_0}\right)-\ln^2\left(\frac{z_1}{1-z_1}\right)-\ln\left(\frac{1-z_1}{z_0}\right)\ln\left(\frac{\ktone^2\rxxtp^2R^2}{c_0^2}\right)\right\}\,,\label{eq:R2R2-outdiv}
\end{align}
\begin{align}
    &\left.\frac{\der \sigma^{\gamma_{\rm L}^{\star}+A\to q+X}}{ \der^2 \ktone  \der \eta_1 }\right|_{\rm R2\times R2^*, reg}=\frac{\alpha_{\rm em}e_f^2N_c}{(2\pi)^4}\int\der^2\xt\der^2\xt'\der^2\yt e^{-i\ktone\cdot\rxxtp}\Xi_{\rm LO}(\xt,\yt,\xt')\nonumber\\
    &\times \frac{(-\alpha_s)C_F}{\pi}\int_{0}^{1-z_1}\frac{\der z_g}{z_g}8z_1^3(1-z_1)^2Q^2K_0(\bar Q_{\mathrm{R}2}r_{x'y})\ln\left(\frac{\ktone^2\rxxtp^2R^2z_g^2}{c_0^2z_1^2}\right)\nonumber\\
    &\times\left\{e^{-i\frac{z_g}{z_1}\ktone\rxxtp}\left(1-\frac{z_g}{1-z_1}\right)^2\left(1+\frac{z_g}{z_1}\right)^2\left(1+\frac{z_g}{z_1}+\frac{z_g^2}{2z_1^2}\right)K_0(\bar Q_{\mathrm{R}2}r_{xy})-K_0(\bar Qr_{xy})\right\} \,.
    \label{eq:R2R2-outreg}
\end{align}
One can already notice that the double logarithmic term in $\ln^2(z_0)$ cancels between the ``soft" and the ``in" term, as observed in the inclusive dijet calculation \cite{Taels:2022tza,Caucal:2022ulg}.

\subsubsection{Fermion-jet contributions with UV divergence}
\label{subsub:fermion-real-uv}
An interesting feature of the single-jet semi-inclusive calculation with respect to the inclusive dijet case is that some diagrams that are regular in the latter case become singular in the former. It is the case of the real correction associated with $\rm \overline{R1}\times\overline{R1}^*$ once one integrates over the antiquark and gluon phase space. By symmetry, the same divergence occurs for diagram $\rm R1\times R1^*$ if only the antiquark jet is measured.
Integrating over the anti-quark and gluon 4-momenta in the $q\bar qg$ cross-section obtained in \cite{Caucal:2021ent}, we get
\begin{align}
    &\left.\frac{\der \sigma^{\gamma_{\rm L}^{\star}+A\to q+X}}{ \der^2 \ktone  \der \eta_1 }\right|_{\rm \overline{R1}\times \overline{R1}^*} =\frac{\alpha_{\rm em}e_f^2N_c}{(2\pi)^4}\int\der^2\xt\der^2\xt'\der^2\yt e^{-i\ktone\cdot\rxxtp}\int_{z_0}^{1-z_1}\frac{\der z_g}{z_g}8z_1^3(1-z_1)^2Q^2\nonumber\\
    &\times\frac{\alpha_s}{\pi}\left(1-\frac{z_g}{1-z_1}\right)^2\left(1+\frac{z_g}{1-z_1-z_g}+\frac{z_g^2}{2(1-z_1-z_g)^2}\right)\int\frac{\der^2\zt}{\pi} \frac{1}{\rzyt^2}K_0(QX_R)K_0(QX_R')\nonumber\\
    &\times\Xi_{\rm NLO,4}(\xt,\yt,\zt,\xt') \,,
    \label{eq:barR1barR1}
\end{align}
with
\begin{align}
    \Xi_{\rm NLO,4}(\xt,\yt,\zt,\xt')&\equiv\frac{N_c}{2} \left\langle D_{xx'}-D_{xz}D_{zy}-D_{yz}D_{zx'}+1\right\rangle\nonumber\\
    &-\frac{1}{2N_c}\left\langle D_{xx'} - D_{xy} -  D_{yx'} + 1\right\rangle \,,
\end{align}
and the transverse coordinates
\begin{align}
    X_R^2 &= z_1(1-z_1-z_g)\rxyt^2+z_1z_g \rzxt^2+(1-z_1-z_g)z_g\rzyt^2\,,\\
    X_R'^2 &= z_1(1-z_1-z_g)\rxpyt^2+z_1z_g\rzxpt^2+(1-z_1-z_g)z_g\rzyt^2\,.
\end{align}
This expression is UV divergent because of the $\rzyt\to0$ singularity in the $\zt$ integral. To extract this divergence and provide an expression for $\rm \overline{R1}\times \overline{R1}^*$ which is suitable for numerical evaluation, we follow the same strategy as for the treatment of the UV divergence in coordinate space of diagram $\rm SE1\times LO^*$.
In concrete terms, we write
\begin{align}
    \left.\frac{\der \sigma^{\gamma_{\lambda}^{\star}+A\to q+X}}{ \der^2 \ktone  \der \eta_1 }\right|_{\rm  \overline{R1}\times \overline{R1}^*}&=\underbrace{\left.\frac{\der \sigma^{\gamma_{\lambda}^{\star}+A\to q+X}}{ \der^2 \ktone  \der \eta_1 }\right|_{\rm  \overline{R1}\times \overline{R1}^*}-\left.\frac{\der \sigma^{\gamma_{\lambda}^{\star}+A\to q+X}}{ \der^2 \ktone  \der \eta_1 }\right|_{\rm  \overline{R1}\times \overline{R1}^*,UV}}_{\equiv \rm\overline{R1}\times\overline{R1}^*,reg}\nonumber\\
    &+\left.\frac{\der \sigma^{\gamma_{\lambda}^{\star}+A\to q+X}}{ \der^2 \ktone  \der \eta_1 }\right|_{\rm  \overline{R1}\times \overline{R1}^*,UV} \,,
\end{align}
with the UV subtraction term defined by
\begin{align}
 &\left.\frac{\der \sigma^{\gamma_{\rm L}^{\star}+A\to q+X}}{ \der^2 \ktone  \der \eta_1 }\right|_{\rm \overline{R1}\times \overline{R1}^*,UV}\equiv\frac{\alpha_{\rm em}e_f^2N_c}{(2\pi)^4}\int\der^2\xt\der^2\xt'\der^2\yt e^{-i\ktone\cdot\rxxtp}\int_{z_0}^{1-z_1}\frac{\der z_g}{z_g}8z_1^3(1-z_1)^2Q^2\nonumber\\
    &\times\frac{\alpha_s}{\pi}\left(1-\frac{z_g}{1-z_1}\right)^2\left(1+\frac{z_g}{1-z_1-z_g}+\frac{z_g^2}{2(1-z_1-z_g)^2}\right)\int\frac{\der^2\zt}{\pi} \frac{e^{-\frac{\rzyt^2}{2\xi}}}{\rzyt^2}K_0(\bar Qr_{xy})K_0(\bar Qr_{x'y})\nonumber\\
    &\times\left\langle D_{xx'}-D_{xy}-D_{yx'}+1\right\rangle \,,
\end{align}
which allows for the cancellation of the UV divergence, without introducing any further IR divergence as $\rzyt\to\infty$ thanks to the exponential $\exp(-\rzyt^2/(2\xi))$ suppression at large $\rzyt$. In this exponential, $\xi$ is an arbitrary transverse scale squared, which will be fixed at the end of our calculation to simplify the extraction of the BK-JIMWLK evolution of the CGC dipole operators.
Thanks to its simple form, the UV counter-term can be computed analytically in dimensional regularization  $\der^2\zt\to\der^{2+\varepsilon}\zt$, such that:
\begin{align}
    &\left.\frac{\der \sigma^{\gamma_{\lambda}^{\star}+A\to q+X}}{ \der^2 \ktone  \der \eta_1 }\right|_{\rm \overline{R1}\times \overline{R1}^*,UV}=\frac{\alpha_{\rm em}e_f^2N_c}{(2\pi)^4}\int\der^2\xt\der^2\xt'\der^2\yt e^{-i\ktone\cdot\rxxtp}\Hcal_{\mathrm{LO}}^{\lambda}(z_1,Q^2,\rxyt,\rxpyt)\nonumber\\
    \times &\Xi_{\rm LO}(\xt,\yt,\xt')\times\frac{\alpha_sC_F}{\pi}\left\{\left(\ln\left(\frac{z_2}{z_0}\right)-\frac{3}{4}\right)\left(\frac{2}{\varepsilon}+\ln(2\pi\mu^2\xi)\right)-\frac{1}{4}\right\} \,, 
    \label{eq:R1barR1barUV}
\end{align}
where the last $-1/4$ term comes again from the product between the UV pole and the $\mathcal{O}(\varepsilon)$ term in the Dirac trace of $\rm \overline{R1}\times \overline{R1}^*$.

The regular term reads:
\begin{align}
    &\left.\frac{\der \sigma^{\gamma_{\rm L}^{\star}+A\to q+X}}{ \der^2 \ktone  \der \eta_1 }\right|_{\rm \overline{R1}\times \overline{R1}^*,reg} =\frac{\alpha_{\rm em}e_f^2N_c}{(2\pi)^4}\int\der^2\xt\der^2\xt'\der^2\yt e^{-i\ktone\cdot\rxxtp}\int_{z_0}^{1-z_1}\frac{\der z_g}{z_g}8z_1^3(1-z_1)^2Q^2\nonumber\\
    &\times\frac{\alpha_s}{\pi}\left(1-\frac{z_g}{1-z_1}\right)^2\left(1+\frac{z_g}{1-z_1-z_g}+\frac{z_g^2}{2(1-z_1-z_g)^2}\right) \int\frac{\der^2\zt}{\pi} \frac{K_0(QX_R)K_0(QX_R')}{\rzyt^2}  \nonumber \\
    & \times \left[ \Xi_{\rm NLO,4}(\xt,\yt,\zt,\xt') - e^{-\frac{\rzyt^2}{2\xi}}\frac{K_0(\bar Qr_{xy})K_0(\bar Qr_{x'y})}{K_0(QX_R)K_0(QX_R')} C_F \Xi_{\rm LO}(\xt,\yt,\xt')  \right]\,.
    \label{eq:barR1barR1reg}
\end{align}

\subsubsection{Regular fermion-jet contributions}

We finally gather here all the other real corrections with the quark-jet tagged which are neither collinear nor UV divergent in coordinate space. As mentioned at the beginning of section \ref{sub:real-graphs}, for these diagrams, one can freely integrate over the antiquark and gluon phase space without being worried about configurations where the gluon and the quark, the gluon and the antiquark or the two fermions lie in the same jet as such configurations are power of $R^2$ suppressed in the narrow jet limit.

The first regular diagram is $\rm R2\times\overline{R2}^*$ (and its complex conjugate). We obtain, after a straightforward integration of Eq.\,(B.11) in \cite{Caucal:2022ulg}, the following result:
\begin{align}
    &\left.\frac{\der \sigma^{\gamma_{\rm L}^{\star}+A\to q+X}}{ \der^2 \ktone  \der \eta_1 }\right|_{\rm R2\times \overline{R2}^*}=\frac{\alpha_{\rm em}e_f^2N_c}{(2\pi)^4}\int\der^8\Xt e^{-i\ktone\cdot\rxxtp}8(1-z_1)^2z_1^3Q^2K_0(\bar{Q}r_{x'y'})\nonumber\\
    &\times\frac{(-\alpha_s)}{\pi^2}\int_{z_0}^{1-z_1}\frac{\der z_g}{z_g}e^{-i\ktone\cdot\left(\frac{z_g}{z_1}\rxypt+\frac{1-z_1-z_g}{z_1}\ryytp\right)}\left(1-\frac{z_g}{1-z_1}\right)\left(1+\frac{z_g}{z_1}\right)K_0(\bar{Q}_{\rm R2}r_{xy})\nonumber\\
    &\times\left[1+\frac{z_g}{2z_1}+\frac{z_g}{2(1-z_1-z_g)}\right]\frac{(\ryytp+\frac{z_g}{1-z_1-z_g}\rxypt)\cdot\ryytp}{(\ryytp+\frac{z_g}{1-z_1-z_g}\rxypt)^2\ryytp^2}\Xi_{\rm NLO,3}(\xt,\yt;\xt',\yt')\label{eq:R2xR2'final}
\end{align}
with $\bar Q_{\rm R2}^2=(1-z_1-z_g)(z_1+z_g)Q^2$. The color correlator $\Xi_{\rm NLO,3}$ is identical to the one appearing in the final state vertex correction $\rm V3\times LO^*$ (see Eq.\,\eqref{eq:V3xLO-final}). It will be very natural to combine these two diagrams together in our final result.

The contribution from diagrams $\rm R1\times R2^*$ and $\rm \overline{R1}\times R2^*$ read respectively
\begin{align}
    &\left.\frac{\der \sigma^{\gamma_{\rm L}^{\star}+A\to q+X}}{  \der^2 \ktone \der \eta_1}\right|_{\rm R1 \times R2^*} =-\frac{\alpha_{\rm em}e_f^2N_c}{(2\pi)^4 } \int\der^2\xt \der^2\xt' \der^2\yt e^{-i\ktone \cdot \rxxtp}8z_1^3(1-z_1)^2Q^2\nonumber\\
    &\times \frac{\alpha_s}{\pi}\int_{z_0}^{1-z_1} \frac{\der z_g}{z_g}\left (1-\frac{z_g}{1-z_1}\right)^2\left(1+\frac{z_g}{z_1}\right) \left(1+\frac{z_g}{z_1}+\frac{z_g^2}{2z_1^2} \right) \nonumber\\
     & \times\int\frac{\der^2\zt}{\pi}e^{-i\frac{z_g}{z_1}\ktone\cdot\rzxpt} K_0(\bar{Q}_{\rm R2}r_{x'y}) \frac{\rzxt\cdot \rzxpt}{\rzxt^2 \rzxpt^2} K_0(QX_R)  \Xi_{\rm NLO,1}(\xt,\yt,\zt,\xt') \,,
     \label{eq:R1R2}
\end{align}
\begin{align}
    &\left.\frac{\der \sigma^{\gamma_{\rm L}^{\star}+A\to q+X}}{  \der^2 \ktone \der \eta_1}\right|_{\rm \overline{R1} \times R2^*} =\frac{\alpha_{\rm em}e_f^2N_c}{(2\pi)^4} \int\der^2\xt \der^2\xt' \der^2\yt  e^{-i\ktone \cdot \rxxtp} 8z_1^3(1-z_1)^2Q^2\nonumber\\
    &\times \frac{\alpha_s}{\pi}\int_{z_0}^{1-z_1} \frac{\der z_g}{z_g}\left(1-\frac{z_g}{1-z_1}\right)^2\left(1+\frac{z_g}{z_1}\right) \left(1+\frac{z_g}{2z_1}+\frac{z_g}{2(1-z_1-z_g)} \right) \nonumber\\
     & \times \int\frac{\der^2\zt}{\pi}e^{-i\frac{z_g}{z_1}\ktone\cdot\rzxpt}K_0(\bar{Q}_{\rm R2}r_{x'y}) \frac{\rzyt \cdot\rzxpt}{\rzyt^2 \rzxpt^2} K_0(QX_R)  \Xi_{\rm NLO,1}(\xt,\yt,\zt,\xt') \,.
     \label{eq:barR1R2}
\end{align}
Finally, the two remaining regular real corrections come from diagrams $\rm R1\times R1^*$ and $\rm \overline{R1}\times R1^*$, which respectively read
\begin{align}
        &\left.\frac{\der \sigma^{\gamma_{\rm L}^{\star}+A\to q+X}}{  \der^2 \ktone \der \eta_1}\right|_{\rm R1 \times R1^*} =\frac{\alpha_{\rm em}e_f^2N_c}{(2\pi)^4 } \int\der^2\xt \der^2\xt' \der^2\yt  e^{-i\ktone \cdot \rxxtp}8z_1^3(1-z_1)^2Q^2\nonumber\\
     &\times\frac{\alpha_s}{\pi}\int_{z_0}^{1-z_1} \frac{\der z_g}{z_g} \left(1-\frac{z_g}{1-z_1}\right)^2 \left(1+\frac{z_g}{z_1}+\frac{z_g^2}{2z_1^2} \right)  \nonumber\\
     & \times\int\frac{\der^2\zt}{\pi} \frac{\rzxt \cdot\rzxpt}{\rzxt^2 \rzxpt^2}  K_0(QX_R) K_0(QX_R')  \Xi_{\rm NLO,4}(\xt,\yt,\zt,\xt') \,,
     \label{eq:R1R1}
\end{align}
and
\begin{align}
    &\left.\frac{\der \sigma^{\gamma_{\rm L}^{\star}+A\to q+X}}{  \der^2 \ktone \der \eta_1}\right|_{\rm \overline{R1} \times R1^*} =-\frac{\alpha_{\rm em}e_f^2N_c}{(2\pi)^4 } \int\der^2\xt \der^2\xt' \der^2\yt  e^{-i\ktone \cdot \rxxtp}8z_1^3(1-z_1)^2Q^2\nonumber\\
    &\times\frac{\alpha_s}{\pi}\int_{z_0}^{1-z_1} \frac{\der z_g}{z_g} \left(1-\frac{z_g}{1-z_1}\right)^2\left(1+\frac{z_g}{2z_1}+\frac{z_g}{2(1-z_1-z_g)} \right) \nonumber\\
     & \times\int\frac{\der^2\zt}{\pi} \frac{\rzyt \cdot\rzxpt}{\rzyt^2 \rzxpt^2}  K_0(QX_R) K_0(QX_R')  \Xi_{\rm NLO,4}(\xt,\yt,\zt,\xt') \,.
     \label{eq:barR1R1}
\end{align}

\subsubsection{Gluon-jet contributions}
\label{eq:gluon-jets}

We collect here the real corrections to the NLO semi-inclusive single-jet production cross-section where the measured jet is sourced by the final state gluon. We notice that the quark-antiquark symmetric diagrams give the same expression modulo the replacement $\Xi\to \Xi^*$ in the CGC color correlator. Therefore, we will give the result for the sum of a given diagram plus its $q\bar q$ exchanged counterpart\footnote{We also neglect the diagram where, at amplitude level, the photon splits into quark-antiquark pair that interacts with the shock-wave and later recombines to a final state gluon. This diagram vanishes when summing over the light quark charges.}.

We first consider diagrams $\rm R2\times R2^*$ and $\rm\overline{R2}\times\overline{R2}^*$ which have a collinear singularity. Since we do not want to double count the collinear phase space where the gluon and the quark (respectively, the antiquark) belong to the same jet (already accounted for by Eq.\,\eqref{eq:dijet-NLO-real-collinardiv}) we explicitly exclude this phase space when integrating out the two fermions via the constraint
\begin{equation}
    \Ccal_{gi}^2\ge R^2\kgt^2\frac{z_i^2}{z_g^2} \,,
\end{equation}
with $i=1,2$ for  $\rm R2\times R2^*$ and $\rm \overline{R2}\times\overline{R2}^*$ respectively. This yields
\begin{align}
       &\left.\frac{\der\sigma^{\gamma^*_{\rm L}+A\to g+X}}{\der^2 \kgt\der \eta_g}\right|_{\rm R2\times R2^*+\overline{R2}\times \overline{R2}^*}=\frac{\alpha_{\rm em}e_f^2N_c}{(2\pi)^4}\int\der^2\xt\der^2\xt'\der^2\yt e^{-i\kgt\cdot\rxxtp}2\mathfrak{Re}\left[\Xi_{\rm LO}(\xt,\yt,\xt')\right ]\nonumber\\
        &\times \frac{(-\alpha_s)C_F}{\pi}\int_0^{1-z_g}\frac{\der z_1}{z_g^2} \ e^{-i\frac{z_1}{z_g}\kgt\rxxtp}\left(1-z_1-z_g\right)^2\left(z_1+z_g\right)^2\left(2z_1(z_1+z_g)+z_g^2\right)\nonumber\\
    &\times 4Q^2K_0(\bar Q_{\mathrm{R}2}r_{x'y})K_0(\bar Q_{\mathrm{R}2}r_{xy})\ln\left(\frac{\kgt^2\rxxtp^2R^2z_1^2}{c_0^2z_g^2}\right) \,, \label{eq:gjet-R2R2}
\end{align}
up to power of $R^2$ corrections.
For the other diagrams, we rely on the small $R$ approximation and integrate over the full phase space of the two fermions, as the phase space where one of the fermion and the gluon are clustered together is power of $R^2$ suppressed when the diagram has no collinear divergence. For the cross-diagram $\rm R2\times\overline{R2}^*$, a straightforward integration gives
\begin{align}
    &\left.\frac{\der\sigma^{\gamma^*_{\rm L}+A\to g+X}}{\der^2 \kgt\der \eta_g}\right|_{\rm R2\times\overline{R2}^*+\overline{R2}\times R2^*}=\frac{\alpha_{\rm em}e_f^2N_c}{(2\pi)^4}\int\der^2\xt\der^2\yt\der^2\xt'\der^2\yt' e^{-i\kgt\cdot\rxyt}\frac{\ryytp\cdot\rxxtp}{\ryytp^2\rxxtp^2}\nonumber\\
    &\times\frac{\alpha_s}{\pi^2}\int_0^{1-z_g}\frac{\der z_1}{z_g^2}e^{-i\frac{\kgt}{z_g}\cdot\left((1-z_1)\ryytp+z_1\rxxtp\right)} \ (1-z_1-z_g)(2z_1(1-z_1-z_g)+z_g(1-z_g))\nonumber\\
    &\times 4z_1(1-z_1)(z_1+z_g)Q^2K_0(\bar Q r_{xy})K_0(\bar Q_{\rm R2}r_{x'y'})2\mathfrak{Re}\left[\Xi_{\rm NLO,3}(\xt,\yt;\xt',\yt')\right] \,. \label{eq:gjet-R2R2'}
\end{align}
The diagrams where the tagged gluon crosses the shock-wave either in the amplitude or in the complex conjugate amplitude give 
\begin{align}
     &\left.\frac{\der \sigma^{\gamma_{\rm L}^{\star}+A\to g+X}}{  \der^2 \kgt \der \eta_g}\right|_{\rm R1 \times R2^*+\overline{R1}\times\overline{R2}^*} =-\frac{\alpha_{\rm em}e_f^2N_c}{(2\pi)^4 }\int\der^2\zt\der^2\zt'\der^2\yt  e^{-i\kgt \cdot \rzzpt}\nonumber\\
    &\times \frac{\alpha_s}{\pi}\int_0^{1-z_g}\der z_1 \ 8z_1^2(1-z_1)^2Q^2\left (1-\frac{z_g}{1-z_1}\right)^2\left(1+\frac{z_1}{z_g}\right) \left(1+\frac{z_g}{z_1}+\frac{z_g^2}{2z_1^2} \right) \nonumber\\
     & \times\int\frac{\der^2\xt}{\pi}e^{i\frac{z_1}{z_g}\kgt\cdot\rzpxt} K_0(\bar{Q}_{\rm R2}r_{z'y}) \frac{\rzxt\cdot \rzpxt}{\rzxt^2 \rzpxt^2} K_0(QX_{R})  2\mathfrak{Re}\left[\Xi_{\rm NLO,1}(\xt,\yt,\zt,\zt')\right] \,, \label{eq:gjet-R1R2}
\end{align}
\begin{align}
     &\left.\frac{\der \sigma^{\gamma_{\rm L}^{\star}+A\to g+X}}{  \der^2 \kgt \der \eta_g}\right|_{\rm \overline{R1} \times R2^*+R1\times\overline{R2}^*} =\frac{\alpha_{\rm em}e_f^2N_c}{(2\pi)^4 }\int\der^2\zt\der^2\zt'\der^2\yt  e^{-i\kgt \cdot \rzzpt}\nonumber\\
    &\times \frac{\alpha_s}{\pi}\int_0^{1-z_g}\der z_1 \ 8z_1^2(1-z_1)^2Q^2\left (1-\frac{z_g}{1-z_1}\right)^2\left(1+\frac{z_1}{z_g}\right) \left(1+\frac{z_g}{2z_1}+\frac{z_g}{2(1-z_1-z_g)} \right) \nonumber\\
     & \times\int\frac{\der^2\xt}{\pi}e^{i\frac{z_1}{z_g}\kgt\cdot\rzxpt} K_0(\bar{Q}_{\rm R2}r_{z'y}) \frac{\rzyt\cdot \rzpxt}{\rzyt^2 \rzpxt^2} K_0(QX_{R})  2\mathfrak{Re}\left[\Xi_{\rm NLO,1}(\xt,\yt,\zt,\zt')\right] \,. \label{eq:gjet-R1R2'}
\end{align}
Finally, the formula for the gluon-tagged single-jet semi-inclusive cross-section coming from diagrams where the gluon interacts with the shock-wave both in the amplitude and in the complex conjugate amplitude are
\begin{align}
        &\left.\frac{\der \sigma^{\gamma_{\rm L}^{\star}+A\to g+X}}{  \der^2 \kgt \der \eta_g}\right|_{\rm R1 \times R1^*+\overline{R1}\times\overline{R1}^*} =\frac{\alpha_{\rm em}e_f^2N_c}{(2\pi)^4 }\int\der^2\zt\der^2\zt' \der^2\yt  e^{-i\kgt \cdot \rzzpt}\nonumber\\
     &\times\frac{\alpha_s}{\pi}\int_{0}^{1-z_g} \der z_1 8z_1^2(1-z_1)^2Q^2 \left(1-\frac{z_g}{1-z_1}\right)^2 \left(1+\frac{z_g}{z_1}+\frac{z_g^2}{2z_1^2} \right)  \nonumber\\
     & \times\int\frac{\der^2\xt}{\pi} \frac{\rzxt \cdot\rzpxt}{\rzxt^2 \rzpxt^2}  K_0(QX_{R}) K_0(QX_{R}')  2\mathfrak{Re}\left[\Xi_{\rm NLO,4g}(\xt,\yt,\zt,\zt')\right] \,, \label{eq:gjet-R1R1}
\end{align}
\begin{align}
    &\left.\frac{\der \sigma^{\gamma_{\rm L}^{\star}+A\to g+X}}{  \der^2 \kgt \der \eta_g}\right|_{\rm \overline{R1} \times R1^*+R1\times\overline{R1}^*} =-\frac{\alpha_{\rm em}e_f^2N_c}{(2\pi)^4 } \int \der^2\zt\der^2\zt'\der^2\yt  e^{-i\kgt \cdot \rzzpt}\nonumber\\
    &\times\frac{\alpha_s}{\pi}\int_{0}^{1-z_1}\der z_1 \ 8z_1^2(1-z_1)^2Q^2\left(1-\frac{z_g}{1-z_1}\right)^2\left(1+\frac{z_g}{2z_1}+\frac{z_g}{2(1-z_1-z_g)} \right) \nonumber\\
    &\times\int\frac{\der^2\xt}{\pi} \frac{\rzyt \cdot\rzpxt}{\rzyt^2 \rzpxt^2}  K_0(QX_{R}) K_0(QX_{R}') 2\mathfrak{Re}\left[\Xi_{\rm NLO,4g}(\xt,\yt,\zt,\zt')\right] \,, \label{eq:gjet-R1R1'}
\end{align}
with the CGC color correlator
\begin{align}
    \Xi_{\rm NLO,4g}(\xt,\yt,\zt,\zt')&=\frac{N_c}{2}\left\langle 1-D_{xz}D_{zy}-D_{yz}D_{zx}+D_{z'z}D_{zz'}\right\rangle\nonumber-\frac{1}{2N_c}\left\langle2-D_{xy}-D_{yx}\right\rangle \,,
\end{align}
which only appears in the gluon-tagged jet cross-section.

\subsection{Cancellation of UV, collinear and soft divergences}

In this subsection, we discuss the cancellation of the various divergences in the real and virtual diagrams. First, the gluon-jet real contributions that we have just computed are finite: they do not possess any divergence neither in the transverse coordinate integration nor in the longitudinal momentum integration over $z_1$. 
Hence, we shall focus on the cancellation of divergences between the virtual diagrams and the real fermion-jet corrections. 

\paragraph{The $1/\varepsilon$ poles.} First, we show that the $1/\varepsilon$ pole in transverse dimensional regularization cancels between the real and virtual cross-section (see Eqs.\,\eqref{eq:dijet-NLO-finite-SE1-V1-SE2uv},\,\eqref{eq:dijet-NLO-real-collinardiv} and \eqref{eq:R1barR1barUV}), as expected in the calculation of a collinear safe observable. In dimensional regularization with the convention that scaleless integrals vanish (see Eq.\,\eqref{eq:R2'xR2'}), we do not distinguish infrared from UV poles. Therefore, we combine the UV poles from both virtual and real diagrams with the infrared pole coming from the collinear singularity of the real cross-section:
\begin{align}
     &\left(\left.\frac{\der \sigma^{\gamma_{\lambda}^{\star}+A\to q+X}}{  \der^2 \ktone \der \eta_1}\right|_{\rm UV}+c.c.\right)+\left.\frac{\der \sigma^{\gamma_{\lambda}^{\star}+A\to q +X}}{ \der^2 \ktone  \der \eta_1 }\right|_{\rm R2\times R2^*,in}+\left.\frac{\der \sigma^{\gamma_{\lambda}^{\star}+A\to q+X}}{ \der^2 \ktone  \der \eta_1 }\right|_{\rm \overline{R1}\times \overline{R1}^*,UV}\nonumber\\
     &=\left.\frac{\alpha_sC_F}{\pi}\frac{\der \sigma^{\gamma_{\lambda}^{\star}+A\to q +X}}{ \der^2 \ktone  \der \eta_1 }\right|_{\rm LO}\times\frac{2}{\varepsilon}\left\{\left(\ln\left(\frac{z_1}{z_0}\right)-\ln\left(\frac{z_2}{z_0}\right)\right)+\left(\frac{3}{4}-\ln\left(\frac{z_1}{z_0}\right)\right)\right.\nonumber\\
     &\left.+\left(\ln\left(\frac{z_2}{z_0}\right)-\frac{3}{4}\right)+\mathcal{O}(\varepsilon)\right\} \,.
\end{align}
The $1/\varepsilon$ pole cancels at the cross-section level for the single-jet semi-inclusive cross-section. One notices that this cancellation separately works for the longitudinal $\der \sigma^{\gamma_{\rm L}^{\star}+A\to q+X}$ and transverse $\der \sigma^{\gamma_{\rm T}^{\star}+A\to \bar q+X}$ cross-sections.

\paragraph{Soft divergences.} The second kind of divergences are soft divergences which appear in our regularization scheme as double logarithms of the cut-off $\Lambda^-=z_0q^-$ in longitudinal momentum. Unlike other regularization schemes, such as dimensional regularization in four dimensions, these divergences cancel out separately among virtual and real corrections \cite{Taels:2022tza,Caucal:2022ulg}. As seen by inspection of Eq.\,\eqref{eq:dijet-NLO-finite-SE1-V1-SE2uv}, the sum of of diagrams $\rm V2\times LO^*$, $\rm SE2\times LO^*$, $\rm\overline{SE2}\times LO^*$ does not have a double logarithmic divergence in $z_0$. Similarly, one observes that 
\begin{align}
    &\left.\frac{\der \sigma^{\gamma_{\lambda}^{\star}+A\to q +X}}{ \der^2 \ktone  \der \eta_1 }\right|_{\rm R2\times R2^*,in}+\left.\frac{\der \sigma^{\gamma_{\lambda}^{\star}+A\to q +X}}{ \der^2 \ktone  \der \eta_1 }\right|_{\rm R2\times R2^*,soft}=\nonumber\\
    &\left.\frac{\alpha_sC_F}{\pi}\frac{\der \sigma^{\gamma_{\lambda}^{\star}+A\to q +X}}{ \der^2 \ktone  \der \eta_1 }\right|_{\rm LO}\times\left\{-\ln^2(z_0)+\ln^2\left(\frac{z_1}{z_0}\right)+\mathcal{O}(\ln(z_0))+\mathcal{O}(1)\right\} \,,
\end{align}
so that the double logarithmic divergence in $z_0$ cancels between the term where the gluon and quark are clustered within the same jet and the term where the gluon forms its own jet but is soft.

\subsection{Rapidity divergences: BK-JIMWLK factorization} 

Last, but not least, we have to address how the single logarithmic divergence in $z_0$ cancels in our calculation. We do not expect complete cancellation of the $z_0$ dependence, as this dependence provides the high energy or rapidity evolution of the dipole operators in the LO cross-section. However, we notice interesting grouping between diagrams, where the $z_0$ dependence cancels. We define the following sum
\begin{align}
    \left.\frac{\der \sigma^{\gamma_{\lambda}^{\star}+A\to q +X}}{ \der^2 \ktone  \der \eta_1 }\right|_{\rm NLO,0}&\equiv  \left(\left.\frac{\der \sigma^{\gamma_{\lambda}^{\star}+A\to q +X}}{ \der^2 \ktone  \der \eta_1 }\right|_{\rm UV}+c.c.\right)+ \left.\frac{\der \sigma^{\gamma_{\lambda}^{\star}+A\to q +X}}{ \der^2 \ktone  \der \eta_1 }\right|_{\rm R2\times R2^*}\nonumber\\
    &+\left.\frac{\der \sigma^{\gamma_{\lambda}^{\star}+A\to q+X}}{ \der^2 \ktone  \der \eta_1 }\right|_{\rm \overline{R1}\times \overline{R1}^*,UV} \,,
\end{align}
where $\rm R2\times R2^*$ is itself the sum of three terms, labeled, ``in", ``soft" and ``reg" given in Eqs.\,\eqref{eq:dijet-NLO-real-collinardiv},\,\eqref{eq:R2R2-outdiv} and \eqref{eq:R2R2-outreg} respectively. This ``NLO,0" term gathers all NLO corrections which share the same color correlator $\Xi_{\rm LO}$ as the LO cross-section. Anticipating our discussion of the rapidity divergence of the NLO contribution associated with the CGC correlator $\Xi_{\rm NLO,4}$ which imposes $\xi=e^{\gamma_E}\rxxtp^2/2$ in the UV regulator of $\rm\overline{R1}\times\overline{R1}^*$, the calculation of the above term yields
\begin{align}
    &\left.\frac{\der \sigma^{\gamma_{\rm L}^{\star}+A\to q+X}}{ \der^2 \ktone  \der \eta_1 }\right|_{\rm NLO,0} =\frac{\alpha_{\mathrm{em}} e_f^2 N_c}{(2\pi)^4}\left[8z_1^3(1-z_1)^2Q^2\right]\int\der^6\Xt e^{-i\ktone\cdot\rxxtp}K_0(\bar Q r_{xy})K_0(\bar Qr_{x'y})\nonumber\\
    &\times\frac{\alpha_sC_F}{\pi}\left\{\ln\left(\frac{z_1}{1-z_1}\right)\ln\left(\frac{\ktone^2r_{xy}r_{x'y}R^2}{c_0^2}\right)-\frac{1}{2}\ln^2\left(\frac{z_1}{1-z_1}\right)-\frac{3}{4}\ln\left(\frac{4\ktone^2\rxxtp^2 R^2}{c_0^2}\right)+7-\frac{2\pi^2}{3}\right.\nonumber\\
    &-\int_0^{1-z_1}\frac{\der z_g}{z_g}\ln\left(\frac{\ktone^2\rxxtp^2R^2z_g^2}{c_0^2z_1^2}\right)\nonumber\\
    &\left.\times\left[e^{-i\frac{z_g}{z_1}\ktone\rxxtp}\left(1-\frac{z_g}{1-z_1}\right)^2\left(1+\frac{z_g}{z_1}\right)^2\left(1+\frac{z_g}{z_1}+\frac{z_g^2}{2z_1^2}\right)\frac{K_0(\bar Q_{\mathrm{R}2}r_{xy})K_0(\bar Q_{\mathrm{R}2}r_{x'y})}{K_0(\bar Qr_{xy})K_0(\bar Q r_{x'y})}-1\right]\right\}\nonumber\\
    &\times  \left \langle D_{xx'} - D_{xy} -  D_{yx'} + 1 \right \rangle \,,
\end{align}
which is free of any divergence as $z_g\to 0$. Unlike the inclusive dijet case \cite{Caucal:2021ent}, the NLO corrections proportional to $\Xi_{\rm LO}$ do not contribute to the high energy evolution of $\Xi_{\rm LO}$.

The same observation can be made for the NLO contributions proportional to $\Xi_{\rm NLO,3}$. This color structure appears in final state gluon exchange such as $\rm V3\times LO^*$ and $\rm R2\times\overline{R2}^*$. A close inspection of the $z_g\to 0$ behaviour of the $z_g$ integrand in Eqs.\,\eqref{eq:V3xLO-final} and \eqref{eq:R2xR2'final} shows that
\begin{align}
    \left.\frac{\der \sigma^{\gamma_{\rm L}^{\star}+A\to q+X}}{ \der^2 \ktone  \der \eta_1 }\right|_{\rm V3\times LO^*}
    &=\frac{\alpha_{\mathrm{em}} e_f^2 N_c}{(2\pi)^4}\int\der^2\xt\der^2\xt'\der^2\yt\der^2\yt' e^{-i\ktone\cdot\left(\rxxtp+\frac{1-z_1}{z_1}\ryytp\right)}\nonumber\\
    &\times\Hcal_{\mathrm{LO}}^{\lambda=\rm L}(z,Q^2,\rxyt,\rxytp)\Xi_{\rm NLO,3}(\xt,\yt,\xt')\nonumber\\
    &\times\frac{\alpha_s}{\pi^2}\int_{z_0}\frac{\der z_g}{z_g}\left\{\frac{1}{\ryytp^2}+\mathcal{O}(z_g)\right\} \,,
\end{align}
\begin{align}
    \left.\frac{\der \sigma^{\gamma_{\rm L}^{\star}+A\to q+X}}{ \der^2 \ktone  \der \eta_1 }\right|_{\rm R2\times \overline{R2}^*}
    &=-\frac{\alpha_{\mathrm{em}} e_f^2 N_c}{(2\pi)^4}\int\der^2\xt\der^2\xt'\der^2\yt\der^2\yt' e^{-i\ktone\cdot\left(\rxxtp+\frac{1-z_1}{z_1}\ryytp\right)}\nonumber\\    &\times\Hcal_{\mathrm{LO}}^{\lambda=\rm L}(z,Q^2,\rxyt,\rxytp)\Xi_{\rm NLO,3}(\xt,\yt,\xt')\nonumber\\
    &\times\frac{\alpha_s}{\pi^2}\int_{z_0}\frac{\der z_g}{z_g}\left\{\frac{1}{\ryytp^2}+\mathcal{O}(z_g)\right\} \,.
\end{align}
Consequently, the logarithmic divergence in $z_0$ cancels between $\rm V3\times LO^*$ and $\rm R2\times\overline{R2}^*$, and likewise for their complex conjugate counterpart. One can also notice that the UV divergence as $\ryytp\to 0$ arising in the "slow" gluon limit cancels as well. It is then natural to group all these terms together, in a contribution that we call $\rm NLO,3$ in reference to the label of the CGC color correlator $\Xi_{\rm NLO,3}$ which is common to all these diagrams:
\begin{align}
    \left.\frac{\der \sigma^{\gamma_{\lambda}^{\star}+A\to q +X}}{ \der^2 \ktone  \der \eta_1 }\right|_{\rm NLO,3}&\equiv  \left.\frac{\der \sigma^{\gamma_{\lambda}^{\star}+A\to q +X}}{ \der^2 \ktone  \der \eta_1 }\right|_{\rm V3\times LO^*}+ \left.\frac{\der \sigma^{\gamma_{\lambda}^{\star}+A\to q +X}}{ \der^2 \ktone  \der \eta_1 }\right|_{\rm R2\times \overline{R2}^*}+c.c. \,.\label{eq:NLO3-def}
\end{align}
The analytic expression for this term is provided in the next subsection, which gathers our final result in the most compact way.

In the end, only the contributions proportional to the CGC correlators $\Xi_{\rm NLO,1}$ and $\Xi_{\rm NLO,4}$ have an intrinsic $\ln(z_0)$ divergence which does not cancel between diagrams. Let us define 
\begin{align}
    \left.\frac{\der \sigma^{\gamma_{\lambda}^{\star}+A\to q +X}}{ \der^2 \ktone  \der \eta_1 }\right|_{\rm NLO,1}&\equiv\left.\frac{\der \sigma^{\gamma_{\rm L}^{\star}+A\to q+X}}{  \der^2 \ktone \der \eta_1}\right|_{\rm SE1\times LO^*,reg}+\left.\frac{\der \sigma^{\gamma_{\rm L}^{\star}+A\to q+X}}{  \der^2 \ktone \der \eta_1}\right|_{\rm V1\times LO^*} \nonumber\\
    &+\left.\frac{\der \sigma^{\gamma_{\rm L}^{\star}+A\to q+X}}{  \der^2 \ktone \der \eta_1}\right|_{\rm R1 \times R2^*}+\left.\frac{\der \sigma^{\gamma_{\rm L}^{\star}+A\to q+X}}{  \der^2 \ktone \der \eta_1}\right|_{\rm \overline{R1} \times R2^*}+c.c.\,,\\
    \left.\frac{\der \sigma^{\gamma_{\lambda}^{\star}+A\to q +X}}{ \der^2 \ktone  \der \eta_1 }\right|_{\rm NLO,4}&\equiv \left.\frac{\der \sigma^{\gamma_{\rm L}^{\star}+A\to q+X}}{ \der^2 \ktone  \der \eta_1 }\right|_{\rm \overline{R1}\times \overline{R1}^*,reg}+\left.\frac{\der \sigma^{\gamma_{\rm L}^{\star}+A\to q+X}}{  \der^2 \ktone \der \eta_1}\right|_{\rm R1 \times R1^*}\nonumber\\
    &+\left(\left.\frac{\der \sigma^{\gamma_{\rm L}^{\star}+A\to q+X}}{  \der^2 \ktone \der \eta_1}\right|_{\rm \overline{R1} \times R1^*}+c.c.\right) \,.
\end{align}
Their complete analytic expressions are provided in the next subsection. To properly isolate the leading logarithmic divergence as $z_0\to 0$, we introduce an arbitrary rapidity factorization scale $z_f=k_f^-/q^-$ for the upper bound of the singular $z_g$ integral. Isolating the leading logarithmic $z_g\to 0$ divergence in the integrals over $z_g$ defining these two terms, we get
\begin{align}
    &\left.\frac{\der \sigma^{\gamma_{\lambda}^{\star}+A\to q +X}}{ \der^2 \ktone  \der \eta_1 }\right|_{\rm NLO,1}=\frac{\alpha_{\rm em}e_f^2N_c}{(2\pi)^4} \int\der^2\xt \der^2\xt' \der^2\yt \ e^{-i\ktone \cdot \rxxtp} \Hcal(z_1,Q,\rxyt,\rxpyt)\nonumber\\
     & \times\frac{\alpha_s}{\pi}\int_{z_0}^{z_f} \frac{\der z_g}{z_g}\int\frac{\der^2\zt}{\pi} \left\{\left[ -\frac{\rzxt\cdot \rzyt}{\rzxt^2 \rzyt^2}  +\frac{1}{\rzxt^2}-\frac{\rzxt\cdot\rzxpt}{\rzxt^2\rzxpt^2}+\frac{\rzyt\cdot\rzxpt}{\rzyt^2\rzxpt^2}\right]\Xi_{\rm NLO,1}(\xt,\yt,\zt,\xt')\right.\nonumber\\
     &\left.-\frac{1}{\rzxt^2}\exp\left(-\frac{\rzxt^2}{e^{\gamma_E}\rxyt^2}\right)C_F\Xi_{\rm LO}(\xt,\yt,\xt')\right\}+c.c. \,,
\end{align}
\begin{align}     
     &\left.\frac{\der \sigma^{\gamma_{\lambda}^{\star}+A\to q +X}}{ \der^2 \ktone  \der \eta_1 }\right|_{\rm NLO,4}=\frac{\alpha_{\rm em}e_f^2N_c}{(2\pi)^4} \int\der^2\xt \der^2\xt' \der^2\yt \ e^{-i\ktone \cdot \rxxtp} \Hcal(z_1,Q,\rxyt,\rxpyt)\nonumber\\
     & \times\frac{\alpha_s}{\pi}\int_{z_0}^{z_f} \frac{\der z_g}{z_g}\int\frac{\der^2\zt}{\pi} \left\{\left[\frac{1}{\rzyt^2}+\frac{\rzxt\cdot\rzxpt}{\rzxt^2\rzxpt^2}-\frac{\rzyt\cdot\rzxpt}{\rzyt^2\rzxpt^2}-\frac{\rzyt\cdot\rzxt}{\rzyt^2\rzxt^2}\right]\Xi_{\rm NLO,4}(\xt,\yt,\zt,\xt')\right.\nonumber\\
     &\left.-\frac{1}{\rzyt^2}\exp\left(-\frac{\rzyt^2}{2\xi}\right)C_F\Xi_{\rm LO}(\xt,\yt,\xt')\right\} \,.
\end{align}
We remind that the transverse coordinate scale squared $\xi$ is arbitrary since the $\xi$-dependent term is added and subtracted to the total cross-section. After some straightforward algebra, making use of the identity
\begin{equation}
    \frac{\rzxt\cdot\rzyt}{\rzxt^2\rzyt^2}=\frac{1}{2}\left[-\frac{\rxyt^2}{\rzxt^2\rzyt^2}+\frac{1}{\rzxt^2}+\frac{1}{\rzyt^2}\right] \,,
\end{equation}
the two previous equations simplify to
\begin{align}
    &\left.\frac{\der \sigma^{\gamma_{\lambda}^{\star}+A\to q +X}}{ \der^2 \ktone  \der \eta_1 }\right|_{\rm NLO,1}=\frac{\alpha_{\rm em}e_f^2N_c}{(2\pi)^4} \int\der^2\xt \der^2\xt' \der^2\yt \ e^{-i\ktone \cdot \rxxtp} \Hcal(z_1,Q,\rxyt,\rxpyt)\nonumber\\
    & \times\frac{\alpha_s}{\pi}\int_{z_0}^{z_f} \frac{\der z_g}{z_g}\int\frac{\der^2\zt}{\pi} \left\{\left[ \frac{\rxyt^2}{\rzxt^2 \rzyt^2} + \frac{\rxxtp^2}{\rzxt^2 \rzxpt^2} - \frac{\ryxpt^2}{\rzyt^2 \rzxpt^2} \right]\Xi_{\rm NLO,1}(\xt,\yt,\zt,\xt')\right.\nonumber\\
     &\left.-\frac{1}{\rzxt^2}\exp\left(-\frac{\rzxt^2}{e^{\gamma_E}\rxyt^2}\right)C_F\Xi_{\rm LO}(\xt,\yt,\xt')\right\}+c.c. \,,
\end{align}
\begin{align}     
     &\left.\frac{\der \sigma^{\gamma_{\lambda}^{\star}+A\to q +X}}{ \der^2 \ktone  \der \eta_1 }\right|_{\rm NLO,4}=\frac{\alpha_{\rm em}e_f^2N_c}{(2\pi)^4} \int\der^2\xt \der^2\xt' \der^2\yt \ e^{-i\ktone \cdot \rxxtp} \Hcal(z_1,Q,\rxyt,\rxpyt)\nonumber\\
     & \times\frac{\alpha_s}{\pi}\int_{z_0}^{z_f} \frac{\der z_g}{z_g}\int\frac{\der^2\zt}{\pi} \left\{\left[ \frac{\rxyt^2}{\rzxt^2 \rzyt^2} - \frac{\rxxtp^2}{\rzxt^2 \rzxpt^2} + \frac{\ryxpt^2}{\rzyt^2 \rzxpt^2} \right]\Xi_{\rm NLO,4}(\xt,\yt,\zt,\xt')\right.\nonumber\\
     &\left.-\frac{1}{\rzyt^2}\exp\left(-\frac{\rzyt^2}{2\xi}\right)C_F\Xi_{\rm LO}(\xt,\yt,\xt')\right\} \,.
\end{align}
Then we find that the sum of these two contributions is
\begin{align}
    &\left.\frac{\der \sigma^{\gamma_{\lambda}^{\star}+A\to q +X}}{ \der^2 \ktone  \der \eta_1 }\right|_{\rm NLO,1}+\left.\frac{\der \sigma^{\gamma_{\lambda}^{\star}+A\to q +X}}{ \der^2 \ktone  \der \eta_1 }\right|_{\rm NLO,4}&\nonumber\\
    &=\frac{\alpha_{\rm em}e_f^2N_c}{(2\pi)^4} \int\der^2\xt \der^2\xt' \der^2\yt \ e^{-i\ktone \cdot \rxxtp} \Hcal(z_1,Q,\rxyt,\rxpyt)\ln\left(\frac{z_f}{z_0}\right)\frac{\alpha_sN_c}{2\pi}\int\frac{\der^2\zt}{\pi} \nonumber\\
     & \times\left[ \frac{\rxxtp^2}{\rzxpt^2\rzxt^2} \left\langle D_{zx}D_{zx'} - D_{xx'} \right\rangle -\frac{\rxyt^2}{\rzxt^2\rzyt^2} \left\langle  D_{zx}D_{zy} - D_{xy}  \right\rangle -\frac{\rxpyt^2}{\rzxpt^2\rzyt^2} \left\langle D_{zy}D_{zx'} - D_{yx'}  \right\rangle \right.\nonumber\\
     &\left.+\frac{C_F}{N_c}\Xi_{\rm LO} \left(\frac{\rxxtp^2}{\rzxpt^2\rzxt^2} +\frac{\rxyt^2}{\rzxt^2\rzyt^2}+\frac{\rxpyt^2}{\rzxpt^2\rzyt^2}-\frac{2e^{-\frac{\rzxt^2}{e^{\gamma_E}\rxyt^2 }}}{\rzxt^2} - \frac{2e^{-\frac{\rzxpt^2}{e^{\gamma_E}\rxpyt^2} }}{\rzxpt^2} - \frac{2e^{-\frac{\rzyt^2}{2 \xi }}}{\rzyt^2}\right)\right] \,.
\end{align}
In the first three terms inside the square bracket, one recognizes the BK kernel applied to the dipoles $D_{xx'}$, $-D_{xy}$ and $-D_{yx'}$. For the particular choice $\xi=e^{\gamma_E}\rxxtp^2/2$ one can rely on the identity
\begin{align}
    \int\der^2\zt\left[\frac{\rxxtp^2}{\rzxpt^2\rzxt^2} +\frac{\rxyt^2}{\rzxt^2\rzyt^2}+\frac{\rxpyt^2}{\rzxpt^2\rzyt^2}-\frac{2e^{-\frac{\rzxt^2}{e^{\gamma_E}\rxyt^2 }}}{\rzxt^2} - \frac{2e^{-\frac{\rzxpt^2}{e^{\gamma_E}\rxpyt^2} }}{\rzxpt^2} - \frac{2e^{-\frac{\rzyt^2}{e^{\gamma_E}\rxxtp^2} }}{\rzyt^2}\right]=0 \,,
\end{align}
to cancel the last term which would not be naturally associated with the small-$x$ evolution of the leading order cross-section. The latter identity can be proven using dimensional regularization along the lines of appendices E in \cite{Hanninen:2017ddy,Caucal:2021ent}.

In the end, we find, to leading logarithmic accuracy,
\begin{align}
    \left.\frac{\der \sigma^{\gamma_{\lambda}^{\star}+A\to j +X}}{ \der^2 \kt  \der \eta }\right|_{\rm NLO}=\ln\left(\frac{z_f}{z_0}\right)K_{\rm JIMWLK}\otimes\left.\frac{\der \sigma^{\gamma_{\lambda}^{\star}+A\to j +X}}{ \der^2 \kt \der \eta }\right|_{\rm LO}+\mathcal{O}(\alpha_s) \,,
\end{align}
where the action of the JIMWLK Hamiltonian on the dipole correlator is given by
\begin{align}
        K_{\rm JIMWLK}\otimes \langle D_{xy}\rangle \equiv \frac{\alpha_sN_c}{2\pi^2}\int\der^2\zt\frac{\rxyt^2}{\rzxt^2\rzyt^2}\left\langle D_{zx}D_{zy}-D_{xy}\right\rangle\,.
\end{align}
The dependence on the arbitrary cut-off $z_0$ is then interpreted following a renormalization group approach. The CGC dipole operator in the LO cross-section is regarded as a bare dipole operator $\langle D_{xy}^0\rangle$ and the rapidity divergence $z_0\to 0$ is absorbed into this bare operator. We then define the renormalized CGC dipole operator at the rapidity factorization scale $Y_f=\ln(z_f)$ such that 
\begin{equation}
    \langle D_{xy}\rangle_{Y_f}=\langle D_{xy}^0\rangle+\frac{\alpha_sN_c}{2\pi^2}\ln\left(\frac{z_f}{z_0}\right)\int\der^2\zt\frac{\rxyt^2}{\rzxt^2\rzyt^2}\left\langle D^0_{zx}D^0_{zy}-D^0_{xy}\right\rangle+\mathcal{O}(\alpha_s^2)\label{eq:Yf-dep-dipole}
\end{equation}
and we re-express the SIDIS cross-section in terms of this renormalized CGC dipole operator. The $z_f$ or $Y_f$ independence of the cross-section translates into the following renormalization group equation
\begin{align}
    \frac{\partial \langle D_{xy}\rangle_{Y_f}}{\partial Y_f}=\frac{\alpha_sN_c}{2\pi^2}\int\der^2\zt\frac{\rxyt^2}{\rzxt^2\rzyt^2}\left\langle D_{zx}D_{zy}-D_{xy}\right\rangle_{Y_f}
\end{align}
which is nothing but the first equation of the Balitsky-JIMWLK~\cite{Balitsky:1995ub,JalilianMarian:1997dw,JalilianMarian:1997jx,Iancu:2000hn,Iancu:2001md} hierarchy. It reduces to the well-known BK equation~\cite{Balitsky:1995ub,Kovchegov:1999yj} in the mean-field or large $N_c$ approximations where $\langle D_{zx}D_{zy}\rangle_{Y_f}\approx \langle D_{zx}\rangle_{Y_f}\langle D_{zy}\rangle_{Y_f}$.

\subsection{NLO impact factor for the double differential cross-section}
\label{sub:final-dblediff}

This section summarizes our analytic expressions for the SIDIS NLO impact factor in the small $R$ approximation, in the case of a longitudinally polarized photon. The expressions for transversely polarized photons are given in appendix~\ref{app:transverse}.
We write our final result as 
\begin{align}
      \frac{\der \sigma^{\gamma_{\lambda}^{\star}+A\to j +X}}{ \der^2 \kt  \der \eta }=\left.\frac{\der \sigma^{\gamma_{\lambda}^{\star}+A\to j +X}}{ \der^2 \kt  \der \eta }\right|_{\rm LO}+ \left.\frac{\der \sigma^{\gamma_{\lambda}^{\star}+A\to j_{f} +X}}{ \der^2 \kt  \der \eta }\right|_{\textrm{NLO}}+ \left.\frac{\der \sigma^{\gamma_{\lambda}^{\star}+A\to j_g +X}}{ \der^2 \kt  \der \eta }\right|_{\rm NLO} \,. \label{eq:qg-decomposition}
\end{align}
The first NLO contribution labeled $\der \sigma^{\gamma_{\lambda}^{\star}+A\to j_{f} +X}|_{\rm NLO}$ comes from quark/antiquark-tagged jets. With respect to the formulae shown in the previous section, we include the contribution $\der \sigma^{\gamma_{\lambda}^{\star}+A\to \bar q +X}$ where the antiquark jet is measured, which amounts to replacing $\Xi$ by $2\mathfrak{Re} \Xi$ as we discussed in the text near Eq.\,\eqref{eq:SIDIS-LO-cross-section}. We further decompose the fermion-tagged NLO contribution according to the CGC correlator upon which they depend:
\begin{align}
    \left.\frac{\der \sigma^{\gamma_{\lambda}^{\star}+A\to j_{f} +X}}{ \der^2 \kt  \der \eta }\right|_{\textrm{NLO}}&=\left.\frac{\der \sigma^{\gamma_{\lambda}^{\star}+A\to j_{f} +X}}{ \der^2 \kt  \der \eta }\right|_{\rm NLO,0}+\left.\frac{\der \sigma^{\gamma_{\lambda}^{\star}+A\to j_{f} +X}}{ \der^2 \kt  \der \eta }\right|_{\rm NLO,3}+\left.\frac{\der \sigma^{\gamma_{\lambda}^{\star}+A\to j_{f} +X}}{ \der^2 \kt  \der \eta }\right|_{\rm NLO,1}\nonumber\\
    &+\left.\frac{\der \sigma^{\gamma_{\lambda}^{\star}+A\to j_{f} +X}}{ \der^2 \kt  \der \eta }\right|_{\rm NLO,4} \,. \label{eq:final-decomposition}
\end{align}
To aid the reader, the CGC color correlators that appear in these terms are summarized in Table~\ref{tab:CGC-correlator}. Note that contrary to the inclusive dijet cross-section \cite{Caucal:2021ent}, the CGC correlator $\Xi_{\rm NLO,2}$ does not contribute (explaining why there is no ``NLO,2" term in Eq.\,\eqref{eq:final-decomposition}) thanks to the cancellations discussed in section~\ref{sub:cancellation}.
Implicit in our notations, these CGC correlators are evaluated at the factorization scale $Y_f$. Indeed, small $x$ kinematics imply that $\alpha_s\ln(z_f/z_0)\sim 1$ in Eq.\,\eqref{eq:Yf-dep-dipole}; hence, the $Y_f$ dependence of the CGC correlators in the LO and NLO impact factors are respectively $\mathcal{O}(1)$ and $\mathcal{O}(\alpha_s)$ corrections which must be taken into account in a complete NLO calculation in the Regge limit\footnote{Based on the same power counting argument, the $Y_f$ dependence of the LO impact factor must be computed using the NLO BK-JIMWLK equation~\cite{Balitsky:2007feb,Kovchegov:2006vj,Balitsky:2013fea,Kovner:2013ona,Kovner:2014lca,Lublinsky:2016meo} in a NLO small-$x$ calculation.}. We also systematically set the longitudinal momentum cut-off $\Lambda^-=z_0q^-$ to 0 when the integral over the longitudinal momentum fraction of the gluon $z_g$ is convergent as $z_g\to 0$. This amounts to neglect $\mathcal{O}(\alpha_sz_0)$ corrections which are power suppressed at small $x$ since $z_0$ is physically of order of $x_{\rm Bj}$ \cite{Caucal:2021ent}.

\begin{table}[tbh]
    \centering
    \begin{tabular}{|c|c|}
    \hline
    $\Xi_{\rm LO}(\xt,\yt,\xt')$ & $\left \langle D_{xx'} - D_{xy} -  D_{yx'} + 1 \right \rangle$\\
    \hline
    $\Xi_{\rm NLO,1}(\xt,\yt,\zt,\xt')$ & \makecell{$\frac{N_c}{2}\left\langle 1-D_{yx'}-D_{xz}D_{zy}+D_{zx'}D_{xz} \right\rangle$\\ $- \frac{1}{2N_c}\left\langle 1-D_{xy}-D_{yx'}+D_{xx'} \right\rangle$}\\
    \hline
    $\Xi_{\rm NLO,3}(\xt,\yt,\xt',\yt')$ & \makecell{$\frac{N_c}{2}\left\langle 1-D_{xy}-D_{y'x'}+D_{xy}D_{y'x'}\right\rangle$ \\ $-\frac{1}{2N_c}\left\langle1-D_{xy}-D_{y'x'}+Q_{xy;y'x'}\right\rangle$} \\
    \hline 
    $\Xi_{\rm NLO,4}(\xt,\yt,\zt,\xt')$ & \makecell{$\frac{N_c}{2} \left\langle D_{xx'}-D_{xz}D_{zy}-D_{yz}D_{zx'}+1\right\rangle$ \\ $-\frac{1}{2N_c}\left\langle D_{xx'} - D_{xy} -  D_{yx'} + 1\right\rangle$}\\
    \hline  
     $\Xi_{\rm NLO,4g}(\xt,\yt,\zt,\zt')$ & \makecell{$\frac{N_c}{2}\left\langle 1-D_{xz}D_{zy}-D_{yz}D_{zx}+D_{z'z}D_{zz'}\right\rangle$ \\ $-\frac{1}{2N_c}\left\langle2-D_{xy}-D_{yx}\right\rangle$}\\
    \hline 
    \end{tabular}
    \caption{Color correlators contributing to the next-to-leading order SIDIS cross-section. $1/N_c^2$~suppressed corrections are displayed in the second line of each row.}\label{tab:CGC-correlator}
\end{table}

Let us quote now our relatively compact expressions for these various terms:
\begin{align}
    &\left.\frac{\der \sigma^{\gamma_{\rm L}^{\star}+A\to j_f+X}}{ \der^2 \kt  \der \eta }\right|_{\rm NLO,0} =\frac{\alpha_{\mathrm{em}} e_f^2 N_c}{(2\pi)^4}\int\der^2\xt\der^2\xt'\der^2\yt e^{-i\kt\cdot\rxxtp}8z^3(1-z)^2Q^2K_0(\bar Q r_{xy})\nonumber\\
    &\times K_0(\bar Qr_{x'y})2\mathfrak{Re}\left[\Xi_{\rm LO}(\xt,\yt,\xt')\right]\times\frac{\alpha_sC_F}{\pi}\left\{-\frac{1}{2}\ln^2\left(\frac{z}{1-z}\right)+7-\frac{2\pi^2}{3}\right.\nonumber\\
    &\left.+\ln\left(\frac{z}{1-z}\right)\ln\left(\frac{\kt^2r_{xy}r_{x'y}R^2}{c_0^2}\right)-\frac{3}{4}\ln\left(\frac{4\kt^2\rxxtp^2 R^2}{c_0^2}\right)\right.-\int_0^{1-z}\frac{\der z_g}{z_g}\ln\left(\frac{\kt^2\rxxtp^2R^2z_g^2}{c_0^2z^2}\right)\nonumber\\
    &\left.\times\left[e^{-i\frac{z_g}{z}\kt\cdot\rxxtp}\frac{(1-z-z_g)^2(z+z_g)^2(2z(z+z_g)+z_g^2)}{2z^4(1-z)^2}\frac{K_0(\bar Q_{\mathrm{R}2}r_{xy})K_0(\bar Q_{\mathrm{R}2}r_{x'y})}{K_0(\bar Qr_{xy})K_0(\bar Q r_{x'y})}-1\right]\right\} \,, \label{eq:NLO0-final}
\end{align}
which is valid up to power of the jet radius $\alpha_s R^2$ corrections. We recall that $c_0=2e^{-\gamma_E}$. At small $R$, the leading $\mathcal{O}(\alpha_s)$ corrections come from the $\ln(R)$ terms. One recognizes in the coefficient of the $\alpha_s\ln(R)$ term a soft logarithm $\ln(z/(1-z))$ (which will be further discussed in the next section) and the finite part $-3/2$ of the quark DGLAP splitting function associated with hard but collinear gluon emissions. In the last term, the $z_g$ integral has no rapidity divergence since the square bracket vanishes when $z_g\to 0$ by construction. 

For $\rm NLO,3$ we obtain after adding Eq.\,\eqref{eq:V3xLO-final} and Eq.\,\eqref{eq:R2xR2'final}:
\begin{align}
    &\left.\frac{\der \sigma^{\gamma_{\rm L}^{\star}+A\to j_f+X}}{ \der^2 \kt  \der \eta }\right|_{\rm NLO,3} =\frac{\alpha_{\mathrm{em}} e_f^2 N_c}{(2\pi)^4}\int\der^2\xt\der^2\xt'\der^2\yt\der^2\yt' e^{-i\kt\cdot\left(\rxxtp+\frac{1-z}{z}\ryytp\right)}4z(1-z)Q^2\nonumber\\
    &\times2\mathfrak{Re}\left[\Xi_{\rm NLO,3}(\xt,\yt,\xt',\yt')\right] \frac{\alpha_s}{\pi^2}\int_{z-1}^{z}\frac{\der z_g}{z_g}(1-z+z_g)(z-z_g)^2K_0(\bar Qr_{x'y'})K_0(\bar Q_{\rm V3}r_{xy})\nonumber\\
    &\times e^{i\frac{z_g}{z}\kt\cdot\rxyt}\frac{(1-z)(1+z_g)\ryytp^2+\rxyt\cdot\ryytp(2z(1-z+z_g)-z_g(1+z_g))}{\ryytp^2\left[(1-z)\ryytp^2+2(1-z)(z-z_g)\ryytp\cdot\rxyt-z_g(z-z_g)\rxyt^2\right]}+c.c. \,.\label{eq:NLO3-final}
\end{align}
The analytic expression Eq.\,\eqref{eq:NLO3-final} is one of our new results, which is remarkably simple given that it gathers the sum of the diagrams $\rm V3\times LO^*$ and $\rm R2\times \overline{R2}^*$ plus their complex conjugate.
The integrals in the two previous equations are convergent, both in longitudinal momentum space as $z_g\to 0$ and in transverse coordinate space. On the contrary, the terms $\rm NLO,1$ and $\rm NLO,4$ have a $z_g\to 0$ singularity, which is cured by BK-JIMWLK evolution~\cite{Balitsky:1995ub,JalilianMarian:1997dw,JalilianMarian:1997jx,Iancu:2000hn,Iancu:2001md} as discussed in the previous section. We thus define the shorthand notation
\begin{equation}
    \int_0^{a}\frac{\der z_g}{z_g}\left\{f(z_g)\right\}_{z_f}\equiv \int_0^a\frac{\der z_g}{z_g}\left[f(z_g)-\Theta(z_f-z_g)f(0)\right] \,,
\end{equation}
for any function $f(z_g)$ which has a well-defined limit as $z_g\to 0$.

NLO,1 is obtained by adding Eqs.\,\eqref{eq:V1LO},\,\eqref{eq:SE1LOreg},\,\eqref{eq:R1R2},\eqref{eq:barR1R2} and their complex conjugates, we find
\begin{align}
    &\left.\frac{\der \sigma^{\gamma_{\rm L}^{\star}+A\to j_f+X}}{ \der^2 \kt  \der \eta }\right|_{\rm NLO,1} =\frac{\alpha_{\mathrm{em}} e_f^2 N_c}{(2\pi)^4}\int\der^2\xt\der^2\xt'\der^2\yt\der^2\zt e^{-i\kt\cdot\rxxtp}8z^3 (1-z)^2Q^2\nonumber\\
    &\times\frac{\alpha_s}{\pi^2}\int_0^{1}\frac{\der z_g}{z_g}\left\{\left[K_0(\bar Q r_{x'y})K_0(QX_V)e^{-i\frac{z_g}{z}\kt \cdot \rzxt}\Theta(z-z_g)\left(-\frac{\rzxt\cdot \rzyt}{\rzxt^2 \rzyt^2}\left(1+\frac{z_g}{1-z}\right)\right.\right.\right.\nonumber\\
    &\left.\times\left(1-\frac{z_g}{z} \right) \left(1-\frac{z_g}{2z} - \frac{z_g}{2(1-z+z_g)} \right)+\frac{1}{\rzxt^2}\left(1-\frac{z_g}{z}+\frac{z_g^2}{2z^2} \right)\right)+ K_0(\bar{Q}_{\rm R2}r_{x'y})K_0(QX_R)\nonumber\\
    &\times e^{-i\frac{z_g}{z}\kt \cdot \rzxpt}\Theta(1-z-z_g)\left (1-\frac{z_g}{1-z}\right)^2\left(1+\frac{z_g}{z}\right)\left(\frac{\rzyt \cdot\rzxpt}{\rzyt^2 \rzxpt^2}\left(1+\frac{z_g}{2z}+\frac{z_g}{2(1-z-z_g)} \right)\right.\nonumber\\
    &\left.\left.-\frac{\rzxt\cdot \rzxpt}{\rzxt^2 \rzxpt^2}  \left(1+\frac{z_g}{z}+\frac{z_g^2}{2z^2} \right)\right)\right]\times 2\mathfrak{Re}\left[\Xi_{\rm NLO,1}(\xt,\yt,\zt,\xt')\right]-K_0(\bar Q r_{x'y})K_0(\bar Q r_{xy})\nonumber\\
    &\left.\times \Theta(z-z_g)\frac{e^{-\rzxt^2/(\rxyt^2e^{\gamma_E}})}{\rzxt^2}\left(1-\frac{z_g}{z}+\frac{z_g^2}{2z^2} \right) 2\mathfrak{Re}\left[C_F\Xi_{\rm LO}(\xt,\yt,\xt')\right]\right\}_{z_f}+c.c. \,,\label{eq:NLO1-final}
\end{align}
and NLO,4 is obtained by adding Eqs.\,\eqref{eq:R1R1},\,\eqref{eq:barR1R1} (and complex conjugate),  and \eqref{eq:barR1barR1reg}, we find
\begin{align}
    &\left.\frac{\der \sigma^{\gamma_{\rm L}^{\star}+A\to j_f+X}}{ \der^2 \kt  \der \eta }\right|_{\rm NLO,4} =\frac{\alpha_{\mathrm{em}} e_f^2 N_c}{(2\pi)^4}\int\der^2\xt\der^2\xt'\der^2\yt\der^2\zt e^{-i\kt\cdot\rxxtp}8z^3 (1-z)^2Q^2\nonumber\\
    &\times \frac{\alpha_s}{2\pi^2}\int_0^{1-z}\frac{\der z_g}{z_g}\left\{\left(1-\frac{z_g}{1-z}\right)^2K_0(QX_R)K_0(QX_R')\left[\frac{\rzxt \cdot\rzxpt}{\rzxt^2 \rzxpt^2} \left(1+\frac{z_g}{z}+\frac{z_g^2}{2z^2} \right)\right.\right.\nonumber\\
    &\left. +\frac{1}{\rzyt^2} \left(1+\frac{z_g}{1-z-z_g}+\frac{z_g^2}{2(1-z-z_g)^2}\right)-2\frac{\rzyt \cdot\rzxpt}{\rzyt^2 \rzxpt^2}   \left(1+\frac{z_g}{2z}+\frac{z_g}{2(1-z-z_g)} \right)\right] \nonumber\\
    &\times2\mathfrak{Re}\left[\Xi_{\rm NLO,4}(\xt,\yt,\zt,\xt')\right]-K_0(\bar Q r_{xy})K_0(\bar Q r_{x'y})\left(1-\frac{z_g}{1-z}\right)^2\frac{e^{-\rzyt^2/(\rxxtp^2e^{\gamma_E}})}{\rzyt^2}\nonumber\\
    &\left.\times 
    \left(1+\frac{z_g}{1-z-z_g}+\frac{z_g^2}{2(1-z-z_g)^2} \right) 2\mathfrak{Re}\left[C_F\Xi_{\rm LO}(\xt,\yt,\xt')\right]\right\}_{z_f}+c.c. \,.\label{eq:NLO4-final}
\end{align}
The table~\ref{tab:notations} gathers our shorthand notations for the effective virtualities and dipole sizes at NLO.
\begin{table}[tbh]
    \centering
    \begin{tabular}{|c|c|}
    \hline
    $\bar Q^2$ & $z(1-z)Q^2$\\
    \hline
    $\bar Q_{\rm R2}^2$  & $(1-z-z_g)(z+z_g)Q^2$ \\
    \hline
    $\bar Q_{\rm V3}^2$ & $(1-z+z_g)(z-z_g)Q^2$ \\
    \hline
    $X_V^2$ & $(1-z)(z-z_g)\rxyt^2 + z_g(z-z_g)\rzxt^2 +(1-z)z_g\rzyt^2$\\
    \hline
    $X_R^2$& $z(1-z-z_g)\rxyt^2+zz_g \rzxt^2+(1-z-z_g)z_g\rzyt^2$ \\
    \hline 
    $X_R'^2$ & $z(1-z-z_g)\rxpyt^2+zz_g \rzxpt^2+(1-z-z_g)z_g\rzyt^2$\\
    \hline 
    \end{tabular}
    \caption{Summary of our notations for the kinematic variables appearing in the NLO SIDIS cross-section in coordinate space.}\label{tab:notations}
\end{table}

Finally, the gluon-tagged SIDIS cross-section is also decomposed as
\begin{align}
    \left.\frac{\der \sigma^{\gamma_{\lambda}^{\star}+A\to j_{g} +X}}{ \der^2 \kt  \der \eta }\right|_{\textrm{NLO}}&=\left.\frac{\der \sigma^{\gamma_{\lambda}^{\star}+A\to j_{g} +X}}{ \der^2 \kt  \der \eta }\right|_{\rm NLO,0}+\left.\frac{\der \sigma^{\gamma_{\lambda}^{\star}+A\to j_{g} +X}}{ \der^2 \kt  \der \eta }\right|_{\rm NLO,3}+\left.\frac{\der \sigma^{\gamma_{\lambda}^{\star}+A\to j_{g} +X}}{ \der^2 \kt  \der \eta }\right|_{\rm NLO,1}\nonumber\\
    &+\left.\frac{\der \sigma^{\gamma_{\lambda}^{\star}+A\to j_{g} +X}}{ \der^2 \kt  \der \eta }\right|_{\rm NLO,4g} \,,
\end{align}
with ${\rm NLO,0}$ given by Eq.\,\eqref{eq:gjet-R2R2}, ${\rm NLO,3}$ given by Eq.\,\eqref{eq:gjet-R2R2'},  ${\rm NLO,1}$ given by the sum of Eqs.\,\eqref{eq:gjet-R1R2}-\eqref{eq:gjet-R1R2'} plus their complex conjugate and  ${\rm NLO,4g}$ given by the sum of Eqs.\,\eqref{eq:gjet-R1R1}-\eqref{eq:gjet-R1R1'}. For the sake of self-containment of this summary section, we quote below the expressions of these four terms. The NLO,0 term reads
\begin{align}
     &\left.\frac{\der\sigma^{\gamma^*_{\rm L}+A\to j_g+X}}{\der^2 \kt\der \eta}\right|_{\rm NLO,0}=\frac{\alpha_{\rm em}e_f^2N_c}{(2\pi)^4}\int\der^2\xt\der^2\xt'\der^2\yt e^{-i\kt\cdot\rxxtp}2\mathfrak{Re}\left[\Xi_{\rm LO}(\xt,\yt,\xt')\right ]\nonumber\\
        &\times \frac{(-\alpha_s)C_F}{\pi}\int_0^{1-z}\frac{\der z_1}{z^2} \ e^{-i\frac{z_1}{z}\kgt\rxxtp}\left(1-z_1-z\right)^2\left(z_1+z\right)^2\left(2z_1(z_1+z)+z^2\right)\nonumber\\
    &\times 4Q^2K_0(\bar Q_{\mathrm{R}2}r_{x'y})K_0(\bar Q_{\mathrm{R}2}r_{xy})\ln\left(\frac{\kt^2\rxxtp^2R^2z_1^2}{c_0^2z^2}\right)\,,  \label{eq:gjet-NLO0}
\end{align}
where, obviously, $\bar Q_{\rm R2}^2=(1-z_1-z)(z_1+z)Q^2$ since $z_g=z$ here. The NLO,3 term reads
\begin{align}
    &\left.\frac{\der\sigma^{\gamma^*_{\rm L}+A\to j_g+X}}{\der^2 \kt\der \eta}\right|_{\rm NLO,3}=\frac{\alpha_{\rm em}e_f^2N_c}{(2\pi)^4}\int\der^2\xt\der^2\yt\der^2\xt'\der^2\yt' e^{-i\kt\cdot\rxyt}\frac{\ryytp\cdot\rxxtp}{\ryytp^2\rxxtp^2}\nonumber\\
    &\times\frac{\alpha_s}{\pi^2}\int_0^{1-z}\frac{\der z_1}{z^2}e^{-i\frac{\kt}{z}\cdot\left((1-z_1)\ryytp+z_1\rxxtp\right)} \ (1-z_1-z)(2z_1(1-z_1-z)+z(1-z))\nonumber\\
    &\times 4z_1(1-z_1)(z_1+z)Q^2K_0(\bar Q r_{xy})K_0(\bar Q_{\rm R2}r_{x'y'})2\mathfrak{Re}\left[\Xi_{\rm NLO,3}(\xt,\yt;\xt',\yt')\right]\,, \label{eq:gjet-NLO3}
\end{align}
with the effective virtuality $\bar Q^2=z_1(1-z_1)Q^2$ depending on $z_1$ which is integrated over. The NLO,1 term is given by
\begin{align}
     &\left.\frac{\der \sigma^{\gamma_{\rm L}^{\star}+A\to j_g+X}}{  \der^2 \kt \der \eta}\right|_{\rm NLO,1} =\frac{\alpha_{\rm em}e_f^2N_c}{(2\pi)^4 }\int\der^2\zt\der^2\zt'\der^2\xt\der^2\yt  e^{-i\kt \cdot \rzzpt}\frac{\alpha_s}{\pi^2}\int_0^{1-z}\der z_1 \ e^{i\frac{z_1}{z}\kt\cdot\rzpxt} \nonumber\\
    &\times  8z_1^2Q^2\left (1-z_1-z\right)^2\left(1+\frac{z_1}{z}\right)K_0(\bar{Q}_{\rm R2}r_{z'y})K_0(QX_{R})2\mathfrak{Re}\left[\Xi_{\rm NLO,1}(\xt,\yt,\zt,\zt')\right]  \nonumber\\
    &\times \left[\left(1+\frac{z}{2z_1}+\frac{z}{2(1-z_1-z)} \right)\frac{\rzyt\cdot \rzpxt}{\rzyt^2 \rzpxt^2}- \left(1+\frac{z}{z_1}+\frac{z^2}{2z_1^2} \right)\frac{\rzxt\cdot \rzpxt}{\rzxt^2 \rzpxt^2}\right] +c.c.\,. \label{eq:gjet-NLO1}
\end{align}
Finally, the formula for the NLO,4g term is
\begin{align}
        &\left.\frac{\der \sigma^{\gamma_{\rm L}^{\star}+A\to j_g+X}}{  \der^2 \kt \der \eta}\right|_{\rm NLO,4g} =\frac{\alpha_{\rm em}e_f^2N_c}{(2\pi)^4 }\int\der^2\zt\der^2\zt' \der^2\xt\der^2\yt  \ e^{-i\kt \cdot \rzzpt}\nonumber\\
     &\times\frac{\alpha_s}{\pi^2}\int_{0}^{1-z} \der z_1 \ 8z_1^2Q^2 \left(1-z_1-z\right)^2K_0(QX_{R}) K_0(QX_{R}')  2\mathfrak{Re}\left[\Xi_{\rm NLO,4g}(\xt,\yt,\zt,\zt')\right]  \nonumber\\
     &\times\left[\left(1+\frac{z}{z_1}+\frac{z^2}{2z_1^2} \right)\frac{\rzxt \cdot\rzpxt}{\rzxt^2 \rzpxt^2}-\left(1+\frac{z}{2z_1}+\frac{z}{2(1-z_1-z)} \right)\frac{\rzyt \cdot\rzpxt}{\rzyt^2 \rzpxt^2}\right] \,. \label{eq:gjet-NLO4g}
\end{align}
We emphasize that our expressions are all UV and IR finite, and therefore suitable for numerical evaluation. Nevertheless, the number of integrals to perform (4 transverse integrals and one longitudinal integral at most) and the presence of phases like $e^{-i\kt\cdot\rxxtp}$ makes the numerical task very challenging. In section~\ref{sec:eta-diff}, we will provide results for the $\kt$-integrated cross-section which partially overcomes these difficulties.

\paragraph{Quadrupole dependence of the NLO SIDIS impact factor.} To close this section, we would like to briefly comment on the contribution labeled $\rm NLO,3$ which physically comes from final state (real or virtual) gluon exchanged between the quark and the antiquark. Because of the non-trivial exchange of color after the interaction of the quark-antiquark dipole with the shock-wave, this term exhibits a dependence on a higher point correlator, the quadrupole $Q_{xyy'x'}$. This was not expected given that the leading order cross-section only depends on the dipole operator. Note also that in the case of inclusive dijet production, the NLO impact factor does not involve higher point correlators than those appearing at leading order \cite{Caucal:2021ent} (for instance, the NLO impact factor for inclusive dijet does not depend on the sextupole).

It is important to emphasize that this dependence does not affect the leading $\ln(z_f/z_0)\sim \ln(1/x_{\rm Bj})$ term in the NLO result, as final state gluon radiations do not contribute to the rapidity evolution of the color dipoles. Otherwise, the universality of JIMWLK factorization would have been broken. The dependence on the quadrupole appears then only in the NLO impact factor. We nevertheless expect a rather small contribution from this correlator as it is $1/N_c^2$  suppressed relatively to the dipole operators.

The NLO impact factor for the fully inclusive DIS cross-section does not depend on the quadrupole operator though. Therefore, this dependence must cancel when integrating Eq.\,\eqref{eq:NLO3-final} over the transverse momentum and the rapidity of the jet. We will explain how this cancellation occurs in the last section of this paper.

The appearance of the quadrupole in the NLO impact factor is troublesome from a numerical point of view. Indeed, evaluating the quadrupole at a given rapidity scale $Y_f$ is a difficult task as it requires solving the JIMWLK equation. Several approximations are possible. One can rely on the Gaussian approximation~\cite{Blaizot:2004wv,Dumitru:2011vk,Iancu:2011nj,Metz:2011wb,Dominguez:2011br} to express the quadrupole in terms of the dipole, and use the BK equation to evolve the dipole in rapidity. One can also simply ignore this term by relying on the large $N_c$ approximation, in line with the use of the closed form of the BK equation.

\section{The very forward rapidity regime}
\label{sec:discuss}

As mentioned in the introduction, the very forward rapidity regime of the SIDIS process has recently been shown to be strongly sensitive to saturation in the regime $Q^2\gg Q_s^2$ but $\bar Q^2\sim Q_s^2$ \cite{Iancu:2020jch}. This kinematic domain is interesting because it is perturbative --- the virtuality $Q$ of the photon playing the role of the hard scale is much larger than $\Lambda_{\rm QCD}$ --- but nevertheless sensitive to the saturation scale. This regime corresponds to $z$ close to one, since 
\begin{equation}
    1-z\sim \frac{Q_s^2}{Q^2}\ll 1\,.
\end{equation}
The study in \cite{Iancu:2020jch} makes use of the LO result for the SIDIS cross-section at small $x$. In this section, we analyze the impact of QCD radiative corrections on the very forward jet rapidity regime $z\to 1$. We therefore assume $1-z\ll 1$ without any specific hierarchy between $Q$, $k_\perp$ and $Q_s$. In particular, our calculation also holds in the regime $Q^2\gg Q_s^2$.

A first look at Eq.\,\eqref{eq:NLO0-final} reveals the presence of double and single logarithmic NLO corrections of the form
\begin{equation}
    \alpha_s\ln^2(1-z)\,,\quad \alpha_s\ln(1-z)\,,
\end{equation}
which are therefore large in the very forward rapidity regime $1-z\ll 1$. Physically, these large logarithms, dubbed ``threshold logarithms", come from the incomplete cancellation between real and virtual soft gluon corrections when the constraint $1-z\ll 1$ is imposed on the final state. While a virtual gluon is only constrained by the condition $ z_g\le z$, a real one must satisfy $z_g\le 1-z$ due to longitudinal momentum conservation. A back of the envelop calculation similar to the one done in \cite{Iancu:2020jch} is actually sufficient to get the double logarithm in $1-z$ for single-jet production. To double logarithmic accuracy as $z$ goes to 1, the dominant real contribution comes from diagrams with a collinear singularity as the gluon becomes collinear to one of the two fermions (diagrams $\rm R2\times R2^*$ and $\rm \overline{R2}\times\overline{R2}^*$). Such real diagrams have an overall prefactor $\alpha_s C_F/\pi$ and are parametrically of order
\begin{equation}
    \frac{\alpha_sC_F}{\pi}\int_0^{1-z}\frac{\der z_g}{z_g}\int_{z_g^2\kt^2R^2}^{z_g\kt^2}\frac{\der \Ccal^2_{qg}}{\Ccal^2_{qg}}\,,
\end{equation}
where $\Ccal_{qg}\approx \kgt-z_g\kt$ is the collinearity vector between the gluon and the quark or antiquark (see Eq.\,\eqref{eq:collinearity-def}). The lower bound of the $\Ccal^2_{qg}$ integration comes from imposing the gluon to be outside the fermion jet. The upper bound is a consequence of the gluon formation time $1/k_g^+=2z_g q^-/\kgt^2\sim 2z_gq^-/\Ccal_{qg}^2$ being larger than the quark or antiquark formation time $1/k^+=2zq^-/\kt^2\sim 2q^-/\kt^2$. On the other hand, the leading virtual correction in the $z\to 1$ limit is of order
\begin{equation}
    -\frac{\alpha_sC_F}{\pi}\int_0^{z}\frac{\der z_g}{z_g}\int_{z_g^2\kt^2R^2}^{z_g\kt^2}\frac{\der \Ccal^2_{qg}}{\Ccal^2_{qg}}\,,
\end{equation}
where the $\Ccal_{qg}$ boundaries are identical to the real case, since the phase-space $\Ccal^2_{qg}<z_g^2\kt^2R^2$ in the virtual term exactly cancels with the real in-cone phase space as demonstrated in section~\ref{subsub:pole-term}.
This leads to the following mismatch
\begin{equation}
    \frac{\alpha_sC_F}{\pi}\int_{z}^{1-z}\frac{\der z_g}{z_g}\int_{z_g^2\kt^2R^2}^{z_g \kt^2}\frac{\der \Ccal^2_{qg}}{\Ccal^2_{qg}}\approx-\frac{\alpha_s C_F}{2\pi}\ln^2\left(\frac{1}{(1-z)R^2}\right)\,,\label{eq:dble-log-bof}
\end{equation}
which exhibits the double logarithmic enhancement in the threshold limit. Note also the $R$-dependent single logarithm in $1-z$.

We now want to validate this back of the envelop calculation and compute the single logarithmic corrections by explicitly extracting the logarithmically enhanced terms in the NLO corrections which factorize from the LO cross-section. These logarithms only come from NLO terms which depend on the LO CGC operator, that is the NLO coefficient labeled $\rm NLO,0$ in our final result and given by Eq.\,\eqref{eq:NLO0-final}. There could be ``hidden" single logarithms in $1-z$ in the NLO,1 and NLO,3 terms as they are also combinations of real and virtual diagrams\footnote{We do not expect factorization breaking single logarithms from the NLO,4 term as this contribution contains only real emissions.}. If so, these single logs would break the factorization in terms of the LO cross-section, at least in the deep saturation regime $Q \sim k_\perp \sim Q_s$ where the color structure $\Xi_{\rm NLO,1}$ and $\Xi_{\rm NLO,3}$ cannot be further simplified. We leave for the future the analysis of these possible contributions, in particular in the TMD limit $Q\gg k_\perp, Q_s$.

In the $\rm NLO,0$ contribution given by Eq.\,\eqref{eq:NLO0-final}, some terms are already explicitly logarithmic as $z$ gets closer to one, such as the double logarithm $-\alpha_s C_F/(2\pi)\ln^2(1-z)$ which is exactly the double log computed in Eq.\,\eqref{eq:dble-log-bof}. 
It is interesting to notice that our back of the envelop calculation above is not in one to one correspondance with the real and virtual terms which are combined in the NLO,0 contribution given by Eq.\,\eqref{eq:NLO0-final}. For instance, the virtual piece of Eq.\,\eqref{eq:NLO0-final} coming from the free self-energies and vertex corrections before the shock-wave (see Eq.\,\eqref{eq:dijet-NLO-finite-SE1-V1-SE2uv}) already contains a double logarithm in $1-z$ with a positive sign. It is only after combining virtual and real diagrams together within Eq.\,\eqref{eq:NLO0-final} that the resulting double log becomes in agreement with Eq.\,\eqref{eq:dble-log-bof}. This is a specificity of the dipole picture, in which the emergence of double logarithms from phase space constraints on soft gluons often comes from subtle combinations of diagrams, as also noticed for Sudakov double logarithmic suppression in back-to-back dijet production in DIS \cite{Taels:2022tza,Caucal:2022ulg,Caucal:2023fsf}.
We have also checked that the final result for the double logarithm in $1-z$ is universal with respect to the virtual photon polarization (see Eq.\,\eqref{eq:NLO0-final-transverse} in the appendix), but interestingly, the coefficient of the single threshold log is polarization dependent.

 This double logarithm is also in agreement with the calculation in appendix F of \cite{Iancu:2020jch}, however, contrary to our observation, the authors argue there that this double logarithm should cancel in the case of a jet measurement. The reason behind this disagreement lies in the definition of the final state. While we define jets using jet clustering algorithms such as anti-$k_t$ with a free jet radius $R$, the jet in appendix F of \cite{Iancu:2020jch} is "dynamically" defined event by event, with an effective radius $R\sim k_\perp/((1-z)q^-)$ set by the angular separation between the $q\bar q$ pair in the dipole frame. When $z$ goes to 1, this effective radius diverges and violates our small $R\ll 1$ assumption. In practice, using very large jet radii comes with additional difficulties, both on the experimental side (contamination from the background, effect of the kinematic coverage of the detectors, etc) and on the theory side (for instance, mono-jet configurations with $q\bar qg$ configurations lying inside the same jet contribute). We therefore believe that one cannot avoid these threshold logarithms in realistic SIDIS measurements at very forward rapidities.

In addition to the explicit double and single logarithmic terms in Eq.\,\eqref{eq:NLO0-final}, we need to estimate the limiting behavior as $z\to 1$ of the following integral, which may hide single logarithms of $1-z$,
\begin{align}
    &\int_0^{1-z}\frac{\der z_g}{z_g}\left[\frac{(1-z-z_g)^2(z+z_g)^2(2z(z+z_g)+z_g^2)}{2z^4(1-z)^2}\right.\nonumber\\
    &\left.\times e^{-i\frac{z_g}{z}\kt\cdot\rxxtp}\frac{K_0(\bar Q_{\mathrm{R}2}r_{xy})K_0(\bar Q_{\mathrm{R}2}r_{x'y})}{K_0(\bar Qr_{xy})K_0(\bar Q r_{x'y})}-1\right] \ln\left(\frac{\kt^2\rxxtp^2R^2z_g^2}{c_0^2z^2}\right)\,.
\end{align}
Since the quantity inside the square bracket vanishes as $z_g\to0$, a logarithmic behaviour in $1-z$ can only arise from the $\ln(z_g^2)$ dependence in the overall factor. Expanding the integrant for $z\sim 1$ and subsequently for $z_g\to 0$, one gets
\begin{align}
    &\int_0^{1-z}\frac{\der z_g}{z_g}\left[\frac{z_g^2}{(1-z)^2}-\frac{2z_g}{1-z}+...\right]\ln\left(z_g^2\right)=3\ln\left(\frac{1}{1-z}\right)+\mathcal{O}(1)\,,\label{eq:-3thrlog}
\end{align}
where we omit the finite terms as $z\to 1$ in this result. In the end, we have
\begin{align}
        &\left.\frac{\der \sigma^{\gamma_{\rm L}^{\star}+A\to j_f+X}}{ \der^2 \kt  \der \eta }\right|_{\rm NLO,0} =\frac{\alpha_{\mathrm{em}} e_f^2 N_c}{(2\pi)^4}\int\der^2\xt\der^2\xt'\der^2\yt e^{-i\kt\cdot\rxxtp}2\mathfrak{Re}\left[\Xi_{\rm LO}(\xt,\yt,\xt')\right]\nonumber\\
        &\times 8z^3(1-z)^2Q^2K_0(\bar Q r_{xy})K_0(\bar Qr_{x'y})\nonumber\\
        &\times\frac{\alpha_sC_F}{\pi}\left\{-\frac{1}{2}\ln^2\left(\frac{1}{1-z}\right)+\left[\ln\left(\frac{\kt^2r_{xy}r_{x'y}R^2}{c_0^2}\right)-3\right]\ln\left(\frac{1}{1-z}\right)+\mathcal{O}(1)\right\}\,,\label{eq:NLO0-zsim1}
\end{align}
where the $\mathcal{O}(1)$ term gathers all contributions in $\rm NLO,0$ which are not enhanced by logarithms of $1-z$. The first term in the square bracket is the double logarithm, while the second term is the single logarithm in $1-z$. In appendix~\ref{app:single-log}, we show that the NLO contribution proportional to 
\begin{equation}
    \frac{\alpha_s C_F}{\pi}\ln\left(\frac{\kt^2r_{xy}r_{x'y}}{c_0^2}\right)\ln\left(\frac{1}{1-z}\right)\,,
\end{equation}
is indeed a single threshold log correction and we compute this term more explicitly in terms of the dipole operator and a single transverse coordinate integration. In the end, we find that for a longitudinally polarized photon, this single log coefficient can effectively be replaced by $-2\alpha_s C_F/\pi$ up to powers of $(1-z)$ corrections, while it vanishes for a transversely polarized photon.

The single threshold log also depends on the jet radius, via the $\ln(R^2)$ term associated with soft gluon emissions close to the jet boundary. Last the $-3$ piece also depends on the photon polarization; we find that it is not present for a transversely polarized virtual photon using Eq.\,\eqref{eq:NLO0-final-transverse} and following the same mathematical steps as the ones used to derive Eq.\,\eqref{eq:-3thrlog}. 

As such, any numerical attempt to evaluate the NLO SIDIS cross-section in the forward rapidity regime would lead to unphysical results such as a negative cross-section caused by the presence of the large negative NLO corrections associated with these threshold logarithms in $1-z$. However, these leading logarithms do factorize from the LO cross-section. This points towards the possibility of performing simultaneously small-$x$ and threshold resummation at leading log in $x$ and single logarithmic accuracy in $1-z$, via a simple exponentiation of the logarithmically enhanced corrections:
\begin{align}
    \left.\frac{\der \sigma^{\gamma_{\lambda}^{\star}+A\to j+X}}{ \der^2 \kt  \der \eta }\right|_{\rm resum} =\frac{\alpha_{\mathrm{em}} e_f^2 N_c}{(2\pi)^4}&\int\der^2\xt\der^2\xt'\der^2\yt e^{-i\kt\cdot\rxxtp}2\mathfrak{Re}\left \langle D_{xx'} - D_{xy} -  D_{yx'} + 1 \right \rangle_{Y_f} \nonumber\\
    &\times\Hcal_{\mathrm{LO}}^{\lambda}(z,Q^2,\rxyt,\rxpyt)S^\lambda_{\rm thr.}(1-z)\label{eq:dijet-LO-cross-section-resummed} \,.
\end{align}
with the ``Sudakov" for threshold resummation in very forward SIDIS at small-$x$ given by
\begin{equation}
    S^\lambda_{\rm thr.}(1-z)\equiv\exp\left\{-\frac{\alpha_sC_F}{\pi}\left[\frac{1}{2}\ln^2\left(1-z\right)+\left(\ln(R^2)+c^\lambda\right)\ln\left(1-z\right)\right]\right\}\,,\label{eq:Sthr-def}
\end{equation}
with the coefficient $c^\lambda$ equal to $c^{\lambda=\rm L}=-5$ for a longitudinally polarized photon and $c^{\lambda=\rm T}=0$ for a transversely polarized photon. Although we include here the resummation of threshold logarithms in the SIDIS cross-section for longitudinally polarized photons, one should note that the $\lambda=\rm L$ LO SIDIS cross-section is power suppressed in the $z\to 1$ limit as compared to the transversely polarized photon case (see Eqs.\,\eqref{eq:SIDIS-NLO-LLO}-\eqref{eq:SIDIS-NLO-TLO}). Hence, the very forward jet limit is dominated by the $\lambda=\rm T$ component and only $S^{\lambda=\rm T}_{\rm thr.}$ truly matters. One notices that in the end, the single logarithm in $1-z$ for $\lambda=\rm T$ is just $-\alpha_s C_F/\pi\ln(R^2)\ln(1-z)$ in agreement with our estimation in Eq.\,\eqref{eq:dble-log-bof} and Eq.\,(F.4) in \cite{Iancu:2020jch} provided one chooses the quark mass $m$ cutting the collinear singularity as $m/Q=R$ in this expression. Finally, it is relatively straightforward to include running coupling effects in $S^\lambda_{\rm thr.}(1-z)$ writing the argument of the exponential in Eq.\,\eqref{eq:Sthr-def} as
\begin{align}
    \ln\left( S^\lambda_{\rm thr.}(1-z)\right)=-\frac{C_F}{\pi}\int_{1-z}^{1}\frac{\der z_g}{z_g}\left[\int_{z_g^2k_\perp^2}^{z_gk_\perp^2}\frac{\der \mu^2}{\mu^2}\alpha_s(\mu^2)-(\ln(R^2)+c^{\lambda})\alpha_s(z_gk_\perp^2)\right]\,,
\end{align}
according to the standard resummation procedure for threshold logarithms \cite{Sterman:1986aj,Catani:1989ne}.

In Eq.\,\eqref{eq:dijet-LO-cross-section-resummed}, the small-$x$ resummation is performed within the LO cross-section via BK-JIMWLK evolution of the dipole operators ($Y_f$ dependence), while the $\alpha_s\ln^2(1-z)$ and $\alpha_s\ln(1-z)$ are resummed in the exponential factor. Although the exponentiation of soft gluons effects is a general property of pQCD \cite{Gatheral:1983cz,Frenkel:1984pz}, a formal all-order proof of this result at small-$x$ is beyond the scope of the present paper. In this paper, we propose this exponentiation as a natural procedure to cure the instability of the fixed order NLO calculation as $z$ gets close to 1. The other NLO corrections which are not logarithmically enhanced in the $z\to 1$ limit can simply be added to Eq.\,\eqref{eq:dijet-LO-cross-section-resummed}. Unlike threshold resummation in forward hadron production in $pA$ \cite{Xiao:2018zxf,Liu:2020mpy} or for SIDIS in the collinear framework \cite{Altarelli:1979kv,Cacciari:2001cw,Stratmann:2001pb,Sterman:2006hu,Anderle:2012rq}, we do not need to go to Mellin space in order to exponentiate the threshold logarithms, as $z$ is the measured longitudinal momentum fraction of the jet which does not appear in a convolution neither with a parton distribution function nor a fragmentation function. 

Eq.\,\eqref{eq:dijet-LO-cross-section-resummed} tells us that soft radiative corrections suppress the SIDIS cross-section in the very forward rapidity regime. Therefore, radiative effects effectively mimic the signal of gluon saturation in this kinematic domain, since gluon saturation also leads to a suppression of the cross-section for producing very forward jets \cite{Iancu:2020jch}. In spirit, one recovers the competing effect between saturation and soft gluon radiations (via Sudakov suppression) which is also at play for inclusive back-to-back dijet production, another ``golden channel" for gluon saturation searches at the EIC. There is yet a crucial difference between the Sudakov suppression for back-to-back dijets and the suppression in Eq.\,\eqref{eq:dijet-LO-cross-section-resummed}. While the former is a convolution in transverse coordinate space, the latter completely factorizes, at least to double logarithmic accuracy, from the LO cross-section:
\begin{equation}
    \left.\frac{\der \sigma^{\gamma_{\lambda}^{\star}+A\to j+X}}{ \der^2 \kt  \der \eta}\right|_{\rm resum.}=\left.\frac{\der \sigma^{\gamma_{\lambda}^{\star}+A\to j+X}}{ \der^2 \kt  \der \eta}\right|_{\rm LO}\times e^{-\frac{\alpha_sC_F}{2\pi}\ln^2(1-z)}\left[1+\mathcal{O}(\alpha_s\ln(1-z))\right]\,.\label{eq:1-z-resummed}
\end{equation}
Since the exponential factor does not depend on the nuclear target, one expects complete cancellation of soft gluon effects (up to potential single log in $1-z$ corrections arising from the $\rm NLO,1$ and $\rm NLO,3$ terms) when performing the ratio of the SIDIS cross-section between two different nuclear species. This is another advantage of single-jet measurement over single hadron production, as we expect the threshold logarithms to appear inside a convolution between the fragmentation function and the LO cross-section. In such convolutions, there is no guarantee of cancellation of the threshold logarithm effects in nuclear modification factors. 

\section{The transverse momentum integrated SIDIS cross-section at NLO}
\label{sec:eta-diff}

The analytic expressions for the double differential SIDIS cross-section at NLO provided in section \ref{sub:final-dblediff} remain challenging to evaluate because of the exponential phase and the large number of integrals. In order to decrease the complexity of future evaluations of the SIDIS cross-section in the forward limit which is relevant for gluon saturation, one may wish to integrate over the transverse momentum of the measured jet. The $\kt$-integrated single-jet semi-inclusive cross-section has one transverse integral less and no phases anymore.

The $\kt$ integration of the double differential distribution leads to new UV divergences term by term in the decomposition Eq.\,\eqref{eq:final-decomposition} of our final result. These new divergences arise from real graphs where the Weizs\"{a}cker-Williams kernel $\rzxt\cdot\rzxpt/(\rzxt^2\rzxpt^2)$ becomes logarithmically divergent $1/\rzxt^2$ as the transverse coordinate of the quark in the amplitude $\xt$ is identified to the one in the complex conjugate amplitude $\xt'$. Some additional work is required to show that these new divergences cancel among the four terms and to reorganize the calculation in a way that is suitable for numerical evaluation.

\subsection{UV divergences in the $k_\perp$ integrated  NLO,4 and NLO,1 terms}

In the NLO,4 term for fermionic jets, given by Eq.\,\eqref{eq:NLO4-final}, the tranvserse momentum $\kt$ of the jet only appears inside the phase $e^{-i\kt\cdot\rxxtp}$. The integral over $\kt$ then yields a $(2\pi)^2\delta^{(2)}(\rxxtp)$ function which fixes $\xt=\xt'$. Consequently, two UV divergences arise in the resulting expression: (i) one from the term proportional to $\rzxt\cdot\rzxpt/(\rzxt^2\rzxpt^2)$ which becomes $1/\rzxt^2$ after $\kt$ integration, (ii) one from the term proportional to $1/\rzyt^2$ because the UV regulator of the divergence as $\zt\to\yt$, proportional to $\exp(-\rzyt^2/(e^{\gamma_E}\rxxtp^2))$, vanishes when $\xt=\xt'$. One ends up with the following expression, which is not anymore mathematically well-defined:
\begin{align}
    &\left.\frac{\der \sigma^{\gamma_{\rm L}^{\star}+A\to j_f+X}}{   \der \eta }\right|_{\rm NLO,4} =\frac{\alpha_{\mathrm{em}} e_f^2 N_c}{(2\pi)^2}8z^3 (1-z)^2Q^2\frac{\alpha_s}{\pi^2}\int\der^2\xt\der^2\yt\der^2\zt\int_0^{1-z}\frac{\der z_g}{z_g}\nonumber\\
    &\times \left\{\left(1-\frac{z_g}{1-z}\right)^2K_0^2(QX_R)\left[\frac{1}{\rzyt^2} \left(1+\frac{z_g}{1-z-z_g}+\frac{z_g^2}{2(1-z-z_g)^2}\right)-2\frac{\rzyt\cdot\rzxt}{\rzyt^2 \rzxt^2}\right.\right.\nonumber\\
     &\left.\times \left(1+\frac{z_g}{2z}+\frac{z_g}{2(1-z-z_g)} \right)\left.+\frac{1}{\rzxt^2}\left(1+\frac{z_g}{z}+\frac{z_g^2}{2z^2} \right) \right]2\mathfrak{Re}[\Xi_{\rm NLO,4}(\xt,\yt,\zt,\xt)]\right\}_{z_f}\,.\label{eq:NLO4-ktintegrated}
\end{align}
Similarly, the integral over $\kt$ of the NLO,1 term given in Eq.\,\eqref{eq:NLO1-final} yields
\begin{align}
    &\left.\frac{\der \sigma^{\gamma_{\rm L}^{\star}+A\to j_f+X}}{\der \eta }\right|_{\rm NLO,1} =\frac{\alpha_{\mathrm{em}} e_f^2 N_c}{(2\pi)^2}16z^3 (1-z)^2Q^2\int\der^2\xt\der^2\yt\der^2\zt \nonumber\\
    &\times\frac{\alpha_s}{\pi^2}\int_0^{1}\frac{\der z_g}{z_g}\left\{\left[K_0(\bar Q r_{w_vy})K_0(QX_V)\Theta(z-z_g)\left(-\frac{\rzxt\cdot \rzyt}{\rzxt^2 \rzyt^2}\left(1+\frac{z_g}{1-z}\right)\left(1-\frac{z_g}{z} \right) \right.\right.\right.\nonumber\\
    &\left.\times\left(1-\frac{z_g}{2z} - \frac{z_g}{2(1-z+z_g)} \right)+\frac{1}{\rzxt^2}\left(1-\frac{z_g}{z}+\frac{z_g^2}{2z^2} \right)\right)\times 2\mathfrak{Re}\left[\Xi_{\rm NLO,1}(\xt,\yt,\zt,\boldsymbol{w}_{\perp,v})\right]\nonumber\\
    &+ K_0(\bar{Q}_{\rm R2}r_{w_ry})K_0(QX_R) \Theta(1-z-z_g)\left (1-\frac{z_g}{1-z}\right)^2 \left(\frac{\rzyt \cdot\rzxt}{\rzyt^2 \rzxt^2}\left(1+\frac{z_g}{2z}+\frac{z_g}{2(1-z-z_g)} \right)\right.\nonumber\\
    &\left.\left.-\frac{1}{\rzxt^2 }  \left(1+\frac{z_g}{z}+\frac{z_g^2}{2z^2} \right)\right)\right]\times 2\mathfrak{Re}\left[\Xi_{\rm NLO,1}(\xt,\yt,\zt,\boldsymbol{w}_{\perp,r})\right]\nonumber\\
    &\left. -K_0^2(\bar Q r_{xy}) \Theta(z-z_g)\frac{e^{-\rzxt^2/(\rxyt^2e^{\gamma_E}})}{\rzxt^2}\left(1-\frac{z_g}{z}+\frac{z_g^2}{2z^2} \right) 2\mathfrak{Re}[\Xi_{\rm LO}(\xt,\yt,\xt)]\right\}_{z_f} \,,\label{eq:NLO1-ktintegrated}
\end{align}
with \begin{align}
    \boldsymbol{w}_{\perp,v}&=\xt+\frac{z_g}{z}\rzxt\,,\\
    \boldsymbol{w}_{\perp,r}&=\xt+\frac{z_g}{z+z_g}\rzxt\,.
\end{align}
In this expression, one recognizes two UV divergent kernels proportional to $1/\rzxt^2$. The first one, in the third line of Eq.\,\eqref{eq:NLO1-ktintegrated} is regulated by the UV counter-term in the last line, proportional to $\exp\left(-\rzxt^2/(e^{\gamma_E}\rxyt^2)\right)$. However, the other $1/\rzxt^2$ kernel, coming from the integration of the real diagram $\rm R1\times R2^*$ is not canceled by any other term in Eq.\,\eqref{eq:NLO1-ktintegrated}. As it is, Eq.\,\eqref{eq:NLO1-ktintegrated} is then also ill-defined.

The pieces of $\rm NLO,4$ and $\rm NLO,1$ that contain the UV singularities are 
\begin{align}
    &\left.\frac{\der \sigma^{\gamma_{\rm L}^{\star}+A\to j_f+X}}{   \der \eta }\right|_{\rm NLO,4,UV} =\frac{\alpha_{\mathrm{em}} e_f^2 N_c}{(2\pi)^2}8z^3 (1-z)^2Q^2\int\der^2\xt\der^2\yt\der^2\zt \frac{\alpha_s}{\pi^2} \int_0^{1-z}\frac{\der z_g}{z_g}\nonumber\\
    &\times \left\{\left(1-\frac{z_g}{1-z}\right)^2 K_0^2(Q X_{R}) \left[\frac{1}{\rzyt^2} \left(1+\frac{z_g}{1-z-z_g}+\frac{z_g^2}{2(1-z-z_g)^2}\right)  \right. \right. \nonumber \\
    & \left. \left. +\frac{1}{\rzxt^2}\left(1+\frac{z_g}{z}+\frac{z_g^2}{2z^2} \right) \right]2\mathfrak{Re}[\Xi_{\rm NLO,4}(\xt,\yt,\zt,\xt)]\right\}_{z_f}\,,\label{eq:NLO4-ktintegrated-UV} 
\end{align}
and 
\begin{align}
    &\left.\frac{\der \sigma^{\gamma_{\rm L}^{\star}+A\to j_f+X}}{\der \eta }\right|_{\rm NLO,1,UV} =-\frac{\alpha_{\mathrm{em}} e_f^2 N_c}{(2\pi)^2} 16 z^3 (1-z)^2Q^2\int\der^2\xt\der^2\yt\der^2\zt \frac{\alpha_s}{\pi^2}\int_0^{1-z}\frac{\der z_g}{z_g}\nonumber\\
    &\times\left\{  \left (1-\frac{z_g}{1-z}\right)^2 \frac{1}{\rzxt^2 }  \left(1+\frac{z_g}{z}+\frac{z_g^2}{2z^2} \right) K_0(\bar{Q}_{\rm R2}r_{w_ry})K_0(QX_R) \right. \nonumber  \\
    & \times 2\mathfrak{Re}\left[\Xi_{\rm NLO,1}(\xt,\yt,\zt,\boldsymbol{w}_{\perp,r})\right]\Bigg\}_{z_f} \,. \label{eq:NLO1-ktintegrated-UV} 
\end{align}
We now demonstrate that these new UV divergences as $\zt\to \xt,\yt$ in the integrals Eqs.\,\eqref{eq:NLO4-ktintegrated}-\eqref{eq:NLO1-ktintegrated} are canceled by divergences arising from the transverse momentum integration of the NLO,0 term.

\subsection{UV divergences in the $k_\perp$ integrated NLO,0 term}
\label{sec:kTintegratedNLO0}

The $\kt$ dependence of the NLO,0 term, given by Eq.\,\eqref{eq:NLO0-final} is slightly more complicated as $\kt$ appears both inside the phase $e^{-i\kt\cdot\rxxtp}$ and in the argument of logarithms such as $\ln(\kt^2r_{xy}r_{x'y})$. To perform the $\kt$ integral, we rely on the identities in appendix\,\ref{app:useful-id} for the Fourier transform of the logarithm. These identities will allows us to deal with logarithmic singularities by expressing it in a form that will simplify its combination with the logarithmic singularities in Eqs.\,\eqref{eq:NLO4-ktintegrated}-\eqref{eq:NLO1-ktintegrated}.

Owing to these identities (Eqs.\,\eqref{eq:identity1} and \eqref{eq:identity2}), we can integrate the NLO,0 contribution over $\kt$:
\begin{align}
    &\left.\frac{\der \sigma^{\gamma_{\rm L}^{\star}+A\to j_f+X}}{\der \eta }\right|_{\rm NLO,0} =\frac{\alpha_{\mathrm{em}} e_f^2 N_c}{(2\pi)^2}\int\der^2\xt\der^2\yt 8z^3(1-z)^2Q^2K_0^2(\bar Q r_{xy})\nonumber\\
    &\times\frac{\alpha_sC_F}{\pi}\left\{-\frac{1}{2}\ln^2\left(\frac{z}{1-z}\right)+\ln\left(\frac{z}{1-z}\right)\ln\left(R^2\right)-\frac{3}{2}\ln\left(2R\right)+7-\frac{2\pi^2}{3}\right.\nonumber\\
    &-\int_0^{1-z}\frac{\der z_g}{z_g}\ln\left(\frac{R^2z_g^2}{z^2}\right)\left[\left(1-\frac{z_g}{1-z}\right)^2\left(1+\frac{z_g}{z}+\frac{z_g^2}{2z^2}\right)\frac{K_0^2(\bar Q_{\mathrm{R}2}r_{xy})}{K_0^2(\bar Qr_{xy})}-1\right]\nonumber\\
    &\left.+2\int_0^{1-z}\frac{\der z_g}{z_g}\ln\left(1+\frac{z_g}{z}\right)\left(1-\frac{z_g}{1-z}\right)^2\left(1+\frac{z_g}{z}+\frac{z_g^2}{2z^2}\right)\frac{K_0^2(\bar Q_{\mathrm{R}2}r_{xy})}{K_0^2(\bar Qr_{xy})}\right\}\nonumber\\
    &\times  4\mathfrak{Re}\left \langle 1 - D_{xy}\right \rangle_{Y_f}\nonumber\\
    &-\frac{\alpha_{\mathrm{em}} e_f^2 N_c}{(2\pi)^2}\left[8z^3(1-z)^2Q^2\right]\times\frac{\alpha_sC_F}{\pi^2}\ln\left(\frac{z}{1-z}\right)\nonumber\\
    &\times\int\der^2\xt\der^2\yt \mathcal{P}_{\rxyt}\left(\frac{1}{\rxxtp^2}\right)\left[K_0(\bar Q r_{xy})K_0(\bar Q r_{x'y})2\mathfrak{Re}\left \langle D_{xx'} - D_{xy} -  D_{yx'} + 1 \right \rangle_{Y_f}\right]\nonumber\\
    &+\left.\frac{\der \sigma^{\gamma_{\rm L}^{\star}+A\to j_f+X}}{\der \eta }\right|_{\rm NLO,0,UV}\,.
\end{align}
We have used the natural choice $\ut=\rxyt$ in the definition of the distribution $\mathcal{P}_{\ut}$, which simplifies the form of the finite terms. The UV divergent contribution formally reads
\begin{align}
    &\left.\frac{\der \sigma^{\gamma_{\rm L}^{\star}+A\to j_f+X}}{\der \eta }\right|_{\rm NLO,0,UV} = \frac{\alpha_{\mathrm{em}} e_f^2 N_c}{(2\pi)^2}\int\der^2\xt\der^2\yt\frac{\der^2\xt'}{\rxxtp^2} 8z^3(1-z)^2Q^2K_0(\bar Q r_{xy})K_0(\bar Q r_{x'y})\nonumber\\
    &\times\frac{\alpha_sC_F}{\pi^2}\left\{\frac{3}{4}+\int_0^{1-z}\frac{\der z_g}{z_g}\left[\left(1-\frac{z_g}{1-z}\right)^2\left(1+\frac{z_g}{z}+\frac{z_g^2}{2z^2}\right)\frac{K_0(\bar Q_{\mathrm{R}2}r_{xy})K_0(\bar Q_{\mathrm{R}2}r_{x'y})}{K_0(\bar Q r_{xy})K_0(\bar Q r_{x'y})}-1\right]\right\}\nonumber\\
    &\times2\mathfrak{Re}\left \langle D_{xx'} - D_{xy} -  D_{yx'} + 1 \right \rangle_{Y_f}\,.\label{eq:NLO0-UVterm-a}
\end{align}
To more easily identify the cancellation of UV divergences, it is convenient to re-write the $3/4$ term using the following identity:
\begin{align}
    \int_0^{1-z}\frac{\der z_g}{z_g}\left[\left(1-\frac{z_g}{1-z}\right)^2\left(1+\frac{z_g}{1-z-z_g}+\frac{z_g^2}{2(1-z-z_g)^2}\right)-1\right]=-\frac{3}{4}\,,
\end{align}
which help us express Eq.\,\eqref{eq:NLO0-UVterm-a} as
\begin{align}
        &\left.\frac{\der \sigma^{\gamma_{\rm L}^{\star}+A\to j_f+X}}{\der \eta }\right|_{\rm NLO,0,UV} =- \frac{\alpha_{\mathrm{em}} e_f^2 N_c}{(2\pi)^2}\int\der^2\xt\der^2\yt\frac{\der^2\xt'}{\rxxtp^2} 8z^3(1-z)^2Q^2K_0(\bar Q r_{xy})K_0(\bar Q r_{x'y})\nonumber\\
    &\times\frac{\alpha_sC_F}{\pi^2} \int_0^{1-z}\frac{\der z_g}{z_g}\left(1-\frac{z_g}{1-z}\right)^2\left(1+\frac{z_g}{1-z-z_g}+\frac{z_g^2}{2(1-z-z_g)^2}\right)2\mathfrak{Re}\Xi_{\rm LO}(\xt,\yt,\xt')\nonumber\\
   &+\frac{\alpha_{\mathrm{em}} e_f^2 N_c}{(2\pi)^2}\int\der^2\xt\der^2\yt\frac{\der^2\xt'}{\rxxtp^2} 8z^3(1-z)^2Q^2\frac{\alpha_sC_F}{\pi^2}\int_0^{1-z}\frac{\der z_g}{z_g}K_0(\bar Q_{\mathrm{R}2}r_{xy})K_0(\bar Q_{\mathrm{R}2}r_{x'y})\nonumber\\
    &\times\left(1-\frac{z_g}{1-z}\right)^2\left(1+\frac{z_g}{z}+\frac{z_g^2}{2z^2}\right)2\mathfrak{Re}\Xi_{\rm LO}(\xt,\yt,\xt')\,.\label{eq:NLO0-UVterm}
\end{align}
The first term in the above expression must be combined with the $1/\rzyt^2$ divergent term of NLO,4, while the second term must be combined with the $1/\rzxt^2$ divergent terms in NLO,1 and NLO,4. One should note that although the $z_g$ integral  in each term of Eq.\,\eqref{eq:NLO0-UVterm} is logarithmically singular near $z_g=0$, the divergence exactly cancels between the two terms since Eq.\,\eqref{eq:NLO0-UVterm-a} has no rapidity divergence in the first place. The result of the sum of these terms will be presented in the next subsection.

Finally, the integration of the NLO,3 term does not yield any particular difficulty and the outcome is presented in the following summary subsection.

\subsection{NLO impact factor for the single differential cross-section}
\label{sec:singlediff-xsec}

Instead of dividing the $\eta$-differential cross-section in terms of color structure, as we did for the double differential cross-section, we decompose it into 4 terms which should be computed independently in numerical evaluations in order to avoid convergence issues:
\begin{align}
    \frac{\der \sigma^{\gamma_{\rm L}^{\star}+A\to j_f+X}}{\der \eta }&=\left.\frac{\der \sigma^{\gamma_{\rm L}^{\star}+A\to j_f+X}}{\der \eta }\right|_{\rm LO}+\left.\frac{\der \sigma^{\gamma_{\rm L}^{\star}+A\to j_f+X}}{\der \eta }\right|_{\rm NLO-a}+\left.\frac{\der \sigma^{\gamma_{\rm L}^{\star}+A\to j_f+X}}{\der \eta }\right|_{\rm NLO-b}\nonumber\\
    &+\left.\frac{\der \sigma^{\gamma_{\rm L}^{\star}+A\to j_f+X}}{\der \eta }\right|_{\rm NLO-c}+\left.\frac{\der \sigma^{\gamma_{\rm L}^{\star}+A\to j_f+X}}{\der \eta }\right|_{\rm UV-reg}\,.\label{eq:ktintegrated-NLO-decomp}
\end{align}
Combining Eqs.\,\eqref{eq:NLO4-ktintegrated-UV},\,\eqref{eq:NLO1-ktintegrated-UV} and \eqref{eq:NLO0-UVterm} in 
\begin{align}
   \left.\frac{\der \sigma^{\gamma_{\rm L}^{\star}+A\to j_f+X}}{\der \eta }\right|_{\rm UV-reg}&\equiv \left.\frac{\der \sigma^{\gamma_{\rm L}^{\star}+A\to j_f+X}}{\der \eta }\right|_{\rm NLO,0,UV}+\left.\frac{\der \sigma^{\gamma_{\rm L}^{\star}+A\to j_f+X}}{\der \eta }\right|_{\rm NLO,4,UV}\nonumber\\
   &+\left.\frac{\der \sigma^{\gamma_{\rm L}^{\star}+A\to j_f+X}}{\der \eta }\right|_{\rm NLO,1,UV}\,,
\end{align}
the resulting integral is UV convergent, and given by
\begin{align}
    &\left.\frac{\der \sigma^{\gamma_{\rm L}^{\star}+A\to j_f+X}}{  \der \eta }\right|_{\rm UV-reg} =\frac{\alpha_{\mathrm{em}} e_f^2 N_c}{(2\pi)^2}\frac{\alpha_s}{\pi^2}\int\der^2\xt\der^2\yt\der^2\zt \ 16z^3 (1-z)^2Q^2\nonumber\\
    &\times \mathfrak{Re}\int_0^{1-z}\frac{\der z_g}{z_g}\Bigg\{\frac{1}{\rzyt^2}\left(1-\frac{z_g}{1-z}\right)^2 \left(1+\frac{z_g}{1-z-z_g}+\frac{z_g^2}{2(1-z-z_g)^2}\right)\nonumber\\
     &\times\Big[K_0^2(QX_R)\Xi_{\rm NLO,4}(\xt,\yt,\zt,\xt)-K_0(\bar Qr_{xy})K_0\left(\bar Q|\rzyt-\rxyt|\right)C_F\Xi_{\rm LO}(\yt,\xt,\zt)\Big]\nonumber\\
    &+\frac{1}{\rzxt^2}\left(1-\frac{z_g}{1-z}\right)^2 \left(1+\frac{z_g}{z}+\frac{z_g^2}{2z^2}\right)\Big[K_0^2(QX_R)\Xi_{\rm NLO,4}(\xt,\yt,\zt,\xt)\nonumber\\
     &-2K_0(QX_R)K_0\left(\bar Q_{\rm R2}\left|\rxyt+\frac{z_g}{z+z_g}\rzxt\right|\right)\Xi_{\rm NLO,1}(\xt,\yt,\zt,\boldsymbol{w}_{\perp,r})\nonumber\\
     &+K_0(\bar Q_{\rm R2}r_{xy})K_0\left(\bar Q_{\rm R2}\left|\rxyt+\rzxt\right|\right)C_F\Xi_{\rm LO}(\xt,\yt,\zt)\Big]\Bigg\}_{z_f}\,.
\end{align}
To get this expression, we have renamed $\xt'=\zt$ in Eq.\,\eqref{eq:NLO0-UVterm} and swapped $(\xt\leftrightarrow \yt)$ in the first term of Eq.\,\eqref{eq:NLO0-UVterm} in order to factor out the $1/\rzxt^2$ and $1/\rzyt^2$ kernels.
Although it is not obvious at first sight, this expression is UV finite as all $\zt\to\xt,\yt$ singularities are canceled by subtraction inside each square bracket multiplying the UV divergent  Weizs\"{a}cker-Williams kernels.

The (a) term is the finite piece of the integral over $\kt$ of $\der \sigma^{\gamma_{\rm L}^{\star}+A\to j_f+X}|_{\rm NLO,0}$:
\begin{align}
    &\left.\frac{\der \sigma^{\gamma_{\rm L}^{\star}+A\to j_f+X}}{\der \eta }\right|_{\rm NLO-a} =\frac{\alpha_{\mathrm{em}} e_f^2 N_c}{(2\pi)^2}\int\der^2\bt\der^2\rxyt 8z^3(1-z)^2Q^2K_0^2(\bar Q r_{xy}) 4\mathfrak{Re}\left \langle 1 - D_{xy}\right \rangle_{Y_f}\nonumber\\
    &\times\frac{\alpha_sC_F}{\pi}\left\{-\frac{1}{2}\ln^2\left(\frac{z}{1-z}\right)+\ln\left(\frac{z}{1-z}\right)\ln\left(R^2\right)-\frac{3}{2}\ln\left(2R\right)+7-\frac{2\pi^2}{3}\right.\nonumber\\
    &-\int_0^{1-z}\frac{\der z_g}{z_g}\ln\left(\frac{R^2z_g^2}{z^2}\right)\left[\left(1-\frac{z_g}{1-z}\right)^2\left(1+\frac{z_g}{z}+\frac{z_g^2}{2z^2}\right)\frac{K_0^2(\bar Q_{\mathrm{R}2}r_{xy})}{K_0^2(\bar Qr_{xy})}-1\right]\nonumber\\
    &\left.+2\int_0^{1-z}\frac{\der z_g}{z_g}\ln\left(1+\frac{z_g}{z}\right)\left(1-\frac{z_g}{1-z}\right)^2\left(1+\frac{z_g}{z}+\frac{z_g^2}{2z^2}\right)\frac{K_0^2(\bar Q_{\mathrm{R}2}r_{xy})}{K_0^2(\bar Qr_{xy})}\right\}\nonumber\\
    &-\frac{\alpha_{\mathrm{em}} e_f^2 N_c}{(2\pi)^2}\left[8z^3(1-z)^2Q^2\right]\times\frac{\alpha_sC_F}{\pi^2}\ln\left(\frac{z}{1-z}\right)\nonumber\\
    &\times\int\der^2\bt\der^2\rxyt\mathcal{P}_{\rxyt}\left(\frac{1}{\rxxtp^2}\right)\left[K_0(\bar Q r_{xy})K_0(\bar Q r_{x'y})2\mathfrak{Re}\left \langle D_{xx'} - D_{xy} -  D_{yx'} + 1 \right \rangle_{Y_f}\right]\,.\label{eq:NLOa-final}
\end{align}
In writing Eq.\,\eqref{eq:NLOa-final}, we have implicitly performed the change of variable $\bt=z\xt+(1-z)\yt$, $\rxyt = \xt-\yt$, with $\bt$ the impact parameter. The transverse coordinates $\xt$ and $\yt$ in the CGC correlators should then be expressed in terms of $\bt$ and $\rxyt$.

The (b) term is simply the integral over $\kt$ of the NLO,3 term given by Eq.\,\eqref{eq:NLO3-final}. Importantly, the impact parameter in this diagram is given by 
\begin{equation}
    \bt=(z-z_g)\xt+(1-z+z_g)\yt\,,\label{eq:bt-NLO3}
\end{equation}
because of the transverse recoil caused by the gluon emission. Using the proper definition of $\bt$ will be crucial when checking that the NLO correction depending on the $\rm NLO,3$ CGC correlator vanishes after integrating over the rapidity of the jet $\eta$ (or its longitudinal momentum fraction $z$).
After some algebra, we end up with
\begin{align}
     &\left.\frac{\der \sigma^{\gamma_{\rm L}^{\star}+A\to j_f+X}}{ \der \eta }\right|_{\rm NLO-b} =\frac{\alpha_{\mathrm{em}} e_f^2 N_c}{(2\pi)^2}\int\frac{\der^2\bt}{2\pi}\der^2\rxxtp\frac{\der^2\ryytp}{\ryytp^2}8z^3(1-z)Q^2 \nonumber\\
     &\times\frac{\alpha_s}{\pi} \int_{z-1}^{z}\frac{\der z_g}{z_g} K_0\left(\bar Q\left|(z-z_g)\rxxtp+(1-z+z_g)\ryytp\right|\right)K_0\left(\bar Q_{\rm V3}\left|z\rxxtp+(1-z)\ryytp\right|\right)\nonumber\\
     &\times(1-z+z_g)(z-z_g)^2\left[1+\frac{((1-z)\ryytp+(z-z_g)\rxxtp)\cdot(z\rxxtp+(1-z+z_g)\ryytp)}{(1-z)(1-z+z_g)\ryytp^2-z(z-z_g)\rxxtp^2}\right]\nonumber\\
     &\times 4\mathfrak{Re}\left[\frac{N_c}{2}\left\langle 1-D_{xy}-D_{y'x'} + D_{xy}D_{y'x'} \right\rangle_{Y_f}- \frac{1}{2N_c}\left\langle 1-D_{xy}-D_{y'x'}+Q_{xyy'x'} \right\rangle_{Y_f}\right]\,,\label{eq:NLO3-ktintegrated}
\end{align}
where the transverse coordinates in the $\rm NLO,3$ correlator must be expressed in terms of $\bt,\rxxtp$ and $\ryytp$ as\footnote{For this contribution, we have also re-scaled the variables $\rxxtp \to z_g \rxxtp$ and $\ryytp \to z_g \ryytp$.}
\begin{align}
    \xt&=\bt+(1-z+z_g)[z\rxxtp+(1-z)\ryytp]\,, \nonumber \\
    \yt&=\bt-(z-z_g)[z\rxxtp+(1-z)\ryytp]\,, \nonumber \\
    \xt'&=\bt+(1-z)[(z-z_g)\rxxtp+(1-z+z_g)\ryytp]\,, \nonumber \\
    \yt'&=\bt-z[(z-z_g)\rxxtp+(1-z+z_g)\ryytp] \,. \label{eq:bdep-1}
\end{align}
Finally, the (c) term gathers the pieces from the NLO,1 and NLO,4 which do not introduce any further UV divergences once integrated over $\kt$:
\begin{align}
    &\left.\frac{\der \sigma^{\gamma_{\rm L}^{\star}+A\to j_f+X}}{  \der \eta }\right|_{\rm NLO-c} =\frac{\alpha_{\mathrm{em}} e_f^2 N_c}{(2\pi)^2}16z^3 (1-z)^2Q^2\times\frac{2\alpha_s}{\pi^2}\mathfrak{Re}\int\der^2\xt\der^2\yt\der^2\zt \int_0^{1}\frac{\der z_g}{z_g}\nonumber\\
    &\times\Bigg\{-\frac{\rzxt\cdot\rzyt}{\rzxt^2\rzyt^2}\left(1-\frac{z_g}{z}\right)\left(1+\frac{z_g}{1-z}\right)\left(1-\frac{z_g}{2z}-\frac{z_g}{2(1-z+z_g)}\right)K_0\left(\bar Q\left|\rxyt+\frac{z_g}{z}\rzxt\right|\right)\nonumber\\
    &\times K_0(QX_V)\Theta(z-z_g)\Xi_{\rm NLO,1}(\xt,\yt,\zt,\boldsymbol{w}_{\perp,v})+\frac{\rzxt\cdot\rzyt}{\rzxt^2\rzyt^2}\left(1-\frac{z_g}{1-z}\right)^2\nonumber\\
    &\times\left(1+\frac{z_g}{2z}+\frac{z_g}{2(1-z-z_g)}\right)K_0\left(\bar Q_{\rm R2}\left|\rxyt+\frac{z_g}{z+z_g}\rzxt\right|\right)K_0(QX_R)\Theta(1-z-z_g)\nonumber\\
    &\times \Xi_{\rm NLO,1}(\xt,\yt,\zt,\boldsymbol{w}_{\perp,r})- \frac{\rzxt\cdot\rzyt}{\rzxt^2\rzyt^2}\left(1-\frac{z_g}{1-z}\right)^2\left(1+\frac{z_g}{2z}+\frac{z_g}{2(1-z-z_g)}\right)\nonumber\\
    &\times K_0^2(QX_R)\Theta(1-z-z_g)\left[\Xi_{\rm NLO,4}(\xt,\yt,\zt,\xt)\right]+\frac{\Theta(z-z_g)}{\rzxt^2}\nonumber\\
    &\times\left(1-\frac{z_g}{z}+\frac{z_g^2}{2z^2}\right)\left[K_0\left(\bar Q\left|\rxyt+\frac{z_g}{z}\rzxt\right|\right)K_0(QX_V)\Xi_{\rm NLO,1}(\xt,\yt,\zt,\boldsymbol{w}_{\perp,v})\right.\nonumber\\
     &\left.-e^{-\frac{\rzxt^2}{\rxyt^2e^{\gamma_E}}}K_0^2(\bar Qr_{xy})C_F\Xi_{\rm LO}(\xt,\yt,\xt)\right]\Bigg\}_{z_f}\,.
\end{align}

For completeness of this section, the simplified CGC color correlators necessary for the evaluation of the $\eta$-differential SIDIS cross-section are provided in table~\ref{tab:CGC-correlator2}. Similar results can be obtained for the gluon-tagged jets. 

\begin{table}[tbh]
    \centering
    \begin{tabular}{|c|c|}
    \hline
    $\Xi_{\rm LO}(\xt,\yt,\xt)$ & $2\mathfrak{Re}\left\langle 1-D_{xy}\right\rangle $\\
    \hline
    $\Xi_{\rm NLO,4}(\xt,\yt,\zt,\xt)$ &  $\frac{N_c}{2}2\mathfrak{Re} \left\langle 1-D_{xz}D_{zy}\right\rangle-\frac{1}{2N_c}2\mathfrak{Re}\left\langle 1 - D_{xy}\right\rangle$\\
    \hline 
    \end{tabular}
    \caption{Color correlators contributing to the next-to-leading order $\kt$ integrated SIDIS cross-section (in addition to those already gathered in table~\ref{tab:CGC-correlator}).}\label{tab:CGC-correlator2}
\end{table}

\section{Recovering the NLO fully inclusive DIS cross-section at small $x$}
\label{sec:total-xs}

In this last section, we demonstrate that by integrating over the two-body phase space in the virtual NLO corrections to inclusive dijet production and the three-body phase space in the real NLO corrections, one recovers known results for the total DIS cross-section at small-$x$ in the dipole picture \cite{Beuf:2016wdz,Beuf:2017bpd,Hanninen:2017ddy}. As mentioned in the introduction of this paper, this section is an independent derivation of the NLO DIS structure functions at small $x$ as we rely neither on the optical theorem nor on light-cone perturbation theory, contrary to the aforementioned papers.

\subsection{Cancellations between real and virtual diagrams}
As in the SIDIS case (see Sec.\,\ref{sub:cancellation}), after integrating over the anti-quark phase space we have
\begin{align}
    \left.\sigma^{\gamma_{\lambda}^{\star}+A\to X}\right|_{\rm SE3\times LO^*} = \left.\sigma^{\gamma_{\lambda}^{\star}+A\to X}\right|_{\rm \overline{SE3}\times LO^*}=\left.\sigma^{\gamma_{\lambda}^{\star}+A\to X}\right|_{\rm \overline{R2}\times \overline{R2}^*}=0\,.
\end{align}
and
\begin{align}
    \left.\sigma^{\gamma_{\lambda}^{\star}+A\to X}\right|_{\rm \overline{R1}\times \overline{R2}^*}=-\left.\sigma^{\gamma_{\lambda}^{\star}+A\to X}\right|_{\rm \overline{SE1}\times LO^*}\,,
\end{align}
\begin{align}
    \left.\sigma^{\gamma_{\lambda}^{\star}+A\to X}\right|_{\rm R1\times \overline{R2}^*}=-\left.\sigma^{\gamma_{\lambda}^{\star}+A\to X}\right|_{\rm \overline{V1}\times LO^*}\,.
\end{align}
Since we have derived the single-jet semi-inclusive cross-section in the small-$R$ approximation (neglecting powers of $R^2$ terms), the integration of the $\eta$-differential single-jet semi-inclusive cross-section over $\eta$ does not exactly yield the total cross-section. It is easier to do a brute force integration over $\ktone$, $\kttwo$, $z_1$ and $z_2$ without introducing any jet clustering at intermediate steps. The benefit of such a brute-forte integration is that new cancellations between diagrams arise. In particular, one gets that diagrams $\rm R2\times R2^*$ and $\rm \overline{R2}\times\overline{R2}^*$ vanish in dimensional regularization. This simple result would have been more complicated to obtain by integrating diagram $\rm R2\times R2^*$ \textit{after} implementation of the jet clustering algorithm, as the jet radius introduces a scale that formally breaks the dimensional regularization argument. We have then
\begin{align}
    \left.\sigma^{\gamma_{\lambda}^{\star}+A\to X}\right|_{\rm R2\times R2^*}=0\,.
\end{align}
Let us now consider the diagrams where the gluon interacts only once with the shock-wave, either in the amplitude or in the complex conjugate amplitude. After integration over the full three-body phase space of $\rm R1\times R2^*$, and the following change of variable $z_1\to z_1+z_g$,
one finds that
\begin{align}
    \left.\sigma^{\gamma_{\lambda}^{\star}+A\to X}\right|_{\rm R1\times R2^*}=-\left.\sigma^{\gamma_{\lambda}^{\star}+A\to X}\right|_{\rm SE1\times LO^*}\,.\label{eq:SE1-inclusive-cancel}
\end{align}
Likewise, a similar manipulation gives
\begin{align}
    \left.\sigma^{\gamma_{\lambda}^{\star}+A\to X}\right|_{\rm \overline{R1}\times R2^*}=-\left.\sigma^{\gamma_{\lambda}^{\star}+A\to X}\right|_{\rm V1\times LO^*}\,.\label{eq:V1-inclusive-cancel}
\end{align}

Particularly important diagrams are the final state gluon emission $\rm R2\times \overline{R2}^*$ and $\rm V3\times LO^*$ and their complex conjugate. It is crucial that the sum of these diagrams cancels, since they give a contribution proportional to the quadrupole $Q_{xyy'x'}$ which is not present in the total cross-section at NLO \cite{Beuf:2016wdz,Beuf:2017bpd,Hanninen:2017ddy}. Indeed, we prove in appendix~\ref{app:final-state-cancellation} that final state emissions cancel among virtual and real contributions:
\begin{align}
    \left. \sigma^{\gamma_{\rm L}^{\star}+A\to X}\right|_{\rm V3\times LO^*+R2\times \overline{R2}^*}+c.c.=-\left(\left. \sigma^{\gamma_{\rm L}^{\star}+A\to X}\right|_{\rm V3\times LO^*+R2\times \overline{R2}^*}+c.c.\right)\,.
\end{align}
We also prove in appendix~\ref{app:final-state-cancellation} that this identity holds for the cross-section differential with respect to the virtual photon impact parameter $\der\sigma^{\gamma_{\rm L}^{\star}+A\to X}/\der^2\bt$ provided $\bt$ is defined by Eq.\,\eqref{eq:bt-NLO3} for these particulars diagrams.

In the end, the only diagrams which contribute to the fully inclusive DIS cross-section are (i) the real diagrams where the gluon is emitted and absorbed after before the shock-wave: $\rm R1\times R1^*$, $\rm \overline{R1}\times\overline{R1}^*$, $\rm R1\times \overline{R1}^*$ and $\rm \overline{R1}\times R1^*$, and (ii) the virtual diagrams with gluon exchanged before the shock-wave, namely $\rm SE2\times LO^*$, $\rm \overline{SE2}\times LO^*$ and $\rm V2\times LO^*$ plus their complex conjugate. 

\subsection{Real diagrams}

We start by discussing the real emissions where the gluon is emitted and absorbed before the shock-wave. We will integrate over the quark phase space in Eqs.\,\eqref{eq:R1R1},\,\eqref{eq:barR1R1} and \eqref{eq:barR1barR1}. These contributions are in one-to-one correspondence with the $q\bar q g$ contributions to the NLO light-cone wave-function of the virtual photon in light cone perturbation theory.
The contribution of diagram $\rm R1\times \rm R1^*$ to the total cross-section is given by
\begin{align}
    &\left. \sigma^{\gamma_{\lambda}^{\star}+A\to X}\right|_{\rm R1 \times R1^*} =\frac{\alpha_{\rm em}e_f^2N_c}{(2\pi)^2 } \int_0^1\der z_1 \int\der^2\xt  \der^2\yt 8z_1^2(1-z_1)^2Q^2\nonumber\\
    &\times\frac{\alpha_s}{\pi}\int_{0}^{1-z_1} \frac{\der z_g}{z_g} \left(1-\frac{z_g}{1-z_1}\right)^2 \left(1+\frac{z_g}{z_1}+\frac{z_g^2}{2z_1^2} \right)  \nonumber\\
    & \times\int\frac{\der^2\zt}{\pi} \frac{1}{\rzxt^2}  K_0^2(QX_R) \frac{N_c}{2} \left\langle 2-D_{xz}D_{zy}-D_{yz}D_{zx}-\frac{1}{N_c^2}(2 - D_{xy} -  D_{yx})\right\rangle\,.
\end{align}
Similarly, the contribution of diagram $\rm \overline{R1}\times\overline{R1}$ reads
\begin{align}
    &\left.\sigma^{\gamma_{\lambda}^{\star}+A\to X}\right|_{\rm \overline{R1}\times \overline{R1}^*} =\frac{\alpha_{\rm em}e_f^2N_c}{(2\pi)^2}\int_0^1\der z_1 \int\der^2\xt\der^2\yt  8z_1^2 (1-z_1)^2 Q^2 \nonumber\\
    &\times\frac{\alpha_s}{\pi^2} \int_0^{1-z_1}\frac{\der z_g}{z_g} \left(1 - \frac{z_g}{1-z_1} \right)^2 \left(1+\frac{z_g}{1-z_1-z_g}+\frac{z_g^2}{2(1-z_1-z_g)^2}\right)\nonumber\\
    &\times \int\der^2\zt \frac{1}{\rzyt^2}K_0^2(QX_R) \frac{N_c}{2}\left\langle 2-D_{xz}D_{zy}-D_{yz}D_{zx}-\frac{1}{N_c^2}(2 - D_{xy} -  D_{yx})\right\rangle\,.
\end{align}
Finally, the cross diagram $\rm \overline{R1}\times R1$ reads
\begin{align}
    &\left.\sigma^{\gamma_{\lambda}^{\star}+A\to X}\right|_{\rm \overline{R1} \times R1^*} =-\frac{\alpha_{\rm em}e_f^2N_c}{(2\pi)^2 } \int_0^1\der z_1\int\der^2\xt  \der^2\yt  8z_1^3(1-z_1)^2Q^2\nonumber\\
    &\times\frac{\alpha_s}{\pi}\int_{0}^{1-z_1} \frac{\der z_g}{z_g} \left(1-\frac{z_g}{1-z_1}\right)^2\left(1+\frac{z_g}{2z_1}+\frac{z_g}{2(1-z_1-z_g)} \right) \nonumber\\
     & \times\int\frac{\der^2\zt}{\pi} \frac{\rzyt \cdot\rzxt}{\rzyt^2 \rzxt^2}  K_0^2(QX_R) \frac{N_c}{2}\left\langle 2-D_{xz}D_{zy}-D_{yz}D_{zx}-\frac{1}{N_c^2}(2 - D_{xy} -  D_{yx})\right\rangle\,.
\end{align}

The terms $\rm R1\times R1^*$ and $\rm\overline{R1}\times\overline{R1}^*$ are UV divergent and must be regulated. As in section~\ref{subsub:fermion-real-uv}, we use the following two regulators that we add and subtract to the cross-section:
\begin{align}
    &\left. \sigma^{\gamma_{\lambda}^{\star}+A\to X}\right|_{\rm R1 \times R1^*,UV} =\frac{\alpha_{\rm em}e_f^2N_c}{(2\pi)^2 } \int_0^1\der z_1 \int\der^2\xt  \der^2\yt  \ 8z_1^2(1-z_1-z_g)^2Q^2\nonumber\\
     &\times\frac{\alpha_s}{\pi}\int_{0}^{1-z_1} \frac{\der z_g}{z_g}  \left(1+\frac{z_g}{z_1}+\frac{z_g^2}{2z_1^2} \right) \int\frac{\der^2\zt}{\pi} \frac{e^{-\frac{\rzxt^2}{2\xi_1}}}{\rzxt^2}  K_0^2(\bar Q_{\rm R2}r_{xy}) C_F 2\mathfrak{Re}\left\langle 1-D_{xy}\right\rangle\\
    &\left.\sigma^{\gamma_{\lambda}^{\star}+A\to X}\right|_{\rm \overline{R1}\times \overline{R1},UV} =\frac{\alpha_{\rm em}e_f^2N_c}{(2\pi)^2}\int_0^1\der z_1\int_0^{1-z_1}\frac{\der z_g}{z_g}8z_1^2(1-z_1-z_g)^2Q^2\int\der^2\xt\der^2\yt \nonumber\\
    &\times\frac{\alpha_s}{\pi^2}\left(1+\frac{z_g}{1-z_1-z_g}+\frac{z_g^2}{2(1-z_1-z_g)^2}\right)\int\der^2\zt \frac{-\frac{\rzyt^2}{2\xi_2}}{\rzyt^2}K_0^2(\bar Q r_{xy})C_F 2\mathfrak{Re}\left\langle 1-D_{xy}\right\rangle\,.
\end{align}
In the first UV counter-term, we perform the change of variable
\begin{align}
    \zeta_1&=z_1+z_g\,, \nonumber \\
    \zeta_g&=z_g\,,
\end{align}
and we get
\begin{align}
    &\left. \sigma^{\gamma_{\lambda}^{\star}+A\to X}\right|_{\rm R1 \times R1^*,UV} =\frac{\alpha_{\rm em}e_f^2N_c}{(2\pi)^2 } \int_0^1\der \zeta_1 \int\der^2\xt  \der^2\yt  \ 8\zeta_1^2(1-\zeta_1)^2Q^2\nonumber\\
     &\times\frac{\alpha_s}{\pi}\int_{0}^{\zeta_1} \frac{\der \zeta_g}{\zeta_g} \left(1-\frac{\zeta_g}{\zeta_1}+\frac{\zeta_g^2}{2\zeta_1^2} \right) \int\frac{\der^2\zt}{\pi} \frac{e^{-\frac{\rzxt^2}{2\xi_1}}}{\rzxt^2}  K_0^2(\bar Qr_{xy}) C_F 2\mathfrak{Re}\left\langle 1-D_{xy}\right\rangle\,.
\end{align}
The computation of these integrals in dimension regularization $\der^2\zt\to \der^{2+\varepsilon}\zt$ leads to
\begin{align}
    &\left. \sigma^{\gamma_{\lambda}^{\star}+A\to X}\right|_{\rm R1 \times R1^*,UV} =\frac{\alpha_{\rm em}e_f^2N_c}{(2\pi)^2 } \int_0^1\der z_1 \int\der^2\xt  \der^2\yt  \ 8z_1^2(1-z_1)^2Q^2K_0^2(\bar Qr_{xy}) \nonumber\\
    &\times \frac{\alpha_sC_F}{\pi}\left[\ln\left(\frac{z_1}{z_0}\right)-\frac{3}{4}\right]\left(\frac{2}{\varepsilon}+\ln(2\pi\mu^2\xi_1)\right)2\mathfrak{Re}\left\langle 1-D_{xy}\right\rangle\label{eq:R1R1-tot-UV}\\
    &\left. \sigma^{\gamma_{\lambda}^{\star}+A\to X}\right|_{\rm \overline{R1} \times \overline{R1}^*,UV} =\frac{\alpha_{\rm em}e_f^2N_c}{(2\pi)^2 } \int_0^1\der z_1 \int\der^2\xt  \der^2\yt  \ 8z_1^2(1-z_1)^2Q^2K_0^2(\bar Qr_{xy}) \nonumber\\
    &\times \frac{\alpha_sC_F}{\pi}\left[\ln\left(\frac{1-z_1}{z_0}\right)-\frac{3}{4}\right]\left(\frac{2}{\varepsilon}+\ln(2\pi\mu^2\xi_2)\right)2\mathfrak{Re}\left\langle 1-D_{xy}\right\rangle\,\label{eq:R1bR1b-tot-UV},
\end{align}
where potential finite pieces from the $\mathcal{O}(\varepsilon)$ terms in Dirac structure have been omitted. These poles will cancel against diagrams $\rm V2$ and diagrams $\rm SE2$ that we now discuss.

\subsection{Virtual diagrams}

The $q\bar q$ phase space integral of the diagrams $\rm V2$ and $\rm SE2$ is straightforward from the expressions obtained in \cite{Caucal:2021ent} (see Eqs.\,(5.40)-(5.78) therein).
Adding together the three virtual diagrams with gluon exchange before the shock-wave and their complex conjugate defined as:
\begin{equation}
    \left. \sigma^{\gamma_{\lambda}^{\star}+A\to X}\right|_{\rm init.}\equiv\left. \sigma^{\gamma_{\lambda}^{\star}+A\to X}\right|_{\rm SE2\times LO}+\left. \sigma^{\gamma_{\lambda}^{\star}+A\to X}\right|_{\rm \overline{SE2}\times LO}+\left. \sigma^{\gamma_{\lambda}^{\star}+A\to X}\right|_{\rm V2\times LO}+c.c.\,,
\end{equation}
we get
\begin{align}
     &\left. \sigma^{\gamma_{\lambda}^{\star}+A\to X}\right|_{\rm init.} \!\!\!\! =\frac{\alpha_{\rm em}e_f^2N_c}{(2\pi)^2 }  \frac{\alpha_sC_F}{\pi}\int_0^1\der z_1 \int\der^2\xt  \der^2\yt 8z_1^2(1-z_1)^2Q^2K_0^2(\bar Qr_{xy}) 2\mathfrak{Re}\left\langle 1-D_{xy}\right\rangle\nonumber\\
     &\times\left[\left(\frac{2}{\varepsilon}+\ln(\pi\mu^2e^{\gamma_E}\rxyt^2)\right)\left(\frac{3}{2}-\ln\left(\frac{z_1}{z_0}\right)-\ln\left(\frac{z_2}{z_0}\right)\right)+\frac{1}{2}\ln^2\left(\frac{1-z_1}{z_1}\right)+\frac{5}{2}-\frac{\pi^2}{6}\right]\,.
     \label{eq:DISqqbarinit}
\end{align}
In this expression, we have removed the finite pieces from the $\mathcal{O}(\varepsilon)$ terms in the Dirac structure, in agreement with our convention for the expression of the UV divergent pieces Eqs\,\eqref{eq:R1R1-tot-UV}-\eqref{eq:R1bR1b-tot-UV} from diagrams $\rm R1\times\rm R1$ and $\rm \overline{R1}\times\overline{R1}$.

\subsection{Final result}
We readily observe the cancellation of UV divergences between Eqs.\,\eqref{eq:R1R1-tot-UV},\,\eqref{eq:R1bR1b-tot-UV} and \,\eqref{eq:DISqqbarinit}, we thus define the $q\bar{q}$ contribution to the inclusive DIS cross-section as
\begin{align}
    \left.\sigma^{\gamma_{\lambda}^{\star}+A\to X}\right|_{q\bar q}\equiv \left.\sigma^{\gamma_{\lambda}^{\star}+A\to X}\right|_{\rm init.}+\left. \sigma^{\gamma_{\lambda}^{\star}+A\to X}\right|_{\rm R1 \times R1,UV} +\left. \sigma^{\gamma_{\lambda}^{\star}+A\to X}\right|_{\rm \overline{R1} \times \overline{R1},UV} \,.
\end{align}
For the choice of regulator:
\begin{equation}
    \xi_1=\xi_2=\frac{e^{\gamma_E}\rxyt^2}{2}\,,
\end{equation}
we have
\begin{align}
    \left.\sigma^{\gamma_{\lambda}^{\star}+A\to X}\right|_{q\bar q}&=\frac{\alpha_{\rm em}e_f^2N_c}{(2\pi)^2 } \int_0^1\der z_1 \int\der^2\xt  \der^2\yt  \ 8z_1^2(1-z_1)^2Q^2K_0^2(\bar Qr_{xy}) 2\mathfrak{Re}\langle1-D_{xy}\rangle\nonumber\\
&\times\frac{\alpha_sC_F}{\pi}\left[\frac{1}{2}\ln^2\left(\frac{1-z_1}{z_1}\right)+\frac{5}{2}-\frac{\pi^2}{6}\right]\,.
\end{align}
The above result agrees with Eq.\,(176) in \cite{Hanninen:2017ddy}.

With these choices, the $q\bar{q}g$ contribution defined as the sum of $\rm R1\times R1$, $\rm\overline{R1}\times\overline{R1}$, $\rm \overline{R1}\times R1$ and $\rm R1\times\overline{R1}$ reads
\begin{align}
   & \left.\sigma^{\gamma_{\lambda}^{\star}+A\to X}\right|_{q\bar qg} \nonumber \\
   & \equiv \left.\sigma^{\gamma_{\lambda}^{\star}+A\to X}\right|_{\rm R1\times R1} +\left.\sigma^{\gamma_{\lambda}^{\star}+A\to X}\right|_{\rm \overline{R1}\times \overline{R1}} +\left.\sigma^{\gamma_{\lambda}^{\star}+A\to X}\right|_{\rm \overline{R1}\times R1} +\left.\sigma^{\gamma_{\lambda}^{\star}+A\to X}\right|_{\rm R1\times \overline{R1}} \,,
\end{align}
with
\begin{align}
  & \left.\sigma^{\gamma_{\rm L}^{\star}+A\to X}\right|_{q\bar qg}=\frac{\alpha_{\rm em}e_f^2N_c}{(2\pi)^2 } \frac{\alpha_sC_F}{\pi}2\mathfrak{Re}\int_0^1\der z_1\int_0^{1-z_1}\frac{\der z_g}{z_g}\int\der^2\xt  \der^2\yt\int\frac{\der^2\zt}{\pi}  \ 8z_1^2z_2^2Q^2 \nonumber\\
     &\left(1+\frac{z_g}{z_1}+\frac{z_g^2}{2z_1^2} \right) \frac{1}{\rzxt^2}  \left[K_0^2(QX_R)\left\langle1-\frac{N_c}{2C_F}D_{xz}D_{zy}+\frac{1}{2N_cC_F}D_{xy}\right\rangle\right.\nonumber\\
     &\hspace{9cm}\left.-K_0^2(\bar Q_{\rm R2}r_{xy})e^{-\frac{\rzxt^2}{e^{\gamma_E}{\rxyt^2}}}\langle1-D_{xy}\rangle\right]\nonumber\\
     &+\left(1+\frac{z_g}{z_2}+\frac{z_g^2}{2z_2^2}\right)\frac{1}{\rzyt^2}  \left[K_0^2(QX_R) \left\langle1-\frac{N_c}{2C_F}D_{xz}D_{zy}+\frac{1}{2N_cC_F}D_{xy}\right\rangle\right.\nonumber\\
     &\hspace{9cm}\left.-K_0^2(\bar Q r_{xy})e^{-\frac{\rzyt^2}{e^{\gamma_E}{\rxyt^2}}}\langle1-D_{xy}\rangle\right]\nonumber\\
     &-2\left(1+\frac{z_g}{2z_1}+\frac{z_g}{2z_2} \right) \frac{\rzxt\cdot\rzyt}{\rzxt^2\rzyt^2}K_0^2(QX_R) \left\langle1-\frac{N_c}{2C_F}D_{xz}D_{zy}+\frac{1}{2N_cC_F}D_{xy}\right\rangle\,,
\end{align}
which exactly agrees with Eq.\,(177) in \cite{Hanninen:2017ddy}. We have noted $z_2=1-z_1-z_g$ for compactness. The fully inclusive DIS cross-section at NLO is given by the sum of the LO, $q\bar{q}$ and $q\bar{q}g$ contributions:
\begin{align}
    \sigma^{\gamma_{\lambda}^{\star}+A\to X} =  \left.\sigma^{\gamma_{\lambda}^{\star}+A\to X}\right|_{\rm LO}+ \left.\sigma^{\gamma_{\lambda}^{\star}+A\to X}\right|_{q\bar q} + \left.\sigma^{\gamma_{\lambda}^{\star}+A\to X}\right|_{q\bar qg} \,.
\end{align}

\section{Summary and Outlook}

In this paper, we computed for the first time the NLO semi-inclusive single \textit{jet} production cross-section in DIS at small $x$, working in the dipole picture of DIS and using the Color Glass Condensate effective field theory. Our results follow from the analytic expressions obtained in \cite{Caucal:2021ent} for the inclusive dijet cross-section by integrating out the phase space of one of the jets. These integrations yield new ultra-violet divergences in the transverse coordinate integrals which cancel at the cross-section level for our infrared and collinear safe jet cross-section. The rapidity divergences in the NLO correction, associated with gluons with $k_g^-\ll q^-$, are cured using a renormalization group approach. The bare dipole operator in the LO cross-section, defined at some initial rapidity scale $z_0=\Lambda^-/q^-$ is replaced by a renormalized one at some rapidity factorization scale $z_f=k_f^-/q^-$, and we explicitly demonstrate that the resulting $Y_f=\ln(z_f)$ dependence of the dipole correlator $\langle D_{xy}\rangle_{Y_f}$ satisfies the BK-JIMWLK equation.

We provide analytic expressions for the NLO impact factor for both the double differential (transverse momentum and rapidity) cross-section and the single differential cross-section (rapidity only) in sections~\ref{sub:final-dblediff} and \ref{sec:singlediff-xsec} respectively (see also appendix \ref{app:transverse}). These results are complementary to those obtained in \cite{Bergabo:2022zhe} which focuses on single hadron production in DIS at small $x$. While some contributions are identical, such as the UV and IR regular virtual or real corrections (diagrams $\rm V1\times LO^*$, $\rm R1\times R2^*$, $\rm \overline{R1}\times R2^*$, $\rm R1\times R1^*$ and $\rm \overline{R1}\times R1^*$ in our notations), our different definition of the final state (in terms of jets) lead to different expressions, in particular for the real corrections with a collinear singularity or the virtual diagrams with a UV pole in dimensional regularization. We also provide the analytic expressions for the transversely polarized virtual photon case in appendix~\ref{app:transverse} and we compute the subleading $N_c$ corrections in the NLO impact factor.

Thanks to our explicit analytic expressions, we were able to identify the leading NLO corrections in the very forward rapidity regime, which is the most interesting one in the search for saturation signal in our process \cite{Iancu:2020jch}. We find that the leading NLO corrections for very forward jets come from the incomplete cancellation between real and virtual soft gluon emissions and are thus double logarithmically enhanced. They can give a negative cross-section at any fixed order, spoiling the predictive power of perturbative calculations in this corner of the phase space. As usual in pQCD, the solution we propose to restore the convergence of the perturbative series is to resum to all orders the logarithmic terms via exponentiation. Using the BK-JIMWLK evolved dipole operator in the LO cross-section and the exponentiation of the $\alpha_s\ln^2(1-z)$, $\alpha_s\ln(1-z)$ corrections enables one to simultaneously resum the $\alpha_s\ln(z_f/z_0)\sim \alpha_s\ln(1/x)$ and threshold logarithms to all orders. A formal proof of the exponentiation at small $x$ is left for future studies, it may require a two-loop computation. Nevertheless, assuming exponentiation holds, the effect of soft gluons in the very forward jet regime is to suppress the cross-section, thereby mimicking the signal of saturation. While this may potentially complicate the extraction of saturation from SIDIS at very forward rapidity, the universality of the double logarithms in $1-z$ with respect to the nuclear target suggests that the nuclear modification factor for the SIDIS cross-section could be a promising observable given that the suppression caused by soft gluons exactly cancel in the ratio. We point out that this exact cancellation occurs for single-jet production, but not for single-hadron production, which is another benefit of using jets in SIDIS measurements at small $x$.

The next step of this project will be to numerically evaluate the NLO SIDIS cross-section at small $x$ based on the formulae derived in this paper, including the resummation of the aforementioned threshold logarithms. We forecast two main difficulties with the numerical evaluation. First, the NLO impact factor depends on the quadrupole correlator, because of the final state gluon emission diagrams and it is challenging to solve the JIMWLK equation for the quadrupole correlator. Our plan is to provide a numerical estimation of the quadrupole by relying on the Gaussian approximation~\cite{Blaizot:2004wv,Dumitru:2011vk,Iancu:2011nj,Metz:2011wb,Dominguez:2011br}. Since the quadrupole dependence is $1/N_c^2$ suppressed in the NLO impact factor, this should be a good approximation to the exact result. The second difficulty is related to our treatment of rapidity factorization, with an implicit ordering in minus longitudinal momentum $k_g^-$. It is now well documented that this naive factorization scheme leads to unphysical NLO results, as the rapidity logarithms are oversubtracted~\cite{Beuf:2014uia,Iancu:2016vyg,Iancu:2020whu,Caucal:2022ulg}. The proper way to resum the high energy logarithms is to use an ordering along the target rapidity variable $k_g^+$, which requires implementing a kinematic constraint when subtracting the rapidity divergence in our calculation~\cite{Andersson:1995ju,Kwiecinski:1996td,Kwiecinski:1997ee,Salam:1998tj,Ciafaloni:1998iv,Ciafaloni:1999yw,Ciafaloni:2003rd,SabioVera:2005tiv,Motyka:2009gi,Beuf:2014uia,Iancu:2015vea}. The implementation of this kinematic constraint (maybe along the lines of \cite{Liu:2022xsc}) and how it affects the NLO impact factor is left for an upcoming paper. 

In addition, to the numerical evaluation of SIDIS at general small-$x$ kinematics, we envision several applications of our results in this manuscript that we leave for future work. For example, replacing the jet clustering algorithm with a measurement function for the energy deposition of the final state, our results for the SIDIS cross-section at NLO can be utilized to promote the calculation of the novel nucleon energy correlators for the CGC \cite{Liu:2023aqb} to higher precision. Furthermore, one can use the single jet production in conjunction with the dijet observable in \cite{Caucal:2023fsf} to define a dijet correlation function normalized by the trigger jet \cite{Zheng:2014vka}. In addition, it would be interesting to test if the correspondence between the dipole picture within the CGC EFT and TMD factorization framework holds at NLO in the regime $k_\perp, Q_s \ll Q$. Lastly, by taking the photo-production limit our results could be easily adapted to the inclusive production of jets in ultra-peripheral collisions.

In the last section \ref{sec:total-xs} of this paper, we provide the first independent cross-check of the NLO corrections to the DIS structure functions obtained in \cite{Beuf:2016wdz,Beuf:2017bpd,Hanninen:2017ddy} thanks to the optical theorem. Our initial formula for inclusive dijet (or $q\bar q$ production) cross-section \cite{Caucal:2021ent} have indeed been obtained in covariant perturbation theory, contrary to the calculations in \cite{Beuf:2016wdz,Beuf:2017bpd,Hanninen:2017ddy} done in light-cone perturbation theory. By integrating the NLO corrections to inclusive $q\bar q$ production over the full phase space of the out-going partons, we recover the results of \cite{Hanninen:2017ddy}, meaning that the optical theorem, a fundamental consequence of unitarity, is valid at NLO in the dipole picture. While this is of course a minimal requirement for the dipole picture to be a consistent framework beyond leading order, we emphasize that checking the validity of the optical theorem, both for the DIS structure functions and the DIS structure functions differential with the impact parameter, rely on subtle cancellations between NLO Feynman diagrams and the use of the physical impact parameter of the virtual photon. Hence, this is also an important crosscheck of the formulae derived in \cite{Caucal:2021ent} and in this paper.

\smallskip

\noindent{\bf Acknowledgments.} We are grateful to Edmond Iancu, Jani Penttala, and Feng Yuan for valuable discussions. P.C. and F.S. thank the EIC theory institute at BNL for its support during the early stages of this work. P.C. would also like to thank the staff at LBNL for their hospitality during the completion of this project. E.F. is thankful to Subatech for financial support. F.S. is supported by the National Science Foundation (NSF) under grant number ACI-2004571 within the framework of the XSCAPE project of the JETSCAPE collaboration, and the U.S. DOE under Grant No. DE-FG02-00ER41132.

\appendix

\section{NLO impact factor for transversely polarized virtual photon}
\label{app:transverse}

This appendix gathers our final expressions for the NLO semi-inclusive single-jet production in DIS at small $x$, following the notations and conventions detailed in the main text.

\subsection{Spin and polarization sums for the real inclusive dijet cross-section}

In this subsection, we first provide the spin and polarization sums that have not been performed in the original paper \cite{Caucal:2021ent}. These sums are associated with diagrams $\rm R1\times R2^*$, $\rm R1\times \overline{R2}^*$, $\rm R1\times R1^*$ and $\rm R1\times \overline{R1}^*$. Let us quote here the results, following the notations of appendix B in \cite{Caucal:2021ent}. In particular, we recall here the definition of the transverse vectors
\begin{align}
    \RtR &= \rxyt+\frac{z_g}{z_1+z_g}\rzxt\,,\\
    \RtRb&=-\rxyt+\frac{z_g}{z_2+z_g}\rzyt\,.
\end{align}
The vector $\RtR'$ and $\RtRb'$ are defined similarly, with the transverse coordinates replaced by the prime ones. Throughout this appendix, cross product between two 2-dimensional transverse vectors is defined by $\at\times\bt = \epsilon^{ij}a^i b^j$.

The spin and polarization sum for diagram $\rm R1\times R2^*$ is
\begin{align}
    S_1&\equiv \sum_{\lambda,\bar{\lambda},\sigma,\sigma'}\mathcal{N}_{\rm R1}^{\lambda\bar\lambda\sigma\sigma'}(\rxyt,\rzxt)\mathcal{N}_{\rm R2}^{\lambda\bar{\lambda}\sigma\sigma',\dagger}(\rwytp,\rzxtp) \\
    &=\frac{(-8\alpha_s)}{\pi}\frac{z_1z_2^2}{(1-z_2)}\frac{Q \bar{Q}_{\rm R2}K_1(QX_R)K_1(\bar Q_{\rm R2}r_{w'y'})}{X_Rr_{w'y'}}\nonumber\\
    &\times\left\{\left[z_1^2+(1-z_2)^2\right]\left(1-2z_2(1-z_2)\right)\frac{\rzxt\cdot\rzxtp}{\rzxt^2\rzxtp^2}\RtR\cdot\rwytp\right.\nonumber\\
    &\left.+z_g(1+z_1-z_2)(1-2z_2)\frac{\rzxt\times\rzxtp}{\rzxt^2\rzxtp^2}\RtR\times\rwytp-\frac{z_1^2z_2z_g}{(1-z_2)}\frac{\rzxtp\cdot\rwytp}{\rzxtp^2}\right\}\,.
\end{align}
For the diagram $\rm R1\times \overline{R2}^*$, one can use the quark-antiquark symmetry to relate the perturbative factor for $\rm \overline{R2}$ to that of diagram $\rm R2$
\begin{align}
    S_2 &\equiv -\sum_{\lambda,\bar{\lambda},\sigma,\sigma'}\mathcal{N}_{\rm R1}^{\lambda\bar\lambda\sigma\sigma'}(\rxyt,\rzxt)\mathcal{N}_{\rm R2}^{\lambda\bar{\lambda}-\sigma-\sigma',\dagger}(\rwbxtp,\rzytp) \\
&=\frac{8\alpha_s}{\pi}\frac{z_1z_2^2}{(1-z_1)}\frac{Q \bar{Q}_{\rm R2'}K_1(QX_R)K_1(\bar Q_{\rm R2'}r_{x'\bar{w}'})}{X_Rr_{x'\bar{w}'}}\nonumber\\
&\times \left\{(z_1(1-z_1)+z_2(1-z_2)(z_1+z_2-2z_1z_2)\frac{\rzxt\cdot\rzytp}{\rzxt^2\rzytp^2}\RtR\cdot\rxwbtp\right.\nonumber\\
&\left.-z_g(z_1-z_2)^2\frac{\rzxt\times\rzytp}{\rzxt^2\rzytp^2}\RtR\times\rxwbtp-\frac{z_1z_g(1-z_1)^2}{1-z_2}\frac{\rzytp\cdot\rxwbtp}{\rzytp^2}\right\}\,.
\end{align}
The spin and polarization sum of diagram $\rm R1\times R1^*$ reads
\begin{align}
    &S_3 \equiv \sum_{\lambda,\bar{\lambda},\sigma,\sigma'}\mathcal{N}_{\rm R1}^{\lambda\bar\lambda\sigma\sigma'}(\rxyt,\rzxt)\mathcal{N}_{\rm R1}^{\lambda\bar{\lambda}\sigma\sigma',\dagger}(\rxytp,\rzxtp) \\
&=\frac{8\alpha_s}{\pi}z_1z_2^3\frac{Q^2K_1(QX_R)K_1(QX_R')}{X_RX_R'}\left\{\left[z_1^2+(1-z_2)^2\right]\left(1-2z_2(1-z_2)\right)\frac{\rzxt\cdot\rzxtp}{\rzxt^2\rzxtp^2}\RtR\cdot\RtR'\right.\nonumber\\
&+z_g(1+z_1-z_2)(1-2z_2)\frac{\rzxt\times\rzxtp}{\rzxt^2\rzxtp^2}\RtR\times\RtR'-\frac{z_1^2z_2z_g}{1-z_2}\left[\frac{\rzxt\cdot\RtR}{\rzxt^2}+\frac{\rzxtp\cdot\RtR'}{\rzxtp^2}\right]\nonumber\\
&+\left.\frac{z_1^2z_2z_g^2}{(1-z_2)^2}\right\}\,.
\end{align}
Finally, the spin-polarization sum of diagram $\rm R1\times \overline{R1}^*$ is given by
\begin{align}
    & S_4 \equiv -\sum_{\lambda,\bar{\lambda},\sigma,\sigma'}\mathcal{N}_{\rm R1}^{\lambda\bar\lambda\sigma\sigma'}(\rxyt,\rzxt)\mathcal{N}_{\rm R1}^{\lambda\bar{\lambda}-\sigma-\sigma',\dagger}(-\rxytp,\rzytp) \\
    &=\frac{8\alpha_s}{\pi}z_1^2z_2^2\frac{Q^2K_1(QX_R)K_1(QX_R')}{X_RX_R'}\nonumber\\
    &\times\left\{(z_1(1-z_1)+z_2(1-z_2))(z_1+z_2-2z_1z_2)\frac{\rzxt\cdot\rzytp}{\rzxt^2\rzytp^2}\RtR\cdot\RtRb'\right.\nonumber\\
&\left.-z_g(z_1-z_2)^2\frac{\rzxt\times\rzytp}{\rzxt^2\rzytp^2}\RtR\times\RtRb'-z_g\left[\frac{z_1(1-z_1)^2}{1-z_2}\frac{\rzxt\cdot\RtRb'}{\rzxt^2}+\frac{z_2(1-z_2)^2}{1-z_1}\frac{\rzytp\cdot\RtR}{\rzytp^2}\right]\right\}\,.
\end{align}

\subsection{NLO impact factor for the double differential cross-section}

Following the decomposition we used for the longitudinally polarized case, we write the full NLO SIDIS cross-section for transversely polarized photon as Eqs.\,\eqref{eq:qg-decomposition}-\eqref{eq:final-decomposition} with the corresponding terms with fermion-tagged jets given by
\begin{align}
        &\left.\frac{\der \sigma^{\gamma_{\rm T}^{\star}+A\to j_f+X}}{ \der^2 \kt  \der \eta }\right|_{\rm NLO,0} =\frac{\alpha_{\mathrm{em}} e_f^2 N_c}{(2\pi)^4}\int\der^2\xt\der^2\xt'\der^2\yt e^{-i\kt\cdot\rxxtp}2z(z^2+(1-z)^2)\frac{\rxyt\cdot\rxpyt}{r_{xy}r_{x'y}}\nonumber\\
        &\times \bar Q^2K_1(\bar Qr_{x'y})K_1(\bar Qr_{x'y})2\mathfrak{Re}\left[\Xi_{\rm LO}(\xt,\yt,\xt')\right]\times\frac{\alpha_sC_F}{\pi}\left\{-\frac{1}{2}\ln^2\left(\frac{z}{1-z}\right)+7-\frac{2\pi^2}{3}\right.\nonumber\\
        &\left.+\ln\left(\frac{z}{1-z}\right)\ln\left(\frac{\kt^2r_{xy}r_{x'y}R^2}{c_0^2}\right)-\frac{3}{4}\ln\left(\frac{4\kt^2\rxxtp^2 R^2}{c_0^2}\right)\right.-\int_0^{1-z}\frac{\der z_g}{z_g}\ln\left(\frac{\kt^2\rxxtp^2R^2z_g^2}{c_0^2z^2}\right)\nonumber\\
    &\times\left[e^{-i\frac{z_g}{z}\kt\rxxtp}\frac{(1-z-z_g)(z+z_g)((1-z-z_g)^2+(z+z_g)^2)(2z(z+z_g)+z_g^2)}{2z^3(1-z)(z^2+(1-z)^2)}\right.\nonumber\\
    &\times \left.\left.\frac{K_1(\bar Q_{\mathrm{R}2}r_{xy})K_1(\bar Q_{\mathrm{R}2}r_{x'y})}{K_1(\bar Qr_{xy})K_1(\bar Q r_{x'y})}-1\right]\right\}\,,\label{eq:NLO0-final-transverse}
\end{align}
for the term proportional to the LO SIDIS CGC color correlator $\Xi_{\rm LO}$. Here, $c_0=2e^{-\gamma_E}$ with $\gamma_E$ the Euler-Mascheroni constant. The term proportional to $\Xi_{\rm NLO,3}$ reads
\begin{align}
    &\left.\frac{\der \sigma^{\gamma_{\rm T}^{\star}+A\to j_f+X}}{ \der^2 \kt  \der \eta }\right|_{\rm NLO,3} \!\!\!\!\!\!\! =\frac{\alpha_{\rm em}e_f^2N_c}{(2\pi)^4} \int\der^2\xt \der^2\xt' \der^2\yt \der^2\yt' \ e^{-i\ktone \cdot \rxxtp-i\frac{1-z}{z}\ktone \cdot  \ryytp}\frac{\rxyt\cdot\rxytp}{r_{x'y'}r_{xy}} \nonumber\\
     & \times\frac{\alpha_s}{\pi^2} \int_{z-1}^{z} \frac{\der z_g}{z_g}  \frac{e^{i\frac{z_g}{z}\ktone\cdot\rxyt }[2\mathfrak{Re}\Xi_{\rm NLO,3}(\xt,\yt,\xt',\yt')]\bar{Q}\QV K_1( \bar{Q} r_{x'y'})K_1(\QV r_{xy})}{\ryytp^2\left[(1-z)\ryytp^2+2(z-z_g)(1-z)\ryytp\cdot\rxyt-z_g(z-z_g)\rxyt^2\right]}\nonumber\\
     & \times \Bigg\{ (z(z-z_g) + (1-z)(1-z+z_g))\left(z-z_g\right)\left[\rxyt\cdot\ryytp (2z(1-z+z_g)-z_g(1+z_g))\right.\nonumber\\
     &\left.+(1-z)(1+z_g)\ryytp^2\right] +z_g(z_g+1-2z)^2(z-z_g)   \frac{(\rxyt\times\rxytp)(\rxyt\times\ryytp)}{\rxyt\cdot\rxytp}  \Bigg\}\,,
\end{align}
where we recall that $\QV^2 =(1-z+z_g)(z-z_g)Q^2$. The NLO,1 term is given by
\begin{align}
    &\left.\frac{\der \sigma^{\gamma_{\rm T}^{\star}+A\to j_f+X}}{  \der^2 \kt \der \eta}\right|_{\rm NLO,1} =\frac{\alpha_{\rm em}e_f^2N_c}{(2\pi)^4} \int\der^2\xt\der^2\xt'\der^2\yt\der^2\zt e^{-i\ktone \cdot \rxxtp}   \frac{\alpha_s}{\pi^2}Q \int_0^{1} \frac{\der z_g}{z_g} \nonumber\\
    & \times  \Bigg\{ 2\mathfrak{Re}[\Xi_{\rm NLO,1}(\xt,\yt,\zt,\xt')] \times \Bigg[ 2z^2(1-z) \bar{Q} \frac{K_1(\bar{Q}r_{x'y})}{r_{x'y}} \frac{K_1(QX_V)}{X_V} \nonumber \\
    & \times \left( (z^2+(1-z)^2)\left(1-\frac{z_g}{z}+\frac{z_g^2}{2z^2}\right) \frac{\RtS\cdot\rxpyt}{\rzxt^2} \right. \nonumber \\
    & -\frac{z_g(z-z_g)}{2}\left[\frac{1}{1-z+z_g}+\frac{(1-z)(z-z_g)}{z^3}\right] \frac{\rzxt\cdot\rxpyt}{\rzxt^2} - \frac{z_g(z-z_g)(1+z_g-2z)^2}{2z^2(1-z)} \nonumber \\
    & \times \frac{(\RtV\times\rxpyt)(\rzxt\times\rzyt)}{\rzxt^2\rzyt^2} - [z(z-z_g)+(1-z)(1-z+z_g)] \left(1-\frac{z_g}{z} \right) \left(1+\frac{z_g}{1-z}\right)  \nonumber \\
    & \left. \times \left(1-\frac{z_g}{2z}-\frac{z_g}{2(1-z+z_g)}\right) \frac{(\RtV\cdot\rxpyt)(\rzxt\cdot\rzyt)}{\rzxt^2\rzyt^2} \right) e^{-i\frac{z_g}{z}\ktone\cdot\rzxt} \Theta(z-z_g) \nonumber \\
    & + \bar{Q}_{\rm R2} \frac{K_1(\bar{Q}_{\rm R2}r_{x'y})}{r_{x'y}} \frac{K_1(QX_R)}{X_R}  \left(-\bar z(1-2(z+z_g)\bar z)(z^2+(z+z_g)^2)\frac{(\RtR\cdot\rxpyt)(\rzxt\cdot\rzxpt)}{\rzxt^2\rzxpt^2}\right. \nonumber \\
    &+z(2z^2-(2z+z_g)(1-z_g))(1-z_g-2z\bar z)\frac{(\RtRb\cdot\rxpyt)(\rzyt\cdot\rzxpt)}{\rzyt^2\rzxpt^2}\nonumber\\
    &+z_g\bar z(1-2(z+z_g))(2z+z_g)\frac{(\RtR\times\rxpyt)(\rzxt\times\rzxpt)}{\rzxt^2\rzxpt^2}\nonumber \\
    & +zz_g(\bar z-z)^2 \frac{(\RtRb\times\rxpyt)(\rzyt\times\rzxpt)}{\rzyt^2\rzxpt^2}\nonumber\\
    &\left. +\frac{zz_g\bar z(z(1-z)\bar z+(z+z_g)^3)}{(z+z_g)(1-z)} \frac{\rzxpt\cdot\rxpyt}{\rzxpt^2} \right) e^{-i\frac{z_g}{z}\ktone\cdot\rzxpt}\Theta(1-z-z_g) \Bigg] \nonumber \\ 
    &   - 2z^2(1-z)(z^2+(1-z)^2)\left(1-\frac{z_g}{z}+\frac{z_g^2}{2z^2}\right)   QK_1(\bar{Q}r_{xy})K_1(\bar{Q}r_{x'y})   \nonumber \\ 
    & \times \frac{\rxyt\cdot\rxpyt}{r_{xy}r_{x'y}}\frac{e^{-\frac{\rzxt^2}{\rxyt^2e^{\gamma_E}}}}{\rzxt^2}2C_F\mathfrak{Re}[\Xi_{\rm LO}(\xt,\yt,\xt')]\Bigg\}_{z_f}  + c.c.\,,
\end{align}
where the shorthand variables in this expression are summarized in Table~\ref{tab:notations-transverse}.

\begin{table}[tbh]
    \centering
    \begin{tabular}{|c|c|}
    \hline
    $\bar z$ & $1-z-z_g$\\
    \hline
    $\bar Q^2$ & $z(1-z)Q^2$ \\
    \hline
    $\bar Q_{\rm R2}^2$  & $\bar z(z+z_g)Q^2$ \\
    \hline
    $\bar Q_{\rm V3}^2$ & $(1-z+z_g)(z-z_g)Q^2$ \\
    \hline
    $X_V^2$ & $(1-z)(z-z_g)\rxyt^2 + z_g(z-z_g)\rzxt^2 +(1-z)z_g\rzyt^2$\\
    \hline
    $X_R^2$& $z\bar z \rxyt^2+zz_g\rzxt^2+\bar z z_g\rzyt^2$ \\
    \hline 
    $X_R'^2$ & $z\bar z\rxpyt^2+zz_g \rzxpt^2+\bar z z_g\rzyt^2$\\
    \hline 
    $\RtS$ & $\rxyt+\frac{z_g}{z}\rzxt$ \\
    \hline
    $\RtV$ & $ \rxyt+\frac{z_g}{1-z+z_g}\rzyt$ \\
    \hline
    $\RtR$ & $\rxyt+\frac{z_g}{z+z_g}\rzxt$ \\
    \hline
    $\RtRb$ & $-\rxyt+\frac{z_g}{1-z}\rzyt$\\
    \hline
    \end{tabular}
    \caption{Summary of our notations for the NLO impact factor of the SIDIS jet cross-section for transversely polarized virtual photons.}\label{tab:notations-transverse}
\end{table}
The NLO contribution proportional to $\Xi_{\rm NLO,4}$ is given by
\begin{align}
    &\left.\frac{\der \sigma^{\gamma_{\rm T}^{\star}+A\to j_f+X}}{  \der^2 \kt \der \eta}\right|_{\rm NLO,4} =\frac{\alpha_{\rm em}e_f^2N_c}{(2\pi)^4}\int\der^2\xt\der^2\xt'\der^2\yt\der^2\zt e^{-i\ktone \cdot \rxxtp} Q^2 \int_0^{1-z} \frac{\der z_g}{z_g}   \nonumber\\
    & \times \frac{\alpha_s}{2\pi^2} \frac{K_1(QX_R)K_1(QX_R')}{X_R X_R'}  \Bigg\{ \left[z\bar z^2(z^2+(1-\bar z)^2)\left(1-2\bar z(1-\bar z)\right)\frac{\rzxt\cdot\rzxpt}{\rzxt^2\rzxpt^2}\RtR\cdot\RtR'\right.\nonumber\\
    &+z\bar z^2z_g(1+z-\bar z)(1-2\bar z)\frac{\rzxt\times\rzxpt}{\rzxt^2\rzxpt^2}\RtR\times\RtR'-\frac{z^3\bar z^3z_g}{z+z_g}\left[\frac{\rzxt\cdot\RtR}{\rzxt^2}+\frac{\rzxpt\cdot\RtR'}{\rzxpt^2}\right]\nonumber\\
    &+\left.\frac{z^3\bar z^3z_g^2}{(z+z_g)^2}+\frac{\bar z^2z^4z_g^2}{(1-z)^2}+z^3(\bar z^2+(1-z)^2)\left(1-2z(1-z)\right)\frac{\RtRb\cdot\RtRb'}{\rzyt^2}\right.\nonumber\\
    &+z^3z_g(1+\bar z-z)(1-2z)\frac{\RtRb\times\RtRb'}{\rzyt^2}-\frac{\bar z^2z^4z_g}{1-z}\frac{\rzyt\cdot(\RtRb+\RtRb')}{\rzyt^2}\nonumber\\
    &+2z\bar z^2(z(1-z)+\bar z(1-\bar z))(z+\bar z-2z\bar z)\frac{\rzxt\cdot\rzyt}{\rzxt^2\rzyt^2}\RtR\cdot\RtRb' \nonumber \\
    & -2z\bar z^2z_g(z-\bar z)^2\frac{\rzxt\times\rzyt}{\rzxt^2\rzyt^2}\RtR\times\RtRb'\nonumber\\
    &\left.-2z\bar z^2z_g\left(\frac{z(1-z)^2}{1-\bar z}\frac{\rzxt\cdot\RtRb'}{\rzxt^2}+\frac{\bar z(1-\bar z)^2}{1-z}\frac{\rzyt\cdot\RtR}{\rzyt^2}\right)\right]  2\mathfrak{Re}\left[\Xi_{\rm NLO,4}(\xt,\yt,\zt,\xt')\right] \nonumber \\ 
     &  - z^3(\bar z^2+(1-z)^2)\left(1-2z(1-z)\right)\frac{\rxyt\cdot\rxpyt}{\rzyt^2} e^{-\frac{\rzyt^2}{\rxxtp^2e^{\gamma_E}}} \frac{K_0(\bar Q r_{xy})K_0(\bar Q r_{x'y})X_R X_R'}{K_1(QX_R)K_1(QX_R')r_{xy}r_{x'y}} \nonumber \\ 
     &  \times 2C_F\mathfrak{Re}\left[\Xi_{\rm LO}(\xt,\yt,\xt')\right]\Bigg\}_{z_f} + c.c.\,,
\end{align}
with $\RtR'=\rxpyt+\frac{z_g}{z+z_g}\rzxpt$, $\RtRb'=-\rxpyt+\frac{z_g}{1-z}\rzyt$ and $X_R'^2=z\bar z \rxpyt^2+zz_g\rzxpt^2+\bar z z_g\rzyt^2$.

The gluon-tagged jet contributions are also reported here:
\begin{align}
    &\left.\frac{\der\sigma^{\gamma^*_{\rm T}+A\to j_g+X}}{\der^2\kt\der\eta}\right|_{\rm NLO,0}=\frac{\alpha_{\rm em}e_f^2N_c}{(2\pi)^4}\int\der^2\xt\der^2\xt'\der^2\yt e^{-i\kt\cdot\rxxtp}2\mathfrak{Re}\left[\Xi_{\rm LO}(\xt,\yt,\xt')\right]\nonumber\\
    &\times\frac{(-\alpha_s )C_F}{\pi}\int_0^{1-z}\frac{\der z_1}{z^2}e^{-i\frac{z_1}{z}\kt\cdot\rxxtp}[(1-z_1-z)^2+(z_1+z)^2]\left(2z_1(z_1+z)+z_g^2)\right)\nonumber\\
    &\times\bar Q_{\rm R2}^2\frac{\rxyt\cdot\rxpyt}{r_{xy}r_{x'y}}K_1(\bar Q_{\rm R2}r_{xy})K_1(\bar Q_{\rm R2}r_{x'y})\ln\left(\frac{\kt^2\rxxtp^2z_1^2}{c_0^2z^2}\right)\\
     &\left.\frac{\der\sigma^{\gamma^*_{\rm T}+A\to j_g+X}}{\der^2\kt\der\eta}\right|_{\rm NLO,3}=\frac{\alpha_{\rm em}e_f^2N_c}{(2\pi)^4}\int\der^2\xt\der^2\xt'\der^2\yt\der^2\yt' e^{-i\kt\cdot\rxyt} \nonumber\\
     &\times \frac{\alpha_s}{\pi^2}\int_0^{1-z}\frac{\der z_1}{z^2}\left[(2z_1(1-z_1-z)+z(1-z))(1-z-2z_1(1-z_1-z))\frac{\rxyt\cdot\rxytp}{r_{xy}r_{x'y'}}\frac{\rxxtp\cdot\ryytp}{\rxxtp^2\ryytp^2}\right.\nonumber\\
     &\left.-z(1-2z_1-z)^2\frac{\rxyt\times\rxytp}{r_{xy}r_{x'y'}}\frac{\rxxtp\times\ryytp}{\rxxtp^2\ryytp^2}\right] 2\mathfrak{Re}\left[\Xi_{\rm NLO,3}(\xt,\yt,\xt',\yt')\right] \nonumber \\
     & \times e^{-i\frac{\kt}{z}\cdot\left(z_1\rxxtp+(1-z_1)\ryytp\right)}\bar Q_{\rm R2}\bar Q_{\rm R2'}K_1(\bar Q_{\rm R2}r_{xy})K_1(\bar Q_{\rm R2'}r_{x'y'})\,,
\end{align}
where for these terms, $\bar Q_{\rm R2}^2=(1-z-z_1)(z_1+z)Q^2$ and $\bar Q_{\rm R2'}=z_1(1-z_1)Q^2$ in the integrals over $z_1$. For the two remaining terms, we have
\begin{align}
     &\left.\frac{\der\sigma^{\gamma^*_{\rm T}+A\to j_g+X}}{\der^2\kt\der\eta}\right|_{\rm NLO,1}=\frac{\alpha_{\rm em}e_f^2N_c}{(2\pi)^4}\int\der^2\zt\der^2\zt'\der^2\xt\der^2\yt e^{-i\kt\cdot\rzzpt} 2\mathfrak{Re}[\Xi_{\rm NLO,1}(\xt,\yt,\zt,\zt')]\nonumber\\
     &\times\left[\frac{(-\alpha_s)}{\pi^2}\int_0^{1-z}\frac{\der z_1}{z^2} \ e^{i\frac{z_1}{z}\kt\cdot\rzpxt}(1-z_1-z)(z_1+z)\frac{Q \bar{Q}_{\rm R2}K_1(QX_R)K_1(\bar Q_{\rm R2}r_{z'y})}{X_Rr_{z'y}}\right.\nonumber\\
    &\times\left\{\left[z_1^2+(z_1+z)^2\right]\left(1-2(1-z_1-z)(z_1+z)\right)\frac{\rzxt\cdot\rzpxt}{\rzxt^2\rzpxt^2}\RtR\cdot\rzpyt\right.\nonumber\\
    &\left.+z(2z_1+z)(2z_1+2z-1)\frac{\rzxt\times\rzpxt}{\rzxt^2\rzpxt^2}\RtR\times\rzpyt-\frac{z_1^2(1-z_1-z)z}{z_1+z}\frac{\rzpxt\cdot\rzpyt}{\rzpxt^2}\right\}\nonumber\\
    &+\frac{(-\alpha_s)}{\pi^2}\int_0^{1-z}\frac{\der z_1}{z^2}e^{i\frac{1-z_1-z}{z}\kt\cdot\rzpyt}(1-z_1-z)(z_1+z)\frac{Q \bar{Q}_{\rm R2'}K_1(QX_R)K_1(\bar Q_{\rm R2'}r_{z'x})}{X_Rr_{z'x}}\nonumber\\
&\times \left\{(z_1(1-z_1)+(1-z_1-z)(z_1+z))(1-z-2z_1(1-z_1-z))\frac{\rzxt\cdot\rzpyt}{\rzxt^2\rzpyt^2}\RtR\cdot\rzpxt\right.\nonumber\\
&\left.\left.-z(1-2z_1-z)^2\frac{\rzxt\times\rzpyt}{\rzxt^2\rzpyt^2}\RtR\times\rzpxt-\frac{z_1z(1-z_1)^2}{z_1+z}\frac{\rzpyt\cdot\rzpxt}{\rzpyt^2}\right\}\right]+c.c.\,,
\end{align}
and
\begin{align}
     &\left.\frac{\der\sigma^{\gamma^*_{\rm T}+A\to j_g+X}}{\der^2\kt\der\eta}\right|_{\rm NLO,4g}=\frac{\alpha_{\rm em}e_f^2N_c}{(2\pi)^4}\int\der^2\zt\der^2\zt'\der^2\xt\der^2\yt e^{-i\kt\cdot\rzzpt} 2\mathfrak{Re}[\Xi_{\rm NLO,4g}(\xt,\yt,\zt,\zt')]\nonumber\\
     &\times\frac{\alpha_s}{\pi^2}\int_0^{1-z}\der z_1\frac{Q^2K_1(QX_R)K_1(QX_R')}{X_RX_R'}\Bigg\{(1-z_1-z)^2\left[z_1^2+(z_1+z)^2\right]\nonumber\\
     &\times\left(1-2(1-z_1-z)(z_1+z)\right)\frac{\rzxt\cdot\rzpxt}{\rzxt^2\rzpxt^2}\RtR\cdot\RtR'+z(1-z_1-z)^2(2z_1+z)\nonumber\\
     &\times (1-2(1-z_1-z))\frac{\rzxt\times\rzpxt}{\rzxt^2\rzpxt^2}\RtR\times\RtR'-\frac{z_1^2(1-z_1-z)^3z}{z_1+z}\left[\frac{\rzxt\cdot\RtR}{\rzxt^2}+\frac{\rzpxt\cdot\RtR'}{\rzpxt^2}\right]\nonumber\\
&+\frac{z_1^2(1-z_1-z)^3z^2}{(z_1+z)^2}+z_1(1-z_1-z)(z_1(1-z_1)+(1-z_1-z)(z_1+z))\nonumber\\
&\times (1-z-2z_1(1-z_1-z))\frac{\rzxt\cdot\rzpyt}{\rzxt^2\rzpyt^2}\RtR\cdot\RtRb'-zz_1(1-z_1-z)(1-2z_1-z)^2\frac{\rzxt\times\rzpyt}{\rzxt^2\rzpyt^2}\RtR\times\RtRb'\nonumber\\
&\left.-zz_1(1-z_1-z)\left[\frac{z_1(1-z_1)^2}{z_1+z}\frac{\rzxt\cdot\RtRb'}{\rzxt^2}+\frac{(1-z_1-z)(z_1+z)^2}{1-z_1}\frac{\rzpyt\cdot\RtR}{\rzpyt^2}\right]\right\}\,.
\end{align}
Implicit in our notations, the transverse sizes $X_R$, $X_R'$ are evaluated with $z_g$ set to $z$, since $z_g$ is the longitudinal momentum fraction of the tagged jet. Likewise, the vectors $\RtR'$ and $\RtRb'$ should be replaced by their expressions provided in Table~\ref{tab:notations-transverse} with the prime index on the $\zt$ transverse coordinate only.

\subsection{NLO impact factor for the single differential cross-section }

Finally, we quote in this subsection the formulae for the $\kt$-integrated semi-inclusive single-jet cross-section. We again rely on the results established in section~\ref{sec:eta-diff} for a longitudinally polarized virtual photon, and in particular on the decomposition Eq.\,\eqref{eq:ktintegrated-NLO-decomp} of the NLO cross-section. 

\begin{align}
    &\left.\frac{\der \sigma^{\gamma_{\rm L}^{\star}+A\to j_f+X}}{\der \eta }\right|_{\rm NLO-a} =\frac{\alpha_{\mathrm{em}} e_f^2 N_c}{(2\pi)^2}\int\der^2\bt\der^2\rxyt 2z(z^2+(1-z)^2)\bar Q^2K_1^2(\bar Qr_{xy}) 4\mathfrak{Re}\left \langle 1 - D_{xy}\right \rangle_{Y_f}\nonumber\\
    &\times\frac{\alpha_sC_F}{\pi}\left\{-\frac{1}{2}\ln^2\left(\frac{z}{1-z}\right)+\ln\left(\frac{z}{1-z}\right)\ln\left(R^2\right)-\frac{3}{2}\ln\left(2R\right)+7-\frac{2\pi^2}{3}\right.\nonumber\\
    &-\int_0^{1-z}\frac{\der z_g}{z_g}\ln\left(\frac{R^2z_g^2}{z^2}\right)\left[\frac{(1-z-z_g)((1-z-z_g)^2+(z+z_g)^2)(2z(z+z_g)+z_g^2)}{2z(1-z)(z+z_g)(z^2+(1-z)^2)}\frac{K_1^2(\bar Q_{\mathrm{R}2}r_{xy})}{K_1^2(\bar Qr_{xy})}-1\right]\nonumber\\
    &\left.+2\int_0^{1-z}\frac{\der z_g}{z_g}\ln\left(1+\frac{z_g}{z}\right)\frac{(1-z-z_g)((1-z-z_g)^2+(z+z_g)^2)(2z(z+z_g)+z_g^2)}{2z(1-z)(z+z_g)(z^2+(1-z)^2)}\frac{K_1^2(\bar Q_{\mathrm{R}2}r_{xy})}{K_1^2(\bar Qr_{xy})}\right\}\nonumber\\
    &-\frac{\alpha_{\mathrm{em}} e_f^2 N_c}{(2\pi)^2}\left[2z(z^2+(1-z)^2)\bar Q^2\right]\times\frac{\alpha_sC_F}{\pi^2}\ln\left(\frac{z}{1-z}\right)\nonumber\\
    &\times\int\der^2\xt\der^2\yt\mathcal{P}_{\rxyt}\left(\frac{1}{\rxxtp^2}\right)\left[K_1(\bar Q r_{xy})K_1(\bar Q r_{x'y})\frac{\rxyt\cdot\rxpyt}{r_{xy}r_{x'y}}2\mathfrak{Re}\left \langle D_{xx'} - D_{xy} -  D_{yx'} + 1 \right \rangle_{Y_f}\right]\,.\label{eq:NLOa-transverse-final}
\end{align}
\begin{align}
    &\left.\frac{\der \sigma^{\gamma_{\rm T}^{\star}+A\to j_f +X}}{\der \eta }\right|_{\rm NLO-b} =\frac{\alpha_{\rm em}e_f^2N_c}{(2\pi)^2} \int\frac{\der^2\bt}{(2\pi)} \der^2\rxxtp \der^2\ryytp \ 2z^2\bar{Q}\times\frac{\alpha_s}{\pi}\nonumber\\
     & \times \int_{z-1}^{z} \frac{\der z_g}{z_g} \left(z-z_g\right) \QV \frac{K_1( \bar{Q} |(z-z_g)\rxxtp+(1-z+z_g)\ryytp|)}{|(z-z_g)\rxxtp+(1-z+z_g)\ryytp|} \frac{K_1(\QV |z\rxxtp+(1-z)\ryytp|)}{|z\rxxtp+(1-z)\ryytp|} \nonumber\\
     & \times \Bigg\{[z(z-z_g) + (1-z)(1-z+z_g)](z\rxxtp+(1-z)\ryytp)\cdot((z-z_g)\rxxtp+(1-z+z_g)\ryytp) \nonumber\\
     &\times\left[1+\frac{((1-z)\ryytp+(z-z_g)\rxxtp)\cdot(z\rxxtp+(1-z+z_g)\ryytp)}{(1-z)(1-z+z_g)\ryytp^2-z(z-z_g)\rxxtp^2}\right]+z_g^2(z_g+1-2z)^2  \nonumber\\ 
     & \times\frac{(\rxxtp\times\ryytp)^2}{\ryytp^2\left[(1-z)(1-z+z_g)\ryytp^2-z(z-z_g)\rxxtp^2\right]}  \Bigg\}4\mathfrak{Re}[\Xi_{\rm NLO,3}(\xt,\yt,\xt',\yt')]\,.\label{eq:kt1-integrated-V3xLO-transverse}
\end{align}
This expression may be further simplified using
\begin{align}
    (\rxxtp\times\ryytp)^2=\rxxtp^2\ryytp^2-(\rxxtp\cdot\ryytp)^2\,.
\end{align}
The transverse coordinates $\xt$, $\yt$, $\xt'$ and $\yt'$ in the NLO,3 CGC correlator depend on $\bt$, $\rxxtp$ and $\ryytp$ according to Eqs.\,\eqref{eq:bdep-1}.
The UV regular term is given by 
\begin{align}
        &\left.\frac{\der \sigma^{\gamma_{\rm T}^{\star}+A\to j_f+X}}{  \der \eta }\right|_{\rm UV-reg} =\frac{\alpha_{\mathrm{em}} e_f^2 N_c}{(2\pi)^2}\int\der^2\xt\der^2\yt\der^2\zt \frac{\alpha_s}{\pi^2}Q^2 \int_0^{1}\frac{\der z_g}{z_g}\left\{\frac{\Theta(1-z-z_g)}{\rzyt^2}\right.\nonumber\\
        &\times\frac{z^2(\bar z^2+(1-z)^2)(1-2z(1-z))}{1- z}\left[K_1^2(QX_R)\frac{z(1-z)\RtRb^2}{X_R^2}2\mathfrak{Re}\Xi_{\rm NLO,4}(\xt,\yt,\zt,\xt)\right.\nonumber\\
        &\left. -K_1(\bar Qr_{xy})K_1\left(\bar Q|\rxyt-\rzyt|\right)\frac{\rxyt\cdot(\rxyt-\rzyt)}{r_{xy}|\rxyt-\rzyt|}2\mathfrak{Re}C_F\Xi_{\rm LO}(\yt,\xt,\zt)\right]+\frac{\Theta(1-z-z_g)}{\rzxt^2}\nonumber\\
        &\times \frac{z\bar z(z^2+(z+z_g)^2)(1-2\bar z(z+z_g))}{z+z_g}\left[K_1^2(QX_R)\frac{\bar z(1-\bar z)\RtR^2}{X_R^2}2\mathfrak{Re}\Xi_{\rm NLO,4}(\xt,\yt,\zt,\xt)\right.\nonumber\\
     &-K_1(QX_R)K_1\left(\bar Q_{\rm R2}\left|\RtR\right|\right)\frac{\sqrt{\bar z(1-\bar z)}|\RtR|}{X_R}4\mathfrak{Re}\Xi_{\rm NLO,1}(\xt,\yt,\zt,\boldsymbol{w}_{\perp,r})\nonumber\\
     &\left.+K_1(\bar Q_{\rm R2}r_{xy})K_1\left(\bar Q_{\rm R2}\left|\rxyt+\rzxt\right|\right)\frac{\rxyt\cdot(\rxyt+\rzxt)}{r_{xy}|\rxyt+\rzxt|}2\mathfrak{Re}C_F\Xi_{\rm LO}(\xt,\yt,\zt)\right]+\frac{\Theta(z-z_g)}{\rzxt^2}\nonumber\\
        &\times (1-z)(z^2+(1-z)^2)\left(2z(z-z_g)+z_g^2\right)\left[K_1\left(\bar Q\left|\RtS\right|\right)K_1(QX_V)\frac{\sqrt{z(1-z)}|\RtS|}{X_V}\right.\nonumber\\
        &\left.\times 4\mathfrak{Re}\Xi_{\rm NLO,1}(\xt,\yt,\zt,\boldsymbol{w}_{\perp,v})\left.-e^{-\frac{\rzxt^2}{\rxyt^2e^{\gamma_E}}}K_0^2(\bar Qr_{xy})4\mathfrak{Re}C_F\Xi_{\rm LO}(\xt,\yt,\xt)\right]\right\}_{z_f}\,.
\end{align}
Last,
\begin{align}
    &\left.\frac{\der \sigma^{\gamma_{\rm T}^{\star}+A\to j_f+X}}{  \der \eta }\right|_{\rm c} =\frac{\alpha_{\mathrm{em}} e_f^2 N_c}{(2\pi)^2}\int\der^2\xt\der^2\yt\der^2\zt \frac{2\alpha_s}{\pi^2}2\mathfrak{Re}\int_0^{1} \frac{\der z_g}{z_g} \Bigg\{2z^2(1-z)\Theta(z-z_g)  \nonumber \\
     & \times \bar{Q}Q \frac{K_1(\bar{Q}|\RtS|)}{|\RtS|} \frac{K_1(QX_V)}{X_V} \left( -\frac{z_g(z-z_g)}{2}\left[\frac{1}{1-z+z_g}+\frac{(1-z)(z-z_g)}{z^3}\right] \frac{\rzxt\cdot\rxpyt}{\rzxt^2} \right.\nonumber \\
     & - \frac{z_g(z-z_g)(1+z_g-2z)^2}{2z^2(1-z)}\frac{(\RtV\times\RtS)(\rzxt\times\rzyt)}{\rzxt^2\rzyt^2} - [z(z-z_g)+(1-z)(1-z+z_g)] \nonumber \\
     & \left. \times \left(1-\frac{z_g}{z} \right) \left(1+\frac{z_g}{1-z}\right)  \left(1-\frac{z_g}{2z}-\frac{z_g}{2(1-z+z_g)}\right) \frac{(\RtV\cdot\RtS)(\rzxt\cdot\rzyt)}{\rzxt^2\rzyt^2} \right)  \nonumber \\
     & \times \Xi_{\rm NLO,1}(\xt,\yt,\zt,\boldsymbol{w}_{\perp,v})+ \Theta(1-z-z_g)Q\bar{Q}_{\rm R2} \frac{K_1(\bar{Q}_{\rm R2}|\RtR|)}{|\RtR|} \frac{K_1(QX_R)}{X_R}  \nonumber \\
     &\times\left(z(2z^2-(2z+z_g)(1-z_g))(1-z_g-2z\bar z)\frac{(\RtRb\cdot\RtR)(\rzyt\cdot(\rzyt-\RtR))}{\rzyt^2(\rzyt-\RtR)^2}\right.  \nonumber\\
     &\left.+zz_g(\bar z-z)^2 \frac{(\RtRb\times\RtR)(\RtR\times \rzyt)}{\rzyt^2(\rzyt-\RtR)^2}+\frac{zz_g\bar z(z(1-z)\bar z+(z+z_g)^3)}{(z+z_g)(1-z)} \frac{(\rzyt-\RtR)\cdot\RtR}{(\rzyt-\RtR)^2} \right)\nonumber\\
    &\times \Xi_{\rm NLO,1}(\xt,\yt,\zt,\boldsymbol{w}_{\perp,r})+\Theta(1-z-z_g)Q^2 \frac{K_1^2(QX_R)}{X_R^2}   \left(-\frac{z^3\bar z^3z_g}{z+z_g}\frac{\rzxt\cdot\RtR}{\rzxt^2}\right.+\left.\frac{z^3\bar z^3z_g^2}{2(z+z_g)^2}\right.\nonumber\\
&+\frac{\bar z^2z^4z_g^2}{2(1-z)^2}-\frac{\bar z^2z^4z_g}{1-z}\frac{\rzyt\cdot\RtRb}{\rzyt^2}+z\bar z^2(z(1-z)+\bar z(1-\bar z))(z+\bar z-2z\bar z)\frac{\rzxt\cdot\rzyt}{\rzxt^2\rzyt^2}\RtR\cdot\RtRb\nonumber\\
&-z\bar z^2z_g(z-\bar z)^2\frac{\rzxt\times\rzyt}{\rzxt^2\rzyt^2}\RtR\times\RtRb\left.-z\bar z^2z_g\left(\frac{z(1-z)^2}{1-\bar z}\frac{\rzxt\cdot\RtRb}{\rzxt^2}+\frac{\bar z(1-\bar z)^2}{1-z}\frac{\rzyt\cdot\RtR}{\rzyt^2}\right)\right)\nonumber\\
&\times\Xi_{\rm NLO,4}(\xt,\yt,\zt,\xt)\Bigg\}_{z_f}\,.
\end{align}

\section{Calculation of the $k_\perp$-dependent single threshold logarithm}
\label{app:single-log}

In this appendix, we explicitly compute the contribution to the NLO semi-inclusive single-jet production cross-section enhanced by a single logarithm of $1-z$, where $z$ denotes the longitudinal momentum fraction of the jet relative to that of the virtual photon. We detail the calculation for a longitudinally polarized photon but the rationale is identical in the transversely polarized case. Our starting point is Eq.\,\eqref{eq:NLO0-zsim1} and more precisely the term proportional to $\ln(\kt^2r_{xy}r_{x'y}/c_0^2)$, which we label here as ``sl":
\begin{align}
        &\left.\frac{\der \sigma^{\gamma_{\rm L}^{\star}+A\to j_f+X}}{ \der^2 \kt  \der \eta }\right|_{\rm sl} \equiv\frac{\alpha_{\mathrm{em}} e_f^2 N_c}{(2\pi)^4}\int\der^2\xt\der^2\xt'\der^2\yt e^{-i\kt\cdot\rxxtp}2\mathfrak{Re}\left[\Xi_{\rm LO}(\xt,\yt,\xt')\right]\nonumber\\
        &\times 8z^3(1-z)^2Q^2K_0(\bar Q r_{xy})K_0(\bar Qr_{x'y})\frac{\alpha_sC_F}{\pi}\ln\left(\frac{\kt^2r_{xy}r_{x'y}}{c_0^2}\right)\ln\left(\frac{1}{1-z}\right)\,,\label{eq:appB1}
\end{align}
and we aim to demonstrate that this term is truly single logarithmic in $1-z$ (it does not contain ``hidden" double log in $1-z$). This is \textit{a priori} not obvious as one could argue that the $K_0$ Bessel functions parametrically enforce $r_{xy}\sim r_{x'y}\sim 1/\bar Q$ so that the apparent single log coefficient could behave like $\ln(\kt^2/\bar Q^2)$. Since $\bar Q^2\sim (1-z)Q^2$ for $z$ close to 1, one sees that this naive estimation would make the contribution given by Eq.\,\eqref{eq:appB1} double logarithmic in $1-z$.

To do so, we reduce the number of transverse coordinate integrals following appendix~A of \cite{Iancu:2021rup}. Assuming translational invariance in transverse space, and after the change of variables $\rt=\rxyt$, $\rt'=\rxpyt$, $\bt=z\xt+(1-z)\yt$, we first rewrite the LO semi-inclusive single-jet cross-section as
\begin{align}
    \left.\frac{\der \sigma^{\gamma_{\rm L}^{\star}+A\to j_f+X}}{ \der^2 \kt  \der \eta }\right|_{\rm LO}&=\alpha_{\mathrm{em}} e_f^2 N_c16z(1-z)\mathfrak{Re}\int\frac{\der^2\bt}{(2\pi)^2}\int\frac{\der^2\rt}{(2\pi)}\frac{\der^2\rt'}{(2\pi)} e^{-i\kt\cdot(\rt-\rt')}\nonumber\\
        &\times \bar Q^2K_0(\bar Q r_\perp)K_0(\bar Qr')\left[T_{Y_f}(\rt)+T_{Y_f}(-\rt')-T_{Y_f}(\rt-\rt')\right]\,,
\end{align}
with $T_{Y_f}(\rt)=1-\frac{1}{N_c}\langle \Tr[ V(\boldsymbol{0}_\perp)V^\dagger(\rt)]\rangle_{Y_f}$. As shown in \cite{Iancu:2020jch}, this expression can be further simplified into
\begin{align}
    \left.\frac{\der \sigma^{\gamma_{\rm L}^{\star}+A\to j_f+X}}{ \der^2 \kt  \der \eta }\right|_{\rm LO}&=\alpha_{\mathrm{em}} e_f^2 N_c16z(1-z)\mathfrak{Re}\int\frac{\der^2\bt}{(2\pi)^2}\int\frac{\der^2\rt}{(2\pi)}e^{-i\kt\cdot\rt}T_{Y_f}(\rt)\nonumber\\
        &\times \left[\frac{2\bar Q^2}{\kt^2+\bar Q^2}K_0(\bar Q r_\perp)-\frac{\bar Q r_\perp}{2}K_1(\bar Q r_\perp)\right]\,.\label{eq:appLO}
\end{align}
We wish to repeat this calculation with the additional $\ln(\kt^2r_{xy}r_{x'y}/c_0^2)$ present in Eq.\,\eqref{eq:appB1}. The terms depending on $T(\rt)$ and $T(-\rt')=T(\rt')^*$ contribute equally and can be computed as follows. Considering the $T(\rt)$ term alone, one needs the integral
\begin{align}
    \int\frac{\der^2\rt'}{(2\pi)}e^{i\kt\cdot\rt'}K_0(\bar Q r_\perp')\ln\left(\frac{k_\perp r'}{c_0}\right)=\frac{1}{\kt^2+\bar Q^2}\ln\left(\frac{k_\perp \bar Q}{\kt^2+\bar Q^2}\right)\,,
\end{align}
which has been derived in appendix~(A1) of \cite{Caucal:2023nci}. Therefore, the $T(\rt)+T(-\rt')$ terms give
\begin{align}
        \alpha_{\mathrm{em}} e_f^2 N_c16z(1-z)\mathfrak{Re}\frac{\alpha_sC_F}{\pi}&\int\frac{\der^2\bt}{(2\pi)^2}\int\frac{\der^2\rt}{(2\pi)}e^{-i\kt\cdot\rt}T_{Y_f}(\rt)\nonumber\\
        &\times\frac{2\bar Q^2}{\kt^2+\bar Q^2}K_0(\bar Q r_\perp)\ln\left(\frac{\kt^2\bar Qr_\perp}{c_0(\kt^2+\bar Q^2)}\right)\,.
\end{align}
The term proportional to $T(\rt-\rt')$ is more complicated, but can be computed in a similar fashion:
\begin{align}
    \int\frac{\der^2\rt}{(2\pi)}\frac{\der^2\rt'}{(2\pi)}&e^{-i\kt\cdot(\rt-\rt')}T_{Y_f}(\rt-\rt')\bar Q^2K_0(\bar r_\perp)K_0(\bar Qr_\perp')\ln\left(\frac{\kt^2rr'}{c_0^2}\right)=\nonumber\\
    &\int\frac{\der^2\rt}{(2\pi)}e^{-i\kt\cdot\rt}T_{Y_f}(\rt)\int\frac{\der^2\lt}{(2\pi)}\frac{e^{-i\lt\cdot\rt}}{[\lt^2+\bar Q^2]^2}\ln\left(\frac{\lt^2\bar Q^2}{[\lt^2+\bar Q^2]^2}\right)\,.
\end{align}
We now define the following function
\begin{align}
    \kappa_L(x)\equiv\frac{2}{xK_1(x)}\int_0^\infty\der \ell \  \frac{\ell J_0(\ell x)}{(\ell^2+1)^2}\ln\left(\frac{\ell^2}{(\ell^2+1)^2}\right)\,.
\end{align}
where $x=\bar Q r_\perp$ in terms of physical variables.
Without the logarithm inside the integral over $\ell$, one would simply get $\kappa(x)=1$.  When $x$ goes to 0 (meaning $\bar Q r_\perp\sim \sqrt{1-z}Q r_\perp\to 0$ when $z\to 1$), the function $\kappa(x)$ has a finite limit\begin{equation}
    \lim\limits_{x\to 0}\kappa_L(x)=-2\,.\,\label{eq:app-kappalim}
\end{equation}
This function helps us to write our final exact expression for Eq.\,\eqref{eq:appB1} as
\begin{align}
        &\left.\frac{\der \sigma^{\gamma_{\rm L}^{\star}+A\to j_f+X}}{ \der^2 \kt  \der \eta }\right|_{\rm sl}=\alpha_{\mathrm{em}} e_f^2 N_c16z(1-z)\mathfrak{Re}\ln\left(\frac{1}{1-z}\right)\int\frac{\der^2\bt}{(2\pi)^2}\int\frac{\der^2\rt}{(2\pi)}e^{-i\kt\cdot\rt}T_{Y_f}(\rt)\nonumber\\
        &\times \frac{\alpha_s C_F}{\pi}\left[\frac{2\bar Q^2}{\kt^2+\bar Q^2}K_0(\bar Q r_\perp)\ln\left(\frac{\kt^2\bar Qr_\perp}{c_0(\kt^2+\bar Q^2)}\right)-\frac{\bar Q r_\perp}{2}K_1(\bar Q r_\perp)\kappa_L\left(\bar Q r_\perp\right)\right]\label{eq:appNLO-sl} \,.
\end{align}
Comparing Eq.\,\eqref{eq:appLO} with Eq.\,\eqref{eq:appNLO-sl}, it is clear that the integral over $\rt$ does not yield an additional $\ln(1-z)$ dependence when $z\to 1$ so that the single log contribution Eq.\,\eqref{eq:appB1} is indeed single logarithmic in the threshold limit. In fact, thanks to Eq.\,\eqref{eq:app-kappalim}, one has
\begin{align}
     &\left.\frac{\der \sigma^{\gamma_{\rm L}^{\star}+A\to j_f+X}}{ \der^2 \kt  \der \eta }\right|_{\rm sl}\underset{z\to 1}{\sim}\frac{(-2)\alpha_s C_F}{\pi}\ln\left(\frac{1}{1-z}\right)\left.\frac{\der \sigma^{\gamma_{\rm L}^{\star}+A\to j_f+X}}{ \der^2 \kt  \der \eta }\right|_{\rm LO}\,,
\end{align}
which implies that the $\kt$-dependent single log coefficient can effectively by replaced by $-2\alpha_s C_F/\pi$ up to power suppressed terms as $z$ goes to 1.

Likewise, the LO SIDIS cross-section for transversely polarized virtual photons can be expressed in terms of $T_{Y_f}(\rt)$ as
\begin{align}
     \left.\frac{\der \sigma^{\gamma_{\rm T}^{\star}+A\to j_f+X}}{ \der^2 \kt  \der \eta }\right|_{\rm LO}&=\alpha_{\mathrm{em}} e_f^2 N_c4z(z^2+(1-z)^2)\mathfrak{Re}\int\frac{\der^2\bt}{(2\pi)^2}\int\frac{\der^2\rt}{(2\pi)}e^{-i\kt\cdot\rt}T_{Y_f}(\rt)\nonumber\\
        &\times \left[\frac{i\kt\cdot\rt}{\bar Q r_\perp}\frac{2\bar Q^2}{\kt^2+\bar Q^2}K_1(\bar Q r_\perp)-K_0(\bar Q r_\perp)+\frac{\bar Q r_\perp}{2}K_1(\bar Q r_\perp)\right]\,,
\end{align}
while the $\kt$-dependent single log coefficient reads
\begin{align}
     &\left.\frac{\der \sigma^{\gamma_{\rm T}^{\star}+A\to j_f+X}}{ \der^2 \kt  \der \eta }\right|_{\rm sl}=\alpha_{\mathrm{em}} e_f^2 N_c4z(z^2+(1-z)^2)\mathfrak{Re}\frac{\alpha_s C_F}{\pi}\ln\left(\frac{1}{1-z}\right)\int\frac{\der^2\bt}{(2\pi)^2}\nonumber\\
        &\times \int\frac{\der^2\rt}{(2\pi)}e^{-i\kt\cdot\rt}T_{Y_f}(\rt)\left[\frac{i\kt\cdot\rt}{\bar Q r_\perp}\frac{2\bar Q^2}{\kt^2+\bar Q^2}K_1(\bar Q r_\perp)\left(\ln\left(\frac{k_\perp r_\perp}{c_0}\right)-\frac{1}{2}\ln\left(\frac{\kt^2+\bar Q^2}{\kt^2}\right)\right.\right.\nonumber\\
        &\left.\left.+\frac{1}{2}\frac{\bar Q^2}{\kt^2}\ln\left(1+\frac{\kt^2}{\bar Q^2}\right)\right)-\left(K_0(\bar Q r_\perp)-\frac{\bar Q r_\perp}{2}K_1(\bar Q r_\perp)\right)\kappa_T\left(\bar Q r_\perp)\right)\right]\,,
\end{align}
with the function $\kappa_T(x)$ defined by
\begin{align}
    \kappa_T(x)\equiv\frac{2}{2K_0(x)-xK_1(x)}\int_0^\infty\der\ell \ \frac{\ell^3J_0(\ell x)}{(\ell^2+1)^2}\left[\frac{1}{\ell}\ln(1+\ell)-\ln\left(1+\frac{1}{\ell}\right)\right] \,.
\end{align}
The function $\kappa_T(x)$ goes to 0 as $x\to 0$, which implies that the $\kt$-dependent coefficient of the single threshold logarithm vanishes in the $z \to 1$ limit.

\section{The Fourier transform of the logarithm as a distribution}
\label{app:useful-id}

The following identities will be useful for the computations in Sec.\,\eqref{sec:kTintegratedNLO0}: 
\begin{align}
    \int \der^2 \rt f(\rt) \int  \der^2 \kt e^{-i \kt \cdot \rt} \ln \left(\frac{\ut^2 \kt^2}{c_0^2} \right) = - (4\pi) \Pcal_{\ut}\left( \frac{1}{\rt^2} \right) [f] \,,
    \label{eq:identity1}
\end{align}
\begin{align}
    \int \der^2 \rt f(\rt) \int  \der^2 \kt e^{-i \kt \cdot \rt} f(\rt) \ln \left(\frac{\rt^2 \kt^2}{c_0^2} \right) = - 4\pi \int \der^2 \rt \frac{f(\rt)}{\rt^2} \,,
    \label{eq:identity2}
\end{align}
where we defined
\begin{align}
    \Pcal_{\ut}\left(\frac{1}{\rt^2} \right)[f] \equiv \int \der^2 \rt \left[\frac{f(\rt) - f(0) \Theta(\ut^2-\rt^2)}{\rt^2} \right] \,.
\end{align}
We begin with the proof of Eq.\,\eqref{eq:identity1} following section~(2.9) in \cite{vladimirov1976equations} :
\begin{align}
    \Pcal_{\ut}\left( \frac{1}{\rt^2} \right) [f] = \int \der^2 \rt  \int  \frac{\der^2 \kt}{(2\pi)^2} \tilde{f}(\kt) \left[ \frac{e^{i \kt \cdot \rt} -  \Theta(\ut^2 -\rt^2)}{\rt^2} \right] \,,
    \label{eq:proof1}
\end{align}
where we used the definition of $\Pcal_{\ut}$ and we traded the function $f(\rt)$ by its Fourier transform $\tilde{f}(\kt)$ defined as 
\begin{align}
    \tilde{f}(\kt) = \int \der^2 \rt e^{-i \kt \cdot \rt} f(\rt) \,.
    \label{eq:FT1}
\end{align}
Now one can easily perform the integration over $\rt$
\begin{align}
    \int \frac{ \der^2 \rt}{2\pi}  \left[ \frac{e^{i \kt \cdot \rt} -  \Theta(\ut^2 -\rt^2)}{\rt^2} \right] & =  \int  \frac{\der^2 \rt}{2\pi} \left[ \frac{e^{i \kt \cdot \rt} -  \Theta(\ut^2 -\rt^2)}{\rt^2} \right] \nonumber \\
    & =   \int_0^\infty \frac{\der r_\perp}{r_\perp} \left[ J_0( k_\perp r_\perp) -  \Theta(u_\perp -r_\perp) \right] \nonumber \\
    & =  - \frac{1}{2} \ln \left(\frac{\ut^2 \kt^2}{c_0^2} \right) \,,
    \label{eq:proof2}
\end{align}
where in the last equality we used $\int_0^\infty \frac{\der z}{z} \left[ J_0(z) -  \Theta(1 -z) \right] = \ln(c_0)$ with $c_0 = 2 \exp(-\gamma_E)$.

Combining the results in Eqs.\eqref{eq:proof1} and \eqref{eq:proof2}, we find
\begin{align}
    \Pcal_{\ut}\left( \frac{1}{\rt^2} \right) [f] = - \frac{1}{4\pi} \int  \der^2 \kt \tilde{f}(\kt) \ln \left(\frac{\ut^2 \kt^2}{c_0^2} \right) \,.
\end{align}
Eq.\,\eqref{eq:identity1} is readily obtained by inserting Eq.\,\eqref{eq:FT1} into Eq.\,\eqref{eq:proof2}.

Next we prove the identity in Eq.\,\eqref{eq:identity2}. We re-express the identity in Eq.\,\eqref{eq:identity1} by rewriting the logarithm in the integrand:
\begin{align}
    &- (4\pi) \Pcal_{\ut}\left( \frac{1}{\rt^2} \right) [f] \nonumber \\
    & = \int \der^2 \rt f(\rt) \int \der^2 \kt e^{-i \kt \cdot \rt}  \left[ \ln \left(\frac{(\ut^2+\rt^2) \kt^2}{c_0^2} \right) + \ln \left(\frac{\ut^2}{(\ut^2+\rt^2)} \right) \right] \,.
\end{align}
One then observes that the second term does not contribute as the integral over $\kt$  results in a delta function $\delta^{(2)}(\rt)$ and then this term is proportional to $\ln(\ut^2/\ut^2) = 0$. Thus, we arrive at an alternative form of Eq.\,\eqref{eq:identity1}:
\begin{align}
    - (4\pi) \Pcal_{\ut}\left( \frac{1}{\rt^2} \right) [f] = \int \der^2 \rt f(\rt)  \int  \der^2 \kt e^{-i \kt \cdot \rt} \ln \left(\frac{(\ut^2+\rt^2) \kt^2}{c_0^2} \right) \,. 
\end{align}
The result in Eq.\,\eqref{eq:identity2} is obtained by letting $\ut \to 0$ in the equation above.

\section{Cancellation of final state emissions in fully inclusive DIS}
\label{app:final-state-cancellation}

For longitudinal photon, the total cross-section from these diagrams can be obtained by integrating over $\eta$ the expression given by Eq.\,\eqref{eq:NLO3-ktintegrated} (modulo a factor of $1/2$ since Eq.\,\eqref{eq:NLO3-ktintegrated} accounts for both quark and antiquark tagged jets)
\begin{align}
     &\left. \sigma^{\gamma_{\rm L}^{\star}+A\to X}\right|_{\rm V3\times LO^*+R2\times \overline{R2}^*}+c.c. =\frac{\alpha_{\mathrm{em}} e_f^2 N_c}{(2\pi)^2}\int_0^{1}\der z \ \left[8z^2(1-z)Q^2\right]\int\frac{\der^2\bt}{2\pi}\der^2\rxxtp\frac{\der^2\ryytp}{\ryytp^2} \nonumber\\
     &\times \frac{\alpha_s}{\pi}\int_{z-1}^{z}\frac{\der z_g}{z_g} K_0\left(\bar Q\left|(z-z_g)\rxxtp+(1-z+z_g)\ryytp\right|\right)K_0\left(\bar Q_{\rm V3}\left|z\rxxtp+(1-z)\ryytp\right|\right)\nonumber\\
     &\times(1-z+z_g)(z-z_g)^2\left[1+\frac{((1-z)\ryytp+(z-z_g)\rxxtp)\cdot(z\rxxtp+(1-z+z_g)\ryytp)}{(1-z)(1-z+z_g)\ryytp^2-z(z-z_g)\rxxtp^2}\right]\nonumber\\
     &\times 2\mathfrak{Re}\left[\Xi_{\rm NLO}(\xt,\yt,\xt',\yt')\right]\,,\label{eq:NLO3-inclusive}
\end{align}
where the transverse coordinates in the color structure are evaluated at the coordinates $\xt,\yt,\xt',\yt'$ are given by Eqs.\,\eqref{eq:bdep-1} in terms of the impact parameter $\bt$ and the transverse vectors $\rxxtp,\ryytp$.
In Eq.\,\eqref{eq:NLO3-inclusive}, we now perform the change of variable
\begin{align}
    \zeta&=z-z_g\,,\\
    \zeta_g&=-z_g\,.
\end{align}
We have then
\begin{equation}
    \int_0^1\der z\int_{-1}^1\der z_g\Theta(z-z_g)\Theta(z_g+1-z)=\int_0^1\der \zeta\int_{-1}^1\der \zeta_g\Theta(\zeta-\zeta_g)\Theta(\zeta_g+1-\zeta)\,,
\end{equation}
which shows that the longitudinal phase space integrals remains the same. After this change of variable, Eq.\,\eqref{eq:NLO3-inclusive} reads
\begin{align}
     &\left. \sigma^{\gamma_{\rm L}^{\star}+A\to X}\right|_{\rm V3\times LO^*+R2\times \overline{R2}^*}+c.c.=\frac{\alpha_{\mathrm{em}} e_f^2 N_c}{(2\pi)^2}\int_0^{1}\der \zeta \ \left[8\zeta ^2(1-\zeta)Q^2\right]\int\frac{\der^2\bt}{2\pi}\der^2\rxxtp\frac{\der^2\ryytp}{\ryytp^2} \nonumber\\
     &\times\frac{\alpha_s}{\pi} \int_{\zeta-1}^{\zeta}\frac{\der \zeta_g}{(-\zeta_g)} K_0\left(\bar Q\left|(\zeta-\zeta_g)\rxxtp+(1-\zeta+\zeta_g)\ryytp\right|\right)K_0\left(\bar Q_{\rm V3}\left|\zeta\rxxtp+(1-\zeta)\ryytp\right|\right)\nonumber\\
     &\times(1-\zeta+\zeta_g)(\zeta-\zeta_g)^2\left[1+\frac{((1-\zeta)\ryytp+(\zeta-\zeta_g)\rxxtp)\cdot(\zeta\rxxtp+(1-\zeta+\zeta_g)\ryytp)}{(1-\zeta)(1-\zeta+\zeta_g)\ryytp^2-\zeta(\zeta-\zeta_g)\rxxtp^2}\right]\nonumber\\
     &\times 2\mathfrak{Re}\left[\Xi_{\rm NLO,3}( \widetilde\xt, \widetilde\yt, \widetilde\xt', \widetilde\yt')\right]\,,\label{eq:NLO3-inclusive-changeofvariable}
\end{align}
with obviously, $\bar Q^2=\zeta(1-\zeta)Q^2$ and $\bar Q_{\rm V3}^2=(\zeta-\zeta_g)(1-\zeta+\zeta_g)Q^2$. Note the important minus sign in the denominator $\der\zeta_g/(-\zeta_g)$. The CGC correlator is evaluated at the transverse coordinate $\widetilde\xt,\widetilde\yt,\widetilde\xt',\widetilde\yt'$, which are functions of the new variables $\zeta,\zeta_g$:
\begin{align}
    \widetilde\xt&=\bt+(1-\zeta)[(\zeta-\zeta_g)\rxxtp+(1-\zeta+\zeta_g)\ryytp]\,, \nonumber \\
    \widetilde\yt&=\bt-\zeta[(\zeta-\zeta_g)\rxxtp+(1-\zeta+\zeta_g)\ryytp]\,, \nonumber \\
    \widetilde\xt'&=\bt+(1-\zeta+\zeta_g)[\zeta\rxxtp+(1-\zeta)\ryytp]\,, \nonumber \\
    \widetilde\yt'&=\bt-(\zeta-\zeta_g)[\zeta\rxxtp+(1-\zeta)\ryytp]\,.
\end{align}
It is clear that the transverse coordinates $\widetilde\xt,\widetilde\yt,\widetilde\xt'$ and $\widetilde\yt'$ are nothing but $\xt',\yt',\xt$ and $\yt$ expressed in terms of $\zeta$ and $\zeta_g$. Further, the color structure satisfies the identity
\begin{align}
    2\mathfrak{Re}\left[\Xi_{\rm NLO,3}(\xt,\yt,\xt',\yt')\right]=2\mathfrak{Re}\left[\Xi_{\rm NLO,3}(\xt',\yt',\xt,\yt)\right]\,,
\end{align}
which follows from $\Xi_{\rm NLO,3}(\xt',\yt',\xt,\yt)=\Xi_{\rm NLO,3}^*(\xt,\yt,\xt',\yt')$. Hence, one gets
\begin{align}
    \left. \sigma^{\gamma_{\rm L}^{\star}+A\to X}\right|_{\rm V3\times LO^*+R2\times \overline{R2}^*}+c.c.=-\left(\left. \sigma^{\gamma_{\rm L}^{\star}+A\to X}\right|_{\rm V3\times LO^*+R2\times \overline{R2}^*}+c.c.\right)\,,
\end{align}
which implies that the sum of the total cross-section associated with diagrams $\rm V3\times LO^*$, $\rm R2\times \overline{R2}^*$ and their complex conjugate is exactly zero.

We emphasize that it is crucial to use the proper definition of the impact parameter $\bt=(z-z_g)\xt+(1-z+z_g)\yt$ for diagrams $\rm V3\times LO^*$ and $\rm R2\times\overline{R2}^*$ to have the cancellation of these diagrams also differentially in impact parameter space, namely
\begin{align}
     \left. \frac{\der \sigma^{\gamma_{\rm L}^{\star}+A\to X}}{\der^2\bt}\right|_{\rm V3\times LO^*+R2\times \overline{R2}^*}+  \left. \frac{\der \sigma^{\gamma_{\rm L}^{\star}+A\to X}}{\der^2\bt}\right|_{\rm LO\times V3^*+\overline{R2}\times R2^*}= 0\,. \label{eq:opticalV3-bt-diff}
\end{align}
This cancellation is a requirement for the optical theorem to work differentially with the impact parameter $\bt$ (as it should), since a contribution coming from $\rm V3\times LO^*$ and $\rm R2\times\overline{R2}^*$ would contradict the results obtained in \cite{Beuf:2016wdz,Beuf:2017bpd,Hanninen:2017ddy} (indeed, these results does not depend on the quadrupole correlator for instance).

\bibliographystyle{utcaps}
\bibliography{SIDIS_NLO}

\end{document}